\documentclass{elsart}

\usepackage{epsfig}
\usepackage{subfigure}
\usepackage{wrapfig}

\usepackage{amsmath}
\usepackage{amssymb}
\usepackage{cite}
\usepackage{array}

\usepackage{textcomp}

\usepackage{setup} 

\usepackage{fancyhdr}

\numberwithin{equation}{section}
\numberwithin{figure}{section}

\newcommand{\beqs}{\begin{equation*}}
\newcommand{\beq}{\begin{equation}}

\newcommand{\eeqs}{\end{equation*}}
\newcommand{\eeq}{\end{equation}}

\newcommand{\beqas}{\begin{eqnarray*}}
\newcommand{\beqa}{\begin{eqnarray}}

\newcommand{\eeqas}{\end{eqnarray*}}
\newcommand{\eeqa}{\end{eqnarray}}

\newcommand{\blist}{\begin{itemize}}

\newcommand{\elist}{\end{itemize}}

\providecommand{\href}[2]{#2}

\newcommand{\oB}{\vert_{\partial M}}
\newcommand{\ptr}{{\rm tr}_V}
\newcommand{\Cur}{\Omega}
\newcommand{\iM}{\int_M d^nx\sqrt{g}}
\newcommand{\idM}{\int_{\partial M} d^{n-1}x\sqrt{h}}
\newcommand{\Dir}{\FMSlash{D}}
\newcommand{\dvxi}{v}

\begin{document}


\begin{frontmatter}
 
\title{Heat kernel expansion: user's manual}

\author[Leipzig,SPb]{D.V.\ Vassilevich}\ead{vassil@itp.uni-leipzig.de}

\address[Leipzig]{Institut f\"ur Theoretische Physik, Universit\"at Leipzig,
                 Augustusplatz 10, D-04109 Leipzig, Germany}
\address[SPb]{V.A.\ Fock Insitute of Physics, St. Petersburg University, 198904
St.\ Petersburg, Russia}

\begin{abstract}
The heat kernel expansion is a very convenient tool for studying
one-loop divergences, anomalies and various asymptotics of the
effective action. The aim of this report is to collect useful
information on the heat kernel coefficients scattered in
mathematical and physical literature. We present explicit
expressions for these coefficients on manifolds with and without
boundaries, subject to local and non-local boundary conditions, in
the presence of various types of singularities (e.g., domain
walls). In each case the heat kernel coefficients are given in
terms of several geometric invariants. These invariants are
derived for scalar and spinor theories with various interactions,
Yang-Mills fields, gravity, and open bosonic strings. We discuss
the relations between the heat kernel coefficients  and quantum
anomalies, corresponding anomalous actions, and covariant
perturbation expansions of the effective action (both "low-" and
"high-energy" ones).
\end{abstract}

\end{frontmatter}

\newpage
\thispagestyle{plain}
\tableofcontents


\newpage

\section{Introduction} \label{sec1}
It was noted by Fock \cite{Fock:1937} in 1937 that it is convenient
to represent Green functions as integrals over an auxiliary
coordinate (the so-called ``proper time'') of a kernel satisfying
the heat equation. Later on Schwinger \cite{Schwinger:1951} recognised 
that this representations makes many issues related to renormalization
and gauge invariance in external fields
more transparent. These two works introduced the
heat kernel to quantum theory. DeWitt made the heat kernel one of the
main tools of his manifestly covariant approach 
\cite{DeWitt:1965,Dewitt:1967yk,Dewitt:1967ub,Dewitt:1967uc}
to quantum field theory and quantum gravity which became dominant 
for many years. 

Heat kernel is a classical subject in mathematics 
\cite{Hadamard:1932}\footnote{A historical survey of 
the mathematical literature on the heat kernel expansion
can be found in \cite{Gilkey:1995}.}.
Asymptotics of the heat kernel are closely related to the eigenvalue
asymptotics  found by H.~Weyl \cite{Weyl:1912,Weyl:1915} 
and studied further in \cite{Mina,MinaPle}. The problem, as it was
formulated by Kac \cite{Kac:1966}, reads: ``Can one hear the shape of
a drum?''. In other words, this is the problem of recovering geometry
of a manifold from the spectrum of a natural differential operator.
Heat kernel coefficients proved very useful in this context.
On the other hand, the heat kernel is also an adequate tool
to study the index theorem of Atiyah and Singer 
\cite{Atiyah:1963a,Gilkey:1973,AtiyahBP:1973}.

By about 1990 the heat kernel expansion on manifolds without boundaries
or with boundaries and simplest local boundary conditions on them
was well understood. Also, the heat kernel became a standard tool
in calculations of the vacuum polarisation, the Casimir effect, and
in study of quantum anomalies. Later on, progress in theoretical
physics, especially in string theory and related areas, and parallel
developments in mathematics made this field highly specialised.
New results on non-standard boundary conditions (as, e.g., containing
tangential derivatives on the boundary or non-localities) or on
non-standard geometries (domain walls) were scattered in large
amounts of physical and mathematical literature. The aim of this
report is to present a unifying approach to the heat kernel expansion
and to supply the reader with a ``user-friendly'' guide to
the field. The main idea which we shall pursue is the universality of
the heat kernel. A single calculation (though, sometimes, quite
involved) may help in a large variety of applications regardless
of such details as spin, gauge group, etc. As well, just a single
universal object in fact describes counterterms, anomalies,
some asymptotics of the effective action, and much more.

To illustrate the use of the heat kernel in quantum field theory
let us
consider the generating functional for the Green functions of the 
field $\phi$ in the path integral representation
\begin{equation}
Z[J]=\int \mathcal{D} \phi \exp (-\mathcal{L}(\phi,J)) \ . \label{pathint}
\end{equation}
The heat kernel methods are almost exclusively
used for the one-loop calculations. In this approximation it is enough
to expand the action $\mathcal{L}$ to up to the quadratic order in quantum 
fluctuations $\phi$.
\begin{equation}
\mathcal{L}=\mathcal{L}_{\rm cl} + \langle \phi ,J \rangle +
\langle \phi, D\phi \rangle \,,\label{actquad}
\end{equation}
where $\mathcal{L}_{\rm cl}$ is the action on a classical background,
$\langle .,.\rangle$ denotes an inner product on the space of quantum
fields. Usually, this inner product is just an integral over the underlying
space. For real one-component scalar fields it reads
\begin{equation}
\langle \phi_1, \phi_2 \rangle =\int d^n x \sqrt{g} \phi_1(x)\phi_2(x)
\,.\label{inprosca}
\end{equation}
The linear term in (\ref{actquad}) contains in general contributions
from the external sources of the field $\phi$ and from the first 
variation of the classical action. If the background satisfies 
classical equation of motion for the field $\phi$, the latter part
of the linear term vanishes, though the former one (external sources)
should be kept arbitrary if one wishes to study correlation functions
of $\phi$. We stress that the background and quantum fields may be of
completely different nature. For example, it is a meaningful problem
to consider pure quantum scalar fields on the background of pure
classical gravity. $D$ is a differential operator. After a suitable 
number of integrations by part it is always possible to convert the
quadratic part of the action to the form given in (\ref{actquad}).
We postpone discussion of possible boundary terms to the next
sections. In the simplest case of quantum scalar field on the background
of a classical geometry
$D$ is a Laplacian with a mass term:
\begin{equation}
D=D_0:=-\nabla_\mu\nabla^\mu +m^2 \,.\label{simplestD}
\end{equation}
Just this simple example is enough to illustrate the material of this
section. Note, that in this case $J$ has no contribution from
the first variation of the classical action since gravity is not
quantised.

The path integral measure is defined by:
\begin{equation}
1=\int \mathcal{D} \phi \exp (-\langle\phi ,\phi \rangle) \ . \label{measure}
\end{equation}
Strictly speaking the right hand side of (\ref{measure}) is divergent.
The essence of the condition (\ref{measure}) is that
this divergence does not depend on external sources and on the
background geometry and, therefore, may be absorbed in an irrelevant
normalisation constant. 
The Gaussian integral (\ref{pathint}) can be evaluated giving
\begin{equation}
Z[J]=e^{-\mathcal{L}_{\rm cl}}
{\det }^{-\frac 12} (D) \exp \left( \frac 14 J D^{-1} J \right)
 \, . \label{pathdet}
\end{equation}

We stress that the equation (\ref{pathdet}) is true only if
the operator $D$ is self-adjoint. This means that $D$ is 
symmetric or formally self-adjoint with respect to the
scalar product $\langle .,.\rangle$:
\begin{equation}
\langle \phi_1,D\phi_2 \rangle =
\langle D\phi_1,\phi_2 \rangle \label{def-sym-op}
\end{equation}
for any $\phi_1$, $\phi_2$, and that the domains of definition
of $D$ and its adjoint coincide. We will not care about the
second requirement since it involves mathematical machinery
\cite{ReedSimon}
going beyond the scope of the present report. The first requirement
(\ref{def-sym-op}) poses important restrictions on admissible
boundary conditions.

To become convinced that (\ref{def-sym-op}) is really necessary,
one can calculate a finite-dimensional Gaussian integral
with 
\begin{equation}
D=\left( \begin{array}{cc} a & b \\ c& d \end{array} \right)
\end{equation}
and $a$, $b$, $c$ and $d$ being real constants first by completing the
squares in the exponent, and then compare the result with
$\det D$ (of course, one should remember a factor
of $\pi$). The two results coincide if $b=c$.

Let us return to the generating functional (\ref{pathdet}).
To analyse the two multiplies on the right hand side of
(\ref{pathdet}) which depend on the operator $D$ it is
convenient to introduce the heat kernel
\begin{equation}
K(t;x,y;D)=\langle x| \exp (-tD) | y\rangle \,.
\label{hkmatrel}
\end{equation}
This somewhat formal expression means that $K(t;x,y;D)$ should
satisfy the heat conduction equation 
\begin{equation}
(\partial_t+D_x) K(t;x,y;D)=0 \label{heateq}
\end{equation}
with the initial condition
\begin{equation}
K(0;x,y;D)=\delta (x,y)\,.\label{inicon}
\end{equation}

For $D=D_0$ (\ref{simplestD}) on a flat manifold $M=\mathbb{R}^n$ 
the heat kernel reads:
\begin{equation}
K(t;x,y;D_0)= (4\pi t)^{-n/2} \exp\left(
-\frac{ (x-y)^2}{4t} -tm^2 \right) \,.\label{simplestHK}
\end{equation}
The equations (\ref{heateq}) and (\ref{inicon}) can be checked
straightforwardly. Let us consider a more general operator $D$
which contains also a potential term or a gauge field. Then
(\ref{simplestHK}) still describes the leading singularity
 in the heat kernel as $t\to 0$. The subleading terms have a
form of the power-law corrections\footnote{
On a curved space even the leading term must be modified
(cf sec.\ \ref{s4iter}). On manifolds with boundaries also half-integer
powers of $t$ appear in the expansion, and, consequently,
$b_{2j+1}\ne 0$.}:
\begin{equation}
K(t;x,y;D)=K(t;x,y;D_0) \left( 1 +tb_2(x,y)+t^2b_4(x,y)+\dots
\right) \,.\label{sHKexp}
\end{equation}
The coefficients $b_k(x,y)$ are regular in the limit $y\to x$.
They are called the heat kernel coefficients\footnote{We shall
mostly use the coefficients $a_k$ which differ from $b_k$ by a
normalisation factor.}. At coinciding arguments $b_k(x,x)$ 
are local polynomials of background fields and their derivatives.

The propagator $D^{-1}(x,y)$ can be defined through the heat kernel
by the integral representation
\begin{equation}
D^{-1}(x,y)=\int_0^\infty dt\, K(t;x,y;D)\,,
\label{proptoHK}
\end{equation}
which follows from (\ref{hkmatrel}) if we suppose that the heat kernel
vanishes sufficiently fast as $t\to \infty$. We can formally integrate
the expansion (\ref{sHKexp}) to obtain
\begin{equation}
D^{-1}(x,y) \simeq 2(4\pi )^{-n/2} \sum_{j=0}
\left( \frac{ |x-y|}{2m} \right)^{-\frac 12 n+j+1}
K_{-\frac 12 n+j+1} (|x-y|m)b_{2j}(x,y) \,,\label{DBessel}
\end{equation} 
where $b_0=1$. By examining the behaviour of the Bessel function
$K_\nu (z)$ for small argument $z$ \cite{Gradshteyn:2000}
we conclude that the singularities in the propagator at
coinciding points are described by the first several heat kernel
coefficients $b_k$.

Let us consider the part of the generating functional (\ref{pathdet})
which contains $\det (D)$. The functional 
\begin{equation}
W=\frac 12 \ln \det (D) \label{defW}
\end{equation}
is called the one-loop effective action. It describes the quantum 
effects due to the background fields in the one-loop approximation
of quantum fields theory. To relate $W$ to the heat kernel we shall
use the arguments of \cite{Birrell:1982ix}. 

For each positive eigenvalue $\lambda$ of the operator $D$ we may write
an identity
\begin{equation}
\ln \lambda =-\int_0^\infty \frac {dt}t e^{-t\lambda}\,.
\label{lnlam}
\end{equation}
This identity is ``correct'' up to an infinite constant, which does
not depend on $\lambda$ and, therefore, may be ignored in what 
follows\footnote{To ``prove'' this statement one has to differentiate
both sides of eq.\ (\ref{lnlam}) with respect to $\lambda$.
}. Now we use $\ln\det (D)={\rm Tr}\ln (D)$ and extend
(\ref{lnlam}) to the whole operator $D$ to obtain 
\begin{equation}
W=-\frac 12 \int_0^\infty \frac {dt}t K(t,D) \,,\label{WKtD}
\end{equation}
where
\begin{equation}
K(t,D)={\rm Tr}\left( e^{-tD} \right)=\int d^nx \sqrt{g} K(t;x,x;D)
\,.\label{KtDdef}
\end{equation}

Here we have only presented some heuristic arguments in favour of
eq.\ (\ref{WKtD}). A more rigorous treatment of functional determinants
can be found in sec.\ \ref{s2spec}.

The integral in (\ref{WKtD}) may be divergent at both limits.
Divergences at $t=\infty$ are caused by zero or negative eigenvalues
of $D$. These are the infra red divergences. They will not be discussed
in this section. We simply suppose that the mass $m$ is sufficiently
large to make the integral (\ref{WKtD}) convergent at the upper limit.
Divergences at the lower limit cannot be removed in such a way.
Let us introduce a cut off at $t=\Lambda^{-2}$.
\begin{equation}
W_\Lambda = -\frac 12 \int_{\Lambda^{-2}}^\infty \frac {dt}t K(t,D)
\,.\label{WLamreg}
\end{equation}
It is now easy to calculate the part of $W_\Lambda$ which
diverges in the limit $\Lambda\to\infty$:
\begin{eqnarray}
&&W_\Lambda^{\rm div} =-(4\pi )^{-n/2} \int d^n x \sqrt{g} \left[
\sum_{2(j+l)<n} \Lambda^{n-2j-2l} b_{2j}(x,x)
\frac{(-m^2)^l l!}{n-2j-2l} \right. \nonumber\\
&&\qquad\qquad\qquad \left.+ \sum_{2(j+l)=n} \ln (\Lambda) 
(-m^2)^l l!\, b_{2j}(x,x) +\mathcal{O}(\Lambda^0) \right]\,.
\label{divWLam}
\end{eqnarray}
We see that the ultra violet divergences in the one-loop effective
action are defined by the heat kernel coefficients $b_{k}(x,x)$
with $k\le n$.

On non-compact manifolds the integral of $b_0(x,x)$ is divergent.
This divergence is removed by subtracting a ``reference'' heat
kernel (see sec.\ \ref{s6sinpot}).

Contributions from higher heat kernel coefficients $b_k$, $k>n$
to the effective action are not divergent and can be easily
calculated yielding in the limit $\Lambda\to\infty$
\begin{equation}
-\frac 12 (4\pi )^{-n/2} m^n \int d^nx \sqrt{g}
\sum_{2j>n} \frac{b_{2j}(x,x)}{m^{2j}} \Gamma (2j-n)\,.
\label{largemass}
\end{equation}
This is nothing else than the large mass expansion of the effective
action. This expansion is valid for relatively weak and slowly varying
background fields.

We have seen that the heat kernel expansion 
describes
\begin{itemize}
\item short-distance behaviour of the propagator;
\item one-loop divergences and counterterms;
\item $1/m$ expansion of the effective action.
\end{itemize}
We shall see below that heat kernel provides a natural framework
for studying 
\begin{itemize}
\item quantum anomalies (sec.\ \ref{sec7});
\item various perturbative expansions of the effective action
(sec. \ref{sec8});
\item selected non-perturbative relations for the effective
action (sec.\ \ref{sec9}).
\end{itemize}

Of course, in all these applications the heat kernel methods have
to compete with other techniques. The main advantage of the heat
kernel is that it delivers necessary information in terms of just
few geometric invariants. This method does not make distinctions 
between different spins, gauge groups, etc. Even dependence of the
space-time dimensionality is in most cases trivial. Therefore, on
one hand, just a single calculation serves then in many applications.
On the other hand, calculations in simple particular cases give
valuable information on the general structure of the heat kernel.
This property is especially useful when one deals with complicated
geometries (like in the presence of boundaries or singularities).
During the last decade, many models which lead to such complicated
geometries were very actively studied in theoretical physics.
The Dirichlet branes and the brane world scenario are the most 
popular but not the only examples.

We have to mention also the limitations of the heat kernel
formalism. It works less effectively in the presence of spinorial
background fields, i.e. when there is mixing between bosonic
and fermionic quantum fields. This problem is probably of the
technical nature, so that the corresponding formalism may be
developed some time in the future. A more serious drawback is
that the heat kernel expansion is not applicable beyond the
one-loop approximation. It is not clear whether necessary
generalisations to higher loop could be achieved at all.  

It is not possible to write a review paper on heat kernel which
would be complete in all respects, especially in the bibliography.
A more comprehensive treatment of many mathematical problems related
to the heat kernel expansion can be found in 
\cite{Gilkey:1995,Kirsten:2001wz}. The book by Kirsten \cite{Kirsten:2001wz}
considers also specific physical applications as the Casimir energy and
the Bose--Einstein condensation. The recent review paper 
\cite{Bordag:2001qi} is devoted to the Casimir effect
(see also \cite{Milton:2001yy}). The monographs
by Birrell and Davies \cite{Birrell:1982ix} and Fulling 
\cite{Fulling:1989nb} remain standard sources on quantum field theory
in curved space. Quantization of gauge theories is explained in
\cite{Faddeev:1980be}. 
In  \cite{Buchbinder:1992rb,Esposito:1997wt,Esposito:1997mw,
Avramidi:2000bm,Esposito:2001rx}
the heat kernel expansion is treated from the point of view of quantum
gravity and quantum cosmology. Useful information about properties of
the zeta function can be found in \cite{Elizalde:1995zk}.
One may also consult \cite{EORBZ}. The DeWitt approach to the heat kernel
and its' generalisations are described in \cite{Barvinsky:1985an}.
The path integral point of view on the heat kernel can be found
in \cite{Roepstorff:1994ga}.

Our primary goal is local formulae for the heat kernel coefficients.
Therefore, in some cases global aspects will be somewhat neglected.

This report is organised as follows. The next section contains
necessary preliminary information on spectral geometry and
differential geometry. There we define main geometric characteristics
of the manifold and of the boundary. We discuss the zeta function
(which defines the effective action) and the resolvent (which is
a generalisation of the propagator). We relate the asymptotics of these
functions to the heat kernel coefficients. Through this report we
work on Euclidean manifolds. A short remark on the
analytical continuation to the Lorentzian signature is given at the
end of section \ref{sec2}.
In  section \ref{sec3} we consider
the most widely used models of quantum field theory and open bosonic
strings. The one-loop dynamics in each of these models is defined by
a second order differential operator which depends on an effective
connection and on a matrix-valued potential. The connection and
the potential serve as a basis of an invariant description of all
that models in the language of spectral geometry. These quantities
are written down explicitly for each model. We also define suitable
boundary conditions. In section \ref{sec4} we consider the heat kernel
expansion on manifolds without boundary. We introduce a simple and
very powerful method and illustrate it by calculating several leading
terms in the heat kernel expansion. We also briefly discuss some other
methods and non-minimal operators. Section \ref{sec5} is devoted to 
manifolds with boundary. The heat kernel expansion for standard
Dirichlet, Neumann, and mixed boundary conditions is considered in some
detail. We also describe less known oblique and spectral boundary   
conditions (these are the ones which contain tangential derivatives
on the boundary or non-local projectors). We discuss loss of the
so-called strong ellipticity for oblique boundary conditions which
corresponds to the critical value of the electric field in string 
physics. With the example of spectral boundary conditions we
illustrate appearance of non-standard ($\ln t$) asymptotics in the
heat kernel. The results of sections \ref{sec4} and \ref{sec5}
are valid on smooth manifolds with smooth potentials and gauge field.
In section \ref{sec6} we consider the case when either the 
background fields or the manifold itself have singularities.
In particular, conical and domain wall singularities are considered.

Sections \ref{sec7} - \ref{sec9} are devoted to applications.
In section \ref{sec7} we relate conformal and chiral anomalies
to certain heat kernel coefficients and re-derive the anomalies
in several particular models. This is a very spectacular but also 
rather well known application of the heat kernel expansion.
In section \ref{sec8} we go beyond
the power series in the proper time $t$. We consider mainly two
particular cases. The first one is the case when derivatives are more important
than the potentials (so that we sum up leading non-localities).
The second one is the so-called low energy expansion which neglects
derivatives of the background fields starting with certain order
(but treats all powers of background potentials and curvatures 
exactly). In that section we also review some results on the heat
kernel on homogeneous spaces where spectrum of relevant operators
may be found exactly. In section \ref{sec9} we consider two examples
when the heat kernel can be used to obtain exact results for the
effective action. The first one is the famous Polyakov action.
The second example is exact relations between the effective actions
in dual theories. In this section (in contrast to section \ref{sec8})
we don't have to neglect derivatives or powers of
the background fields. Section \ref{sec10} contains concluding
remarks.


\section{Spectral functions: heat kernel, zeta function, resolvent}
\label{sec2}
\subsection{Differential geometry and the operators of Laplace type}
\label{s2geo} 
Let $M$ be a smooth compact Riemannian manifold\footnote{This simply
means that we assume that there is a positive definite metric
tensor $g_{\mu\nu}$ on $M$.}
of dimension $n$ with smooth boundary $\partial M$.
We shall also consider the case when the boundary
$\partial M$ is empty. Let $V$ be a vector 
bundle\footnote{Many physicists strongly dislike vector bundles.
Nevertheless, there are two good reasons for using the fibre
bundles in this paper (in parallel with more familiar notations of matrix-valued
functions, gauge fields, etc). 
First, our simplifying comments and
examples may help the reader to understand mathematical
literature on the subject. Second, one of the main ideas of this
report is to reveal some universal structures behind the heat
kernel expansion. In particular, we shall see that there is not
much difference between different spins and symmetry groups.
The vector bundle language seems to be the most adequate language
to achieve this goal. The reader may consult the excellent review paper 
by Eguchi, Gilkey and Hanson \cite{Eguchi:1980jx}.
} over $M$. This means that there is a vector space
attached to each point of the manifold. For example, this
could be a representation space of a gauge group or of the
space-time symmetry group.
Sections of $V$ are smooth functions bearing
a discrete index which could correspond to internal
or spin degrees of freedom. 

We study differential
operators on $V$. We restrict ourselves to second order
operators of the Laplace type. Locally such operators can
be represented as:
\begin{equation}
D=-(g^{\mu\nu}\partial_\mu\partial_\nu 
+a^\sigma\partial_\sigma +b) \label{D1}
\end{equation}
where $g^{\mu\nu}$ is the inverse metric tensor on $M$;
$a^\sigma$ and $b$ are matrix valued functions on $M$.
There is a unique connection on $V$ and a unique 
endomorphism $E$ of $V$ (another matrix valued function)
so that
\begin{equation}D=-(g^{\mu\nu}\nabla_\mu\nabla_\nu +E)\, ,
\label{laplaceb}\end{equation}
where the covariant derivative $\nabla =\nabla^{[R]} +\omega$
contains both Riemann $\nabla^{[R]}$ and ``gauge'' (bundle) $\omega$
parts.
We may express:
\begin{eqnarray}
    &&\omega_\delta=\frac 12 g_{\nu\delta}(a^\nu+g^{\mu\sigma}\Gamma_{\mu
    \sigma}{}^{\nu }I_{V}) \quad{\rm and}\label{oab}\\
    &&E=b-g^{\nu \mu}(
    \partial_\mu\omega_\nu+\omega_\nu\omega_\mu
     -\omega_\sigma\Gamma_{\nu\mu}{}^\sigma)\,, \label{Eab}
\end{eqnarray}
where
\begin{equation}
\Gamma_{\nu\mu}{}^\sigma =g^{\sigma\rho} \frac 12 \left(
\partial_\mu g_{\nu\rho}+\partial_\nu g_{\mu\rho} -
\partial_\rho g_{\mu\nu} \right) \label{Christ}
\end{equation}
is the Christoffel symbol. 
$I_{V}$ is the unit operator on $V$. 

It is important to understand how general the notion of the Laplace
type operator is. The most obvious restriction on $D$ is that it is
of second order, i.e. it contains second derivatives, but does not
contain higher derivative parts. In this paper we shall also consider
first order operators (Dirac operators, for example). Second, the
operator $D$ is a {\it partial differential operator}. This excludes
negative or fractional powers of the derivatives. The operators containing
such structures are called {\it pseudo-differential operators}.
More information on spectral theory for these operators can be found
in \cite{Seeley:1968,Grubb:1996,Grubb:2002}. 
Third, the operator (\ref{D1}) has a
scalar principal part. This means that the second derivatives in
(\ref{D1}) are contracted with the metric, and the internal index structure
of the second derivative term is trivial. Such operators are also
called {\it minimal}. Non-minimal operators will be briefly
considered in sec\ \ref{s4nonmin}.

We can define local invariants associated with $\omega$
and $g$. Let 
\begin{equation}
{R^\mu}_{\nu\rho\sigma}=\partial_\sigma \Gamma_{\nu\rho}^\mu -
\partial_\rho \Gamma_{\nu\sigma}^\mu +
\Gamma_{\nu\rho}^\lambda \Gamma_{\lambda\sigma}^\mu -
\Gamma_{\nu\sigma}^\lambda \Gamma_{\lambda\rho}^\mu 
\label{Riemann}
\end{equation}
be the Riemann curvature tensor, let $R_{\mu\nu}:={R^\sigma}_{\mu\nu\sigma}$
be the Ricci tensor, and let $R:=R^\mu_\mu$ be the scalar
curvature. 
With our sign
conventions, $R =2$ on the unit sphere $S^2$ in the Euclidean
space. 
Let
Roman indices $i$, $j$, $k$, and $l$ range from 1 through
the dimension $n$ of the manifold and index a local orthonormal frame
(vielbein) $\{ e_1,...,e_n\}$ for the tangent space of the manifold.
In components we have: $e_j^\mu e_k^\nu g_{\mu\nu}=\delta_{jk}$,
$e_j^\mu e_k^\nu \delta^{jk}=g^{\mu\nu}$. The inverse vielbein
is defined by the relation $e_\mu^j e^\mu_k=\delta_k^j$. These two
objects, $e_\mu^k$ and $e^\nu_j$, will be used to transform 
``curved'' indices ($\mu$, $\nu$, $\rho$) to ``flat'' ones
($i$, $j$, $k$) and back. In Euclidean space there is no distinction
between upper and lower flat indices.

As usual, the Riemann part of the covariant derivatives contains the
Christoffel connection so that
\begin{equation}
\nabla_\mu^{[R]}v_\nu = \partial_\mu v_\nu -\Gamma_{\mu\nu}^\rho v_\rho
\label{riemcov}
\end{equation}
for an arbitrary vector $v_\nu$. To extend this derivative to the objects
with flat indices one has to use the spin-connection $\sigma_\mu$:
\begin{equation}
\nabla_\mu v^j = \partial_\mu v^j +\sigma_\mu^{jk}v_k \,.\label{spinder}
\end{equation}
The condition $\nabla_\mu e_\nu^k=0$ yields
\begin{equation}
\sigma_\mu^{kl}=e_l^\nu \Gamma_{\mu\nu}^\rho e_\rho^k -
e^\nu_l \partial_\mu e_\nu^k \,.\label{spincon}
\end{equation} 

Let $\Omega_{\mu\nu}$ be the field strength of the connection
$\omega$:
\begin{equation}
\Omega_{\mu\nu}=\partial_\mu\omega_\nu - \partial_\nu\omega_\mu
+\omega_\mu\omega_\nu -\omega_\nu\omega_\mu \, .
\label{Omega}
\end{equation}
The covariant derivative $\nabla$ acts on both space-time and internal
indices. For example,
\begin{equation}
\nabla_\rho \Omega_{\mu\nu}=\partial_\rho \Omega_{\mu\nu}
-\Gamma_{\rho\mu}^\sigma \Omega_{\sigma\nu}
-\Gamma_{\rho\nu}^\sigma \Omega_{\mu\sigma}
+[\omega_\rho ,\Omega_{\mu\nu}] \,.
\end{equation}

If the boundary $\partial M$ is non-empty we have more invariants.
Let $e_n$ be inward pointing unit vector field. Let Roman indices $a$, $b$,
$c$ and $d$ range from 1 to $n-1$ and index a local orthonormal frame
for the tangent bundle of $\partial M$. Let $L_{ab}:=\Gamma^n_{ab}$
be the second fundamental form (extrinsic curvature) of the boundary.
We use the Levi-Civita (spin) connections and
the connection $\omega$ to covariantly differentiate tensors of all types. Let
\lq ;\rq\ denote multiple covariant differentiation with respect to 
the Levi-Civita connection of $M$ and let
\lq:\rq\ denote multiple tangential covariant differentiation 
on the boundary with respect to the
Levi-Civita connection of the boundary; the difference
 between `;' and `:' is measured by the second
fundamental form. Thus, for example, $E_{;a}=E_{:a}$ since 
there are no tangential indices in $E$ to be
differentiated. On the other hand, 
$E_{;ab}\ne E_{:ab}$ since the index $a$ is also being
differentiated. More precisely $E_{;ab}=E_{:ab}-L_{ab}E_{;n}$.
Since $L$ is only defined on the boundary, this
tensor can only be differentiated tangentially.

Consider an example of the circle $S^1$ in the plane $\mathbb{R}^2$.
The line element has the form
\begin{equation}
(ds)^2 = (dr)^2 +r^2 (d\theta)^2 \,,\label{lineS2}
\end{equation}
where $0\le r<\infty$, $0\le \theta <2\pi$. Then $g_{\mu\nu}=
{\rm diag} (1,r^2)$. $S^1$ is defined by the condition $r=r_0$.
We may choose $e_\theta^1=r$, $e_r^2=-1$
(the minus sign appears because $e^2$ is an {\it inward}
pointing unit vector for the disk with the boundary $S^1$).
Then the second fundamental form of $S^1$ is
\begin{equation}
L_{11}=\left.
e_r^2 e_1^\theta e_1^\theta \Gamma_{\theta\theta}^r\right|_{r=r_0}=
\frac 1{r_0} \,.\label{LonS1}
\end{equation}
In general, on $S^{n-1}$ considered as a boundary of the ball in
$\mathbb{R}^n$ the extrinsic curvature is $L_{ab}=\frac 1{r_0} \delta_{ab}$.

If the boundary $\partial M$ is non-empty, one has to
define boundary conditions for the field $\phi$. A convenient
way to write them down is:
\begin{equation}
\mathcal{B}\,\phi =0 \label{boucon}
\end{equation}
where $\mathcal{B}$ is called the boundary operator. 
In general, the operator $\mathcal{B}$ calculates a linear combination
of the boundary data for any given function $\phi$. If $D$ is of Laplace
type, the boundary data include value of the function at the
boundary and value of it's first normal derivative.
The most frequently used
choices are the Dirichlet and Neumann boundary operators
which we denote $\mathcal{B}^-$ and $\mathcal{B}^+$ respectively:
\begin{eqnarray}
&&\mathcal{B}^-\phi =\phi\vert_{\partial M} \, ,\label{Dirbc} \\
&&\mathcal{B}^+\phi =(\phi_{;n}+S\phi )\vert_{\partial M} \, ,
\label{Neubc}
\end{eqnarray}
where $S$ is a matrix valued function defined on $\partial M$.
The boundary conditions (\ref{Neubc}) are also
called Robin or generalised Neumann. In some literature they
are called mixed boundary conditions. We shall not use this latter
terminology. The name ``mixed'' is reserved for another type
of boundary conditions. Let $\Pi^-$ and $\Pi^+$ be two
complementary projectors defined on $V\oB$, $(\Pi^\pm)^2=\Pi^\pm$,
$\Pi^++\Pi^-=I$.
There is a decomposition
$ V|_{\partial M}=V_N\oplus V_D$, where $V_{N,D}
=\Pi^\pm V|_{\partial M}$. Decompose also 
$\phi=\phi^N\oplus\phi^D$ and set
\begin{equation}
\mathcal{B}\phi:=\phi^D\oplus(\phi^N_{;n}+S\phi^N)
\oB \, .
\label{mixbc}
\end{equation}
The matrix valued function (endomorphism) $S$ acts on $V_N$ only, 
$S=\Pi^+S=S\Pi^+$. In other words, we define Dirichlet boundary
conditions on $V_D$ and generalised Neumann boundary conditions on $V_N$.
For obvious reason the boundary conditions (\ref{mixbc})
will be called mixed. In  sec. \ref{sec3} we shall see that natural
boundary conditions for spinor and vector fields are of this type.
\subsection{Spectral functions}\label{s2spec}
For the boundary conditions considered in this section as well
as on manifolds without a boundary the operator 
$\exp (-tD)$ with positive $t$ is trace class on the space
of square integrable functions $L^2(V)$. This means that for
an auxiliary smooth function $f$ on $M$ 
\begin{equation}
K(t,f,D)={\rm Tr}_{L^2}(f\exp (-tD)) 
\label{heatTr}
\end{equation}
is well defined. We also write
\begin{equation}
K(t,f,D)=\int_M d^nx \sqrt{g} \ptr K(t;x,x;D) f(x) \,,\label{KxxD}
\end{equation}
where $K(t;x,x;D)$ is an $y\to x$ limit of the fundamental
solution
$K(t;x,y;D)$ of the heat equation (\ref{heateq})
with the initial condition (\ref{inicon}).
If there is a boundary, the kernel $K(t;x,y;D)$ should also
satisfy some boundary conditions $\mathcal{B}_xK(t;x,y;D)=0$
in one of the arguments. We stress that $K(t;x,y;D)$ is a matrix
in the internal indices. $\ptr$ denotes the trace over these indices.

Let $D$ be self-adjoint. This implies that in a suitable basis in the
internal space the matrix $\omega_\mu$ is anti-hermitian and $E$ is
hermitian. Let $\{ \phi_\lambda \}$ be a complete basis of orthonormal
eigenfunctions of the operator $D$ corresponding to the eigenvalues
$\{\lambda\}$. Then
\begin{equation}
K(t;x,y;D)=\sum_\lambda \phi^\dag_\lambda (x)\phi_\lambda (y)
e^{-t\lambda} \,.\label{hkeigen}
\end{equation}

We shall almost exclusively work either on manifolds without boundaries, or
on manifolds with boundaries with the fields subject to local boundary
conditions (\ref{Dirbc}), (\ref{Neubc}), or (\ref{mixbc}). In all these
cases there is
an asymptotic expansion as $t\downarrow 0$ 
\cite{Seeley:1966,Seeley:1969,Greiner:1971}\footnote{These papers contain
also a method allowing to calculate the coefficients of the expansion.
The method is, however, too complicated to use it on practice.
}:
\begin{equation}
{\rm Tr}_{L^2} (f\exp (-tD)) 
\cong\sum_{k\ge0}t^{(k-n)/2}a_k(f,D)\,.
\label{asymptotex}
\end{equation}
This expansion is valid for almost all boundary conditions appearing
in applications to physics. There are, however, some exceptions which
will be discussed in sec.\ \ref{s5other}.

The coefficients $a_k$ and the coefficients $b_k$ introduced in the
previous section (cf. (\ref{sHKexp})) are related by the equation
($m=0$ for the simplicity):
\begin{equation}
a_k(f,D)=(4\pi )^{-n/2} \iM b_k(x,x)f(x) \,.\label{akbk}
\end{equation}
Note, that the definition (\ref{sHKexp}) is valid on flat manifolds
without boundary only (although generalisations to other cases
are possible).

The key property of the heat kernel coefficients $a_k$ is that
they are locally computable in most of the cases. This means that
they can be expressed in term of the volume and boundary integrals
of local invariants. 

For a positive operator $D$ one can define the zeta function
\cite{Mina,Seeley:1968}
by the equation:
\begin{equation}
\zeta (s,f,D)= {\rm Tr}_{L^2} (f\, D^{-s}) \,.\label{defzeta}
\end{equation}
The zeta function is related to the heat kernel by the integral
transformation
\begin{equation}
\zeta (s,f,D)=\Gamma (s)^{-1}\int\limits_0^\infty dt\, t^{s-1}
K(t,f,D) \,.\label{zetaheat}
\end{equation}
This relation can be inverted,
\begin{equation}
K(t,f,D)=\frac 1{2\pi i}\oint ds\, t^{-s}\Gamma(s)\zeta (s,f,D)\,,
\label{heatzeta}
\end{equation}
where the integration contour encircles all poles of the integrand.
Residues at the poles can be related to the heat kernel coefficients:
\begin{equation}
a_k(f,D)={\rm Res}_{s=(n-k)/2} \left( \Gamma(s)\zeta (s,f,D) \right)\,.
\label{aRes}
\end{equation}
In particular,
\begin{equation}
a_n(f,D)= \zeta (0,f,D) \,.\label{anz0}
\end{equation}

Zeta-functions can be used to regularize the effective action
\cite{Dowker:1976tf,Hawking:1977ja}. The regularization is achieved
by shifting the power of $t$ in (\ref{WKtD}):
\begin{equation}
W_s=-\frac 12 \tilde\mu^{2s}\int_0^\infty \frac {dt}{t^{1-s}} 
K(t,D) \,,\label{Wzreg}
\end{equation}
where $\tilde\mu$ is a constant of the dimension of mass introduced to keep
proper dimension of the effective action. The regularization is removed
in the limit $s\to 0$. Eq.\ (\ref{Wzreg}) can be considered as a
definition of the regularized effective action without any reference to (\ref{WKtD}).
One can also rewrite (\ref{Wzreg}) in terms of the zeta function:
\begin{equation}
W_s=-\frac 12 \tilde\mu^{2s} \Gamma(s)\zeta (s,D) \,,
\label{WzGam}
\end{equation}  
where $\zeta (s,D):=\zeta (s,1,D)$.

The gamma function has a simple pole at $s=0$:
\begin{equation}
\Gamma (s)=\frac 1s -\gamma_E +\mathcal{O}(s)\,,\label{Gammas}
\end{equation}
where $\gamma_E$ is the Euler constant. The regularized effective action
(\ref{WzGam}) has also a pole at $s=0$:
\begin{equation}
W_s=-\frac 12 \left( \frac 1s -\gamma_E +\ln \tilde\mu^2 \right)
\zeta (0,D) - \frac 12 \zeta'(0,D) \,.\label{Wzpole}
\end{equation}
According to (\ref{anz0}) the divergent term in the zeta function 
regularization is proportional to $a_n(D)$ (cf. (\ref{divWLam}) for another
regularization scheme). The pole term in (\ref{Wzpole}) has to be removed
by the renormalization. The remaining part of $W_s$ at $s=0$ is the 
renormalised effective action:
\begin{equation}
W^{\rm ren}= -\frac 12 \zeta'(0,D)-\frac 12 \ln (\mu^2 )
\zeta (0,D) \,,\label{Wzren}
\end{equation}
where we have introduced a rescaled parameter 
$\mu^2=e^{-\gamma_E}\tilde\mu^2$. In this approach $\mu^2$ describes the
renormalization ambiguity which must be fixed by a suitable
normalisation condition. Let us remind that here we are working on a
compact manifold. On non-compact manifolds $\zeta (s,D)$ may have divergent
contributions proportional to the volume. Such divergences are usually
removed by the subtraction of a ``reference'' heat kernel (see sec.\
\ref{s6sinpot}).

Together with (\ref{defW}) equation (\ref{Wzren}) yields a definition
of the functional determinant for a positive elliptic second order
operator which is frequently used in mathematics:
\begin{equation}
\ln \det (D)=-\zeta'(0,D) -\ln (\mu^2) \zeta (0,D) \,.\label{detzeta}
\end{equation}

Note, that the definitions (\ref{defzeta}), (\ref{detzeta})
are valid for positive operators
only. Elliptic 2nd order differential operators have at most finite number
of zero and negative modes\footnote{On manifolds with boundaries one has to
require {\it strong ellipticity} of the boundary value problem (cf.\
sec.\ \ref{s5other}).} which must be treated separately. However, one 
can extend the definition of the zeta function to operators with negative
modes:
\begin{equation}
\zeta (s,D)=\sum |\lambda |^{-s} \,,\label{zetaneg}
\end{equation}
where the sum extends over all non-zero eigenvalues $\lambda$.
One can also define another spectral function in a similar way:
\begin{equation}
\eta (s,D)=\sum {\rm sign}\, (\lambda ) |\lambda |^{-s} \,.\label{etadef}
\end{equation}
This function is especially useful in spectral theory of Dirac type
operators where $\eta (0,\Dir )$ measures asymmetry of the spectrum.

Another function which is frequently used especially in the mathematical
literature is the {\it resolvent} (or, more precisely, its' powers):
\begin{equation}
{\rm R}^l(z):=(D+z^2)^{-l} \,.\label{resolvent}
\end{equation}
If $D$ is on operator of Laplace type subject to ``good''
boundary conditions\footnote{A more precise meaning of this 
restriction will be explained in sec. \ref{s5other}.}
 and if $l$ sufficiently large ($l>n/2$)
there is a full asymptotic expansion
\begin{equation}
{\rm Tr} ({\rm R}^l(z))=\sum_k 
\frac{\Gamma \left( l+\frac{k-n}2 \right)}{\Gamma (l)}
a_k(D) z^{-2l+n-k} \label{asresol}
\end{equation} 
as $z\to \infty$. The coefficients $a_k$ are the same as in the
heat kernel expansion (\ref{asymptotex}).
\subsection{Lorentzian signature}\label{s2Lorentz}
Locality of the heat kernel coefficients in the Euclidean domain
can be easily understood by examining the free heat kernel
(\ref{simplestHK}). For small $t$ the first term in the exponential
strongly suppresses non-local contributions. For Lorentzian 
metrics the squared distance function $(x-y)^2$ is no longer
positive definite. Therefore, the simple arguments given above
do not work. A partial solution to this problem is to consider
the ``Schr\"{o}dinger'' equation
\begin{equation}
\left( i\partial_\tau -D \right) \tilde K(\tau;x,y;D)=0
\label{Schr}
\end{equation}
for the kernel $\tilde K$ instead of the heat conduction
equation (\ref{heateq}). Then $\tilde K$ oscillates at large
distances. However, even though non-local contributions
to $\tilde K$ oscillate furiously as $\tau\to 0$ they are
not small. Consequently, local asymptotic series do not
exist in many cases. A discussion on this point
can be found in chapter 9 of \cite{Fulling:1989nb}.

Of course, the heat kernel expansion can be used also
on Lorentzian manifolds at least for the renormalization
theory where non-local terms are of less importance.
Counterterms are still defined by the same heat kernel
coefficients with the same functional dependence on
the Lorentzian metric. Explicit definitions with the imaginary
``proper time'' $\tau$ can be found in 
\cite{DeWitt:1965,Birrell:1982ix}. One should note that some
background fields receive an imaginary phase when being
continued to the Euclidean domain (cf. secs.\ \ref{s3string}
and \ref{s3spin}).


\section{Relevant operators and boundary conditions}\label{sec3}
The operator of Laplace type is not necessarily the scalar
Laplacian. In fact, in almost all models of quantum field
theory the one loop effective action is defined by an operator
of this type. This can be demonstrated by bringing relevant
operators to the canonical form (\ref{laplaceb}). 
In this section we give explicit
construction of the connection $\omega$ and the endomorphism
(matrix-valued potential)
$E$ for scalar, spinor, vector and graviton fields. We also
describe appropriate boundary conditions.

\subsection{Scalar fields}\label{s3scalar}
Consider first the example of the multi-component real scalar
field $\Phi^A$ in $n$ dimensions. The action reads
\begin{equation}
\mathcal{L}
=\int_M d^nx\, \sqrt{g} (g^{\mu\nu}\nabla_\mu\Phi^A\nabla_\nu\Phi^A
+U(\Phi ) +\xi R \Phi^A\Phi^A ) \, ,\label{scalaraction}
\end{equation}
where $\xi$ is the conformal coupling parameter, $U$ is a potential.
The covariant derivative $\nabla_\mu$ contains the background
gauge field $G_\mu^{AB}$: $\nabla_\mu\Phi^A=\partial_\mu\Phi^A+
G_\mu^{AB}\Phi^B$. $G_\mu$ is antisymmetric in internal
indices $A,\,B$.
To evaluate the one-loop effective action one should expand 
(\ref{scalaraction}) around a
background field $\bar \Phi$, $\Phi=\bar \Phi +\phi$, and keep the
terms quadratic in fluctuations:
\begin{eqnarray}
&&\mathcal{L}_2
=\int_M d^nx \sqrt{g} \phi^A\left(-(\nabla^\mu\nabla_\mu )^{AB}
+\frac 12(U(\bar\Phi )'')^{AB} +\xi R \delta^{AB} \right)
\phi^B \nonumber \\
&&\qquad -\int_{\partial M}d^{n-1}x\,\sqrt{h} \phi^A \nabla_n \phi^A \,,
\label{S2scalar}
\end{eqnarray}
$h$ is the determinant of the induced metric on the boundary.
The inner product for quantum fields $\phi$ reads
\begin{equation}
<\phi_1,\phi_2>=\int_M d^nx\, \sqrt{g} \phi_1^A \phi_2^A \ .
\label{inn-prod-scal}
\end{equation}

The operator $D$ is defined by the bulk part of the action
(\ref{S2scalar}):
\begin{equation}
D^{AB}:=-(\nabla^\mu\nabla_\mu )^{AB}
+\frac 12(U(\bar\Phi )'')^{AB} +\xi R \delta^{AB}\,.
\label{opDscalar}
\end{equation}
For a special choice of the parameter $\xi$:
\begin{equation}
\xi =\frac{n-2}{4(n-1)} \label{confxi}
\end{equation}
the operator $D$ (\ref{opDscalar}) is conformally covariant
(if also $U(\bar\Phi )''=0$).
To bring the operator (\ref{opDscalar}) to the canonical form
(\ref{laplaceb}) we introduce
\begin{equation}
\omega_\mu^{AB}=G_\mu^{AB}\, ,\qquad
E^{AB}=-\frac 12(U(\bar\Phi )'')^{AB} -\xi R \delta^{AB}\,.
\label{omEscal}
\end{equation}
For this case, $\Omega_{\mu\nu}$ is just the ordinary Yang-Mills
field strength.

The operator (\ref{opDscalar}) is symmetric with respect to the
inner product (\ref{inn-prod-scal})
if  the surface integral
\begin{equation}
\int_{\partial M} d^{n-1}x\,\sqrt{h} (\phi_1^A\nabla_n\phi_2^A -
\phi_2^A\nabla_n\phi_1^A ) \label{symm-cond-scal}
\end{equation}
vanishes for arbitrary $\phi_1$ and $\phi_2$ belonging to its' 
domain of definition. 
This may be achieved if one imposes either
Dirichlet 
\begin{equation}
\phi^A\oB =0 \label{Dirscal}
\end{equation}
or modified Neumann 
\begin{equation}
(\nabla_n\phi^A +S^{AB}\phi^B)\oB =0 \label{neuscal}
\end{equation}
boundary conditions. $S^{AB}$ is an arbitrary symmetric matrix.
Note, that the integral (\ref{symm-cond-scal}) vanishes also if
$S^{AB}$ is an arbitrary symmetric (differential) operator
on the boundary. 

For the Dirichlet conditions (\ref{Dirscal}) the boundary term
in (\ref{S2scalar}) vanishes automatically. To ensure absence
of the surface term for the modified Neumann conditions 
(\ref{neuscal}) one should add to (\ref{scalaraction}) an appropriate
surface action.
\subsection{Bosonic string}\label{s3string}
Our next example is the non-linear sigma model in two dimensions
described by the action
\begin{eqnarray}&&
\mathcal{L}^{[\sigma]}=\int_M d^2 x\left(
\sqrt{g} G_{AB}(X) g^{\mu\nu}
\partial_\mu X^A \partial_\nu X^B +
\epsilon^{\mu\nu} B_{AB}(X) \partial_\mu X^A \partial_\nu X^B \right)
\nonumber\\
&&\qquad\qquad +\int_{\partial M} A_BdX^B\,.
\label{actsigma}
\end{eqnarray}
From the point of view of two-dimensional world sheet the fields
$X^A(x)$ are scalars. In string theory they are interpreted as
coordinates on a $d$-dimensional target manifold with the metric $G_{AB}(X)$.
$\epsilon^{\mu\nu}$ is the Levi-Civita tensor density, 
$\epsilon^{12}=-\epsilon^{21}=1$.
$B_{AB}(X)$ is an antisymmetric tensor field on the target space.
$A_B(X)$ is the electromagnetic vector potential. The action 
(\ref{actsigma}) describes charged open stings. For simplicity,
we absorb the inverse string tension $\alpha'$ into a field redefinition.
We do not include tachyon and dilaton couplings in the bulk or on the
boundary.

Usually the term with the $B$-field gets an imaginary coefficient
in the Euclidean space. Since the physical space-time has Minkowski
signature it is not especially significant which way of continuation
to the Euclidean space has been chosen provided the results are
properly continued back to Minkowski space after the calculations.
As we will see below, real coefficient in front of the $B$-field leads
to a well-defined spectral problem. This situation is in close
analogy with the continuation rules for the axial vector
field in the spinor determinant \cite{Andrianov:1984qj,Andrianov:1984fg}.
The other way to deal with the field $B$ is to keep the coefficient
of the $B$-term in (\ref{actsigma}) imaginary at the expense
of introducing a more sophisticated conjugation operation 
\cite{Osborn:1991gm} containing the sign-reversion of the
$B$-field. The same refers to the electromagnetic potential $A_B$.

The field $X$ enters the action (\ref{actsigma}) at many places making
the background field expansion a quite cumbersome procedure. The most
economic way to arrange such an expansion and to calculate higher
derivatives of the action (\ref{actsigma}) is to introduce the geodesic
coordinates in the target space. A detailed explanation of the method 
as well as further references can be found in \cite{Braaten:1985is}.
Consider the target space geodesics defined
independently at each point of the two-dimensional world surface
and parametrised by the arc length $s$ in the target space. They satisfy
the usual geodesic equation
\begin{equation}
\frac{d^2}{ds^2} X^A(x,s) +{\gamma}_{BC}^A(X)
\frac{d}{ds} X^B(x,s) \frac{d}{ds} X^C(x,s)=0 \,,
\label{geod}
\end{equation}
where ${\gamma}_{BC}^A(X)$ is the Christoffel connection
corresponding to the target space metric $G_{AB}$.
Let us supplement the equation (\ref{geod}) by the initial
conditions
\begin{equation}
X^A(x,0)=\bar X^A(x)\,, \qquad \frac{d}{ds} X^A(x,0)=
\xi^A(x)\,, \label{incondgeo}
\end{equation}
where $\bar X$ is the background field. $\xi^A$ parametrises
deviations from $\bar X$ and, therefore, can be identified with
quantum fluctuations. The
$k$-th order term of the expansion of the action
(\ref{actsigma}) around the background $\bar X$ is given by
\begin{equation}
\mathcal{L}_k=\frac 1{k!} \frac{d^k}{ds^k} \mathcal{L}(X(s)) \vert_{s=0} \,.
\label{Snsig}
\end{equation}
Higher order derivatives of $X(x,s)$ with respect to $s$
can be traded for the first derivatives by means of the
geodesic equation (\ref{geod}) and then replaced by $\xi$
at $s=0$ with the help of the initial conditions (\ref{incondgeo}).

It is convenient to introduce the Riemann curvature
of the target space metric $G_{AB}$, $\mathcal{R}_{ABCD}$,
and a 3-index field strength
\begin{equation}
H_{ABC}=\partial_A B_{BC} +\partial_B B_{CA}
+\partial_C B_{BA} \,.\label{HABC}
\end{equation}
The covariant derivative $\nabla$ is
\begin{equation}
\nabla_\mu \xi^A=\partial_\mu \xi^A +\gamma_{BC}^A(\bar X)
(\partial_\mu {\bar X}^C )\xi^B +\frac 12 {\epsilon_\mu}^\nu
(\partial_\nu {\bar X}^B ){H^A}_{BC}\xi^C \,.
\label{covdsigma}
\end{equation}
The quadratic part of the action (\ref{actsigma}) reads:
\begin{eqnarray}
&&\mathcal{L}_2=\int_M d^2x \sqrt{h} \left( G_{AB}(\bar X)
 \nabla_\mu \xi^A \nabla^\mu \xi^B -\mathcal{R}_{ABCD}
(\partial_\mu \bar X^A)(\partial^\mu \bar X^D) \xi^B\xi^C \right.
\nonumber \\
&&\qquad -\frac 12 (\partial_\mu \bar X^C)(\partial_\nu \bar X^D)
\epsilon^{\mu\nu} \xi^A\xi^B 
D_AH_{BCD} \nonumber \\
&&\qquad \left. +\frac 14 (\partial_\mu \bar X^B)(\partial^\mu \bar X^D )
H_{ABC}{H^C}_{DE}\xi^A\xi^E \right) \label{S2sig} \\
&&\qquad +\int_{\partial M} d\tau \left(
D_\tau \xi^A \xi^B (B_{AB}-F_{AB}) + (\partial_\tau \bar X^B)
D_A (B_{BC}-F_{BC}) \xi^A\xi^C \right) \,,\nonumber
\end{eqnarray}
where $\tau$ is the arc length along the boundary. 
$F_{BC}=\partial_B A_C-\partial_CA_B$. The covariant derivatives
$D_\tau$ and $D_A$ contain the Christoffel connection on the target
space (but not $H_{ABC}$ as 
the full covariant derivative $\nabla$, eq. (\ref{covdsigma})).

The natural inner product in the space of fluctuations $\xi$
reads
\begin{equation}
<\xi_{(1)},\xi_{(2)}>=\int_M d^2x \sqrt{g} G_{AB}(\bar X(x))
\xi_{(1)}^A (x) \xi_{(2)}^B(x) \,.\label{inprxi}
\end{equation}
The volume part of the action (\ref{S2sig}) has now the canonical
form $\langle \xi ,D\xi \rangle$ with the operator $D$ (\ref{laplaceb})
which is obviously of Laplace type.  The connection $\omega$ is defined in 
(\ref{covdsigma}) and
the endomorphism $E$ can be easily extracted from (\ref{S2sig}).
The operator $D$ 
is formally self-adjoint with respect to (\ref{inprxi}) if
we impose the boundary conditions of Neumann type
\begin{equation}
\mathcal{B}\xi =(\nabla_n \xi^A +\mathcal{S}_B^A\xi^B)\oB =0 \label{sigbc1}
\end{equation}
with arbitrary operator $\mathcal{S}$ which should be symmetric
with respect to the restriction of (\ref{inprxi}) to the boundary:
\begin{equation}
\int_{\partial M} d\tau (\xi_{(1)}^A \mathcal{S}_A^B \xi_{(2)B}
-\xi_{(2)}^A \mathcal{S}_A^B \xi_{(1)B}) =0
\end{equation}
There is a preferable choice of the boundary operator.
Let us vary the action (\ref{S2sig}) with respect to
the fluctuation field $\xi$:
\begin{equation}
\frac 12 \delta S_2 =\int_Md^2x\, \sqrt{h} (\delta\xi ) D\xi -
\int_{\partial M} d\tau (\delta \xi ) \mathcal{B}\xi \,.
\label{varS2sig}
\end{equation}
Now we require that the boundary integral in (\ref{varS2sig}) vanishes
for arbitrary $\delta \xi$. Hence we arrive at the boundary conditions
(\ref{sigbc1}) with the operator $\mathcal{S}$ given by
\begin{eqnarray}
&&\mathcal{S}_{A}^B=\frac 12 (\Gamma \nabla_\tau +
\nabla_\tau \Gamma )_A^B +S_A^B \,,\qquad\Gamma_{A}^B={B_A}^B-{F_A}^B \,, 
\label{bopsigma} \\
&& S_A^B=\frac 14 (\partial_n \bar X^C)
\left[ {H^{DA}}_{C}(B_{DB}-F_{DB}) 
+{H^{D}}_{BC}({B_D}^A-{F_D}^A)\right]\nonumber\\
&&\qquad\qquad\qquad
+\frac 12 (\partial_\tau \bar X^C)\left[ D^A (B_{BC}-F_{BC} )
+D_B ({B^A}_C-{F^A}_C)
\right].  \nonumber 
\end{eqnarray}
Note that the operator in (\ref{bopsigma}) is not of the
ordinary Neumann (or Robin) type since it contains tangential
derivatives on the boundary (cf. sec.\ \ref{s5tang}).

The variation (\ref{varS2sig}) vanishes also if we choose Dirichlet
boundary conditions for some of the 
coordinates of the string endpoints. Namely, we can take a projector
$\Pi_+ $ and impose (\ref{sigbc1}) on $(\Pi_+)_B^A \xi^b$
and  $(1-\Pi_+)_B^A \delta\xi^B\oB =0$. 
Physically this means that the endpoints of the bosonic string
are confined to a submanifold in the target space.
Such configurations \cite{Dai:1989ua}
are called the Dirichlet branes.
\subsection{Spinor fields}\label{s3spin}
The action for the spinor fields $\psi$ 
\begin{equation}
\mathcal{L}=\int_M d^n x \sqrt{g} \bar \psi\, \Dir\, \psi
\label{spinS}
\end{equation}
contains a first order operator $\Dir$ of Dirac type.
In Euclidean space the conjugate spinor $\bar \psi$
is just the hermitian conjugate of $\psi$: $\bar \psi=
\psi^\dag$.
By definition, an operator $\Dir$ is of Dirac type if
its square $D=\Dir^2$ is of Laplace type.
Spectral theory of general operators of Dirac type both
on manifolds without boundaries and with local boundary
conditions on manifolds with boundaries can be found
in \cite{Branson:1992a,Branson:1992b}. Here we consider
some physically motivated particular cases only. Let us introduce
the Euclidean Dirac $\gamma$-matrices which satisfy the Clifford
commutation relations:
\begin{equation}
\gamma_\mu\gamma_\nu + \gamma_\nu \gamma_\nu = 2g_{\mu\nu} \,.
\label{Clifford}
\end{equation}
The $\gamma$-matrices defined in this way are hermitian,
$\gamma_\mu^\dag =\gamma_\mu$. We also need the chirality 
matrix which will be denoted $\gamma^5$ independently of
the dimension. It satisfies
\begin{equation}
(\gamma^5)^\dag =\gamma^5 ,\qquad \gamma^5\gamma_\mu =
-\gamma_\mu \gamma^5 \,.\label{gamma5}
\end{equation}
From now on we suppose that the dimension $n$ is even. We fix the sign in
$\gamma^5$ by choosing
\begin{equation}
\gamma^5=\frac{i^{\frac{n(n-1)}2}}{n!}
\epsilon^{\mu\dots\nu}\gamma_\mu \dots \gamma_\nu \,.
\label{sign5}
\end{equation}
Let $\Dir$ be the standard Dirac operator in curved space
with gauge and axial gauge connection
\begin{equation}
\Dir =i\gamma^\mu \left(\partial_\mu +\frac 18 [\gamma_\nu ,\gamma_\rho ]
\sigma_\mu^{\nu\rho} +A_\mu +iA_\mu^5 \gamma^5 \right) \,.
\label{dirop}
\end{equation}
Here $\sigma_\mu^{\nu\rho}$ is the spin-connection (\ref{spincon}). 
$A_\mu$ and
$A_\mu^5$ are vector and axial vector fields respectively
taken in some representation of the gauge group. Both
$A_\mu$ and $A_\mu^5$ are antihermitian in the gauge indices.
The operator $\Dir$ (\ref{dirop}) is formally self-adjoint
in the bulk.
The operator $D=\Dir^2$ is of Laplace type (\ref{laplaceb}) 
with\footnote{The present author is grateful to Valery Marachevsky
for his help in deriving and checking eqs.\ (\ref{oEdir}),
(\ref{Omdir}) and (\ref{spinbc4}), see also \cite{DeBerredo-Peixoto:2001qm}.}
\begin{eqnarray}
&&\omega_\mu =\frac 18 [\gamma_\nu ,\gamma_\rho ]
\sigma_\mu^{\nu\rho} +A_\mu +\frac i2 [\gamma_\mu ,\gamma_\nu ]
A^{5\nu}\gamma^5 \,, \nonumber \\
&&E=-\frac 14 R +\frac 14 [\gamma^\mu ,\gamma^\nu ]F_{\mu\nu}
+i\gamma^5 D^\mu A_\mu^5 -(n-2)A_\mu^5A^{5\mu} \nonumber \\
&&\qquad\qquad\qquad
-\frac 14 (n-3) [\gamma^\mu ,\gamma^\nu ] [A_\mu^5 ,A_\nu^5 ]
\label{oEdir}
\end{eqnarray}
with obvious notations $F_{\mu\nu}=\partial_\mu A_\nu -
\partial_\nu A_\mu +[A_\mu ,A_\nu ]$, 
$D_\mu A_\nu^5 = \partial_\mu A_\nu^5 -
\Gamma_{\mu\nu}^\rho A_\rho^5 +[A_\mu ,A_\nu^5 ]$.
The expression for $\Omega_{\mu\nu}$ is a little bit
lengthy:
\begin{eqnarray}
&&\Omega_{\mu\nu}=F_{\mu\nu}-[A_\mu^5 ,A_\nu^5 ]-\frac 14
\gamma^\sigma \gamma^\rho R_{\sigma\rho\mu\nu} 
-i\gamma^5\gamma^\rho (\gamma_\nu D_\mu A_\rho^5
-\gamma_\mu D_\nu A_\rho^5 ) \nonumber \\
&&\qquad\quad +i\gamma^5 A_{\mu\nu}^5 
+[A_\mu^5,A_\rho^5] \gamma^\rho \gamma_\nu -
[A_\nu^5,A_\rho^5] \gamma^\rho \gamma_\mu \nonumber \\
&&\qquad\quad 
-\gamma^\rho A_\rho^5 \gamma_\mu \gamma^\sigma A_\sigma^5 \gamma_\nu
+ \gamma^\rho A_\rho^5 \gamma_\nu \gamma^\sigma A_\sigma^5 \gamma_\mu
\,,\label{Omdir}
\end{eqnarray}
where
\begin{equation}
A_{\mu\nu}^5=\partial_\mu A_\nu^5-\partial_\nu A_\mu^5
+[A_\mu ,A_\nu^5 ]-[A_\nu ,A_\mu^5 ] \label{amunu}
\end{equation}

Consider now manifolds with boundary. The specific feature of the action
(\ref{spinS}) is that it contains first order derivatives
only. Consequently, boundary conditions should be imposed
on a half of the spinor components. Let these be Dirichlet
boundary conditions
\begin{equation}
\Pi_-\psi \oB =0 \label{spinbc1}
\end{equation}
where $\Pi_-$ is a hermitian projector, $\Pi_-^2=\Pi_-$, 
$\Pi_-^\dag =\Pi_-$. Due to the hermiticity
\begin{equation}
\bar\psi\Pi_- \oB =0 \label{spinbc2}
\end{equation}
Following Luckock \cite{Luckock:1990xr}
let us consider a family of the projectors
\begin{equation}
\Pi_-=\frac 12 \left( 1+\gamma^n \exp (iq\gamma^5) \right)\,,
\label{proluck}
\end{equation}
where $q$ is a scalar which can depend on the coordinate on
the boundary.

To make the operator
$\Dir$ formally self-adjoint including the boundary we must
require that 
\begin{equation}
\int_{\partial M} d^{n-1}x\,\sqrt{h} \bar\psi_1 \gamma_n \psi_2
=0 \label{symdir}
\end{equation}
for all $\psi_{1,2}$ satisfying the boundary conditions
(\ref{spinbc1}). Since $\psi_1$ and $\psi_2$ are arbitrary,
the projector $\Pi_-$ should satisfy
\begin{equation}
(1-\Pi_-)\gamma^n (1-\Pi_-)=0 \,.\label{cpr}
\end{equation}
This condition yields $q=\pm \pi /2$. The projector (\ref{proluck})
takes the form
\begin{equation}
\Pi_-=\frac 12 \left( 1\pm i\gamma^n \gamma^5 \right)\,.
\label{proBG}
\end{equation}

To formulate the spectral problem for the {\it second}
order operator $D=\Dir^2$ we need boundary conditions for
the second half of the spinor components. The relevant functional
space should be spanned by the eigenfunctions of the Dirac
operator $\Dir$. It is clear that on this space the functions
$\Dir\psi$ should satisfy the same boundary conditions
(\ref{spinbc1}) as the $\psi$'s themselves:
\begin{equation}
\Pi_-\Dir\,\psi \oB =0 \,.\label{spinbc3}
\end{equation}
Let us adopt the choice (\ref{proBG}) for $\Pi_-$.
By commuting $\Pi_-$ with $\Dir$ in (\ref{spinbc3}) we obtain
\begin{equation}
(\nabla_n +S)\Pi_+\psi \oB =0,\qquad
S=-\frac 12 L_{aa}\Pi_+ \,,\qquad
\Pi_+=1-\Pi_- \,.\label{spinbc4}
\end{equation}
We remind that $L_{ab}$ is extrinsic curvature of the boundary.
The boundary conditions (\ref{spinbc1}), (\ref{spinbc4})
with (\ref{proBG}) are mixed (cf.\ (\ref{mixbc})).
Spectral geometry of the Dirac operator with these boundary conditions
has been thoroughly studied by Branson and Gilkey 
\cite{Branson:1992b}.

We can generalise the boundary conditions presented above by
considering non-hermitian projectors, $\Pi_-^\dag \ne \Pi_-$.
Then the boundary condition for the conjugated spinors reads:
\begin{equation}
\bar\psi \Pi_-^\dag \oB =0\,.\label{spinbc5}
\end{equation}
Instead of (\ref{cpr}) we have the condition
\begin{equation}
(1-\Pi_-^\dag )\gamma^n (1-\Pi_-)=0 \,,\label{cpr2}
\end{equation}
which yields
\begin{equation}
\Pi_-=\frac 12 \left( 1+i\gamma^n\gamma^5 e^{r(x^a)\gamma^5} \right)
\label{prochir}
\end{equation}
with an arbitrary real function (or even with an arbitrary hermitian
matrix valued function) $r(x^a)$. 

In the Minkowski signature space the spinor conjugation
includes $\gamma^0$. Therefore, the boundary conditions
on the conjugate spinor (\ref{spinbc2}) are changed. 
Roughly speaking, to continue $\Pi_-$ to the Minkowski signature
space time one has to replace $\gamma^n\gamma^5$ by $\gamma^n$
in (\ref{proBG}) and (\ref{prochir}) and to take into account
powers of $i$ which appear in the Dirac gamma matrices.

In particle physics, interest to the
boundary conditions defined by (\ref{proBG}) is due to the 
MIT bag model
of hadrons  proposed in \cite{Chodos:1974pn,Chodos:1974je,DeGrand:1975cf}. 
This model was modified later 
\cite{Chodos:1975ix,Callan:1979bm,Brown:1979ui,
Theberge:1980ye,Rho:1983bh,Goldstone:1983tu}
to include a chiral phase on the boundary in a manner similar to
(\ref{prochir})
(for a review see \cite{Hasenfratz:1978dt}).
Renormalization of quantum field theory with the boundary
conditions defined by (\ref{proBG}) was
considered by Symanzik \cite{Symanzik:1981wd}.

As we have already mentioned above, bag boundary conditions defined
by the projector (\ref{proBG}) are a particular case of mixed boundary
conditions (\ref{mixbc}). They will be considered in detail in sec.\
\ref{s5mixed} (see also sec.\ \ref{s5cases} for further references to
calculations in a ball). Chiral bag boundary conditions defined
by (\ref{prochir}) with $r\ne 0$ are considerably more complicated
because the equation (\ref{spinbc3}) contains a mixture of normal
and tangential derivatives.
We refer to 
\cite{Hrasko:1984sj,Falomir:1990tm,Falomir:1991fy,DeFrancia:1992sf,Wipf:1995dy,
DeFrancia:1996xi,DeFrancia:1998if,Beneventano:2002im,Esposito:2002vz}
for more calculations of spectral functions in this latter case. 

The boundary conditions considered in this section are {\it local},
i.e. they treat the fields at each point of the boundary independently.
One can also define {\it global} boundary conditions for the
Dirac operator (cf. sec.\ \ref{s5spec}).

Due to the fermionic nature of the spinor field the path
integral over $\psi$ and $\bar \psi$ gives determinant
of the operator $\Dir$ to a positive power,
\begin{equation}
Z=\int \mathcal{D}\bar\psi\,\mathcal{D}\psi 
\exp (-\mathcal{L}(\bar \psi ,\psi ))=\det \Dir
\label{pathspin}
\end{equation}
where the action $\mathcal{L}$ is given by (\ref{spinS}) and
$\bar\psi$, $\psi$ are complex Dirac spinors.

\subsection{Vector fields}\label{s3vect}
Consider the Yang--Mills action for the gauge field 
$A_\mu^\alpha$ (Greek letters from the beginning of the alphabet label
generators of the gauge group):
\begin{equation}
\mathcal{L}=\frac 14 \int_M d^nx \sqrt{g} F_{\mu\nu}^\alpha
F^{\mu\nu\alpha} ,\label{YMaction}
\end{equation}
where, as usual, $F_{\mu\nu}^\alpha =
\partial_\mu A_\nu^\alpha -\partial_\nu A_\mu^\alpha
+c^\alpha_{\beta\gamma}A_\mu^\beta A_\nu^\gamma$.
$c^\alpha_{\beta\gamma}$ are totally antisymmetric
structure constants of the
gauge group. Let us introduce the background field $B_\mu^\alpha$
by the shift $A_\mu^\alpha \to B_\mu^\alpha + A_\mu^\alpha$.
From now on $A_\mu^\alpha$ plays the role of quantum
fluctuations. The quadratic part of the action reads:
\begin{eqnarray}
&&\mathcal{L}_2=\frac 12 \int_M d^nx \sqrt{g} \left[ -A^{\alpha\nu}
\nabla^\mu\nabla_\mu A_\nu^\alpha 
+ A_\mu^\alpha \nabla^\mu\nabla^\nu A_\nu^\alpha  +
A_\nu^\alpha A_\mu^\alpha R^{\mu\nu}\right. \nonumber\\
&&\qquad\qquad\qquad\qquad
\left.+2F_{\mu\nu}^\alpha (B) c^\alpha_{\beta\gamma}A^{\beta\mu}
A^{\gamma\nu} \right]
\nonumber \\
&&\qquad +\frac 12 \int_{\partial M}d^{n-1}x \sqrt{h} \left[
A_\nu^\alpha \nabla^\nu A_n^\alpha -
A^{\alpha\nu}\nabla_n A_\nu^\alpha \right] \,.
\label{S2vector}
\end{eqnarray}
The covariant derivative $\nabla$ contains both metric
and gauge parts: $\nabla_\nu A_\mu^\alpha =
\partial_\nu A_\mu^\alpha + B_\nu^\beta c^\alpha_{\beta\gamma}A_\nu^\gamma
-\Gamma_{\nu\mu}^\rho A_\rho^\alpha$. One should impose a gauge
condition on the fluctuation $A_\mu^\alpha$.
We choose
\begin{equation}
\nabla^\mu A_\mu^\alpha = 0 \,.\label{bg-gauge}
\end{equation}
In the gauge (\ref{bg-gauge}) the bulk term in (\ref{S2vector}) defines
an elliptic operator of the Laplace type with
\begin{eqnarray}
&&(\omega_\mu )^{\alpha\rho}_{\nu\beta}=
B_\mu^\gamma c^\alpha_{\gamma\beta} \delta_\nu^\rho
-\Gamma_{\mu\nu}^\rho \delta_\beta^\alpha \,,\label{omegaYM}\\
&&(E)^{\alpha\rho}_{\nu\beta} =-R_\nu^\rho \delta_\beta^\alpha
+2 {F(B)^{\gamma\rho}}_\nu c^\gamma_{\beta\alpha} \,.   \label{EYM}
\end{eqnarray}
The field strength corresponding of the connection (\ref{omegaYM})
reads
\begin{equation}
(\Omega_{\mu\nu})^{\alpha\sigma}_{\rho\beta}=
{R^\sigma}_{\rho\mu\nu}\delta^\alpha_\beta +F(B)^\delta_{\mu\nu}
c^\alpha_{\delta\beta} \delta^\sigma_\rho \,. \label{OmegaYM}
\end{equation}
Note, that all the quantities above (\ref{omegaYM}) - (\ref{OmegaYM})
are matrix functions with both gauge and vector indices.

On a manifold without boundary
the operator $D$ with (\ref{omegaYM}) and (\ref{EYM}) is
symmetric with respect to the standard inner product in the
space of the vector fields
\begin{equation}
<A^{(1)},A^{(2)}>=\int_M \sqrt{g} d^n x\ A^{(1)\alpha\mu}
A^{(2)\alpha}_\mu \,.
\label{inprYM}
\end{equation}

The ghost operator corresponding to the gauge (\ref{bg-gauge})
is just the ordinary scalar Laplacian, $D^{gh}
=-\nabla^{[gh]\mu}\nabla^{[gh]}_\mu$, with the connection
\begin{equation}
(\omega_\mu^{[gh]})^\alpha_\gamma =B_\mu^\beta c^\alpha_{\beta\gamma}\,.
\label{ghostcon}\end{equation}

The one-loop path integral reads:
\begin{equation}
Z(B)=\det (D)^{-\frac 12}_\perp \det (D^{[gh]})^{\frac 12}
\label{p-int-g1}
\end{equation}
where the first determinant is restricted to the fields
satisfying the gauge condition (\ref{bg-gauge}). Note, that
pure gauge fields $A=\nabla \xi$ are zero modes
of the total operator $D+\nabla \nabla $ of the bulk
action (\ref{S2vector}) only on shell, i.e. when the
background field $B$ satisfies the classical equation
of motion $\nabla^\mu F(B)_{\mu\nu}^\alpha =0$. Therefore,
the path integral is gauge-independent on shell, but it
depends on the gauge choice off shell. For
example, the Feynman gauge path integral
\begin{equation}
Z_F(B)=\det (D)^{-\frac 12} \det (D^{[gh]}) \,,
\label{p-int-g2}
\end{equation}
where the first determinant is calculated on the
space of {\it all} vector field, is equal to the
$Z(B)$ defined in (\ref{p-int-g1}) only on shell.
However, physical predictions of the two path integrals
are, of course, equivalent.

The path integral (\ref{p-int-g2}) can be obtained by adding
the gauge fixing term
\begin{equation}
{\mathcal{L}}_{\rm gf}=\frac 1{2\kappa} \iM (\nabla_\mu A^\mu)^2
\label{Svgf}\end{equation}
to (\ref{S2vector}) with $\kappa=1$. The case $\kappa\ne 1$
yields a non-minimal operator on the gauge field fluctuations
(cf sec.\ \ref{s4nonmin}).

One can define ``total'' heat kernel coefficients
for the path integral in a certain gauge $\chi (A)$
\begin{equation}
a^{tot}_k =a_k(D^\chi )-2a_k( D^{[gh]}_\chi ) \,,
\label{atot}
\end{equation}
where $D^\chi$ is defined by the action (\ref{S2vector})
with the gauge fixing term $\chi^2(A)$, and the ghost
operator is $D^{[gh]}_\chi \xi =\chi (\nabla \xi )$.
Even on shell only the coefficient $a_n^{tot}$ is
gauge-independent. Only this coefficient contains
information on the one-loop divergences in a gauge-invariant
regularization like the $\zeta$-function one. 

On a manifold with boundary one should impose 
boundary conditions on gauge fields and  ghosts.
The boundary conditions should be gauge invariant.
Consider a more general set-up when we have some
quantum fields $\Phi$ and a linearised
gauge transformation $\Phi\to\Phi +\delta_\xi \Phi$
with local parameter $\xi$. Boundary operator 
$\mathcal{B}$ defines gauge invariant boundary conditions
\begin{equation}
\mathcal{B}\Phi  =0
\label{ginvbc1}
\end{equation}
if there exist boundary conditions for the gauge
parameter $\xi$ 
\begin{equation}
\mathcal{B}_\xi \xi  =0 
\label{ginvbc2}
\end{equation}
such that 
\begin{equation}
\mathcal{B} \delta_\xi \Phi  =0 \,.
\label{ginvbc3}
\end{equation}
The equation (\ref{ginvbc3}) means that the
functional space defined by the boundary conditions (\ref{ginvbc1})
is invariant under the gauge transformations provided the gauge
parameter $\xi$ satisfies (\ref{ginvbc2}). Upon quantisation
(\ref{ginvbc2}) become boundary conditions for the ghost fields.
The condition (\ref{ginvbc3}) ensures validity of the 
Faddeev-Popov trick on a manifold with boundary and
guarantees gauge-independence of the on-shell path integral
\cite{Vassilevich:1998iz}.

Since the boundary term in (\ref{S2vector}) is diagonal
in the gauge index $\alpha$ we can consider the case
of an abelian gauge group and drop $\alpha$ from the
notations. The general non-abelian case does not offer
considerable complications. There are two sets of local
boundary conditions which satisfy the gauge-invariance
requirements above. The first set is called absolute
boundary conditions. It reads
\begin{eqnarray}
&&A_n\oB =0\,,\qquad \partial_n A_a \oB =
(\nabla_n \delta_{ab} -L_{ab} )A_b\oB =0 \,,
\nonumber \\
&&\nabla_n \xi \oB =0 \,.\label{absbc}
\end{eqnarray}
The second set
\begin{eqnarray}
&&(\nabla_n -L_{aa})A_n \oB =0\,, \qquad A_a\oB =0\,,
\nonumber \\
&& \xi \oB =0 \label{relbc}
\end{eqnarray}
is called relative boundary conditions. 
The projectors on the Dirichlet and Neumann subspaces and the 
endomorphism $S$ are $(\Pi^-)_{ij}=\delta_{in}\delta_{jn}$,
$(\Pi^+)_{ij}=\delta_{ij}-\delta_{in}\delta_{jn}$, 
$S_{ij}=-L_{ab}\delta_{ia}\delta_{jb}$ and
$(\Pi^-)_{ij}=\delta_{ij}-\delta_{in}\delta_{jn}$,
$(\Pi^+)_{ij}=\delta_{in}\delta_{jn}$,
$S_{ij}=-L_{aa}\delta_{in}\delta_{jn}$ for the absolute and
relative boundary conditions respectively.

It is straightforward
to check gauge invariance of the absolute boundary conditions
(\ref{absbc}). A bit more job is required to show that the
condition (\ref{relbc}) for the normal component is also gauge
invariant. The key observation is that near the boundary
the scalar Laplacian can be represented as
$D^{[gh]}=-(\nabla_n -L_{aa})\nabla_n +\mathcal{E}$,
where $\mathcal{E}$ does not contain normal derivatives
and, therefore, $\mathcal{E}\xi \oB =0$ under the boundary
conditions (\ref{relbc}) for the $\xi$. Consequently,
\begin{equation}
(\nabla_n -L_{aa}) \delta_\xi A_n \oB =
(\nabla_n -L_{aa})\nabla_n \xi \oB =
(-D^{[gh]}\xi +\mathcal{E}\xi )\oB=0 
\end{equation}
where $D^{[gh]}\xi \oB=0$ on the eigenfunctions of the
operator $D^{[gh]}$ which is enough for our purposes.

The boundary term in the action (\ref{S2vector}) can be
rewritten as\footnote{The equation to follow contains two types
of covariant derivatives. Definitions can be found in sec.\ \ref{s2geo}.}
\begin{equation}
\frac 12 \int_{\partial M} d^{n-1}x \sqrt h 
[A_a(A_{n:a} +L_{ab}A_b -A_{a;n} )] \,.
\label{btvect}
\end{equation}
This boundary action vanishes for both types of
the boundary conditions (\ref{absbc}) and (\ref{relbc}).
Hence the operator $D$ is symmetric for these
boundary conditions. Another remarkable fact is that
the fields $A^\perp$ satisfying the gauge condition
(\ref{bg-gauge}) are orthogonal to the gauge transformations.
Indeed,
\begin{equation}
<A^\perp ,\nabla \xi > =-\int_{\partial M} d^{n-1}x \sqrt h
A^\perp_n \xi =0
\end{equation}
for both (\ref{absbc}) and (\ref{relbc}). 

We should also check whether the gauge condition (\ref{bg-gauge})
is indeed compatible with the boundary conditions (\ref{relbc})
and (\ref{absbc}). For an arbitrary $A$ there should exist
unique $\xi$ such that
\begin{equation}
\nabla^\mu (A_\mu + \nabla_\mu \xi )=0 .\label{compatib}
\end{equation}
We rewrite (\ref{compatib}) as
\begin{equation}
\nabla^\mu A_\mu =-\Delta \xi \,. \label{compatib2}
\end{equation}
Let us start with relative boundary conditions (\ref{relbc}).
In this case the left hand side of (\ref{compatib2}) satisfies
Dirichlet boundary conditions. The scalar Laplacian
$\Delta$ for these boundary conditions is invertible. 
Therefore, a solution for 
(\ref{compatib2}) always exists, and it is unique. The case of
absolute boundary conditions is a bit more involved.
One should take care of a zero mode in the ghost sector.
We leave this case as an exercise. A more extensive 
discussion of compatibility of gauge and boundary conditions
can be found in \cite{Vassilevich:1998iz}.

\subsection{Graviton}
We start this section with the Einstein-Hilbert action
on a four dimensional Euclidean manifold without boundary:
\begin{equation}
\mathcal{L}=\frac 1{16\pi G_N} \int d^4 x\sqrt{g} (R-2\Lambda )\,,
\label{gract}
\end{equation}
where $R$ is the scalar curvature,
$G_N$ is the Newton constant, $\Lambda$ is the
cosmological constant.
 As usual, let us shift the metric 
$g_{\mu\nu}\to g_{\mu\nu} +h_{\mu\nu}$. From now on
$g_{\mu\nu}$ will denote the background metric, $h_{\mu\nu}$
will be the quantum fluctuations. We can decompose the 
$h_{\mu\nu}$ further in trace, longitudinal and transverse-traceless
part:
\begin{equation}
h_{\mu\nu} =\frac 14 hg_{\mu\nu} +(L\xi )_{\mu\nu} +
h^\perp_{\mu\nu} \,, \label{DYdec}
\end{equation}
where $g^{\mu\nu}h^\perp_{\mu\nu}=\nabla^\nu h^\perp_{\mu\nu}=0$
and
\begin{equation}
(L\xi )_{\mu\nu}=\nabla_\mu \xi_\nu +\nabla_\nu \xi_\nu
-\frac 12 g_{\mu\nu} \nabla^\rho \xi_\rho \,.
\label{opL}
\end{equation}
The decomposition (\ref{DYdec}) is orthogonal with respect to
the inner product:
\begin{eqnarray}
&& <h, h'>=\int d^4x \sqrt {g}
G^{\mu \nu \rho \sigma} h_{\mu \nu}
h'_{\rho \sigma} \,,\label{innprgrav} \\
&&G^{\mu \nu \rho \sigma}=\frac 12 (g^{\mu \rho} g^{\nu \sigma}+
g^{\mu \sigma} g^{\nu \rho} +Cg^{\mu \nu}g^{\rho \sigma})
\,. \nonumber 
\end{eqnarray}
Here $C$ is a constant. For positivity of (\ref{innprgrav}) $C$
should be greater than $-\frac 12$.
Under the action of infinitesimal diffeomorphism generated by
a vector $\epsilon_\mu$ the components of (\ref{DYdec}) transform
as
\begin{equation}
\xi_\mu \to \xi_\mu +\epsilon_\mu \,,\quad
h \to h+2\nabla^\mu \epsilon_\mu\,,\quad
h_{\mu \nu}^\perp \to h_{\mu \nu}^\perp \label{diftr}
\end{equation}
One can fix the gauge freedom (\ref{diftr}) by the condition
\begin{equation}
\xi_\mu =0\,. \label{gfgr}
\end{equation}
If the background admits conformal Killing vectors (these are
the vectors which are annihilated by the operator $L$
(\ref{opL})) the condition (\ref{gfgr}) is not enough and
one should impose one more gauge condition on the trace
part (see e.g. \cite{Vassilevich:1993rk}). We suppose
that conformal Killing vectors are absent.

The Jacobian factor induced by the change of variables
(\ref{DYdec}) $h_{\mu\nu}\to (h,\xi_\mu ,h_{\mu\nu}^\perp )$
is
\begin{equation}
J=\det_V (L^\dag L)^{\frac 12} \,,\label{Jgrav}
\end{equation}
where the determinant is calculated on the space of the vector
fields (excluding the conformal Killing vectors which we do not
take into account). It is convenient to shift the scalar
part of the metric fluctuations by $\nabla^\mu \xi_\mu$ so that
the decomposition (\ref{DYdec}) becomes
\begin{equation}
h_{\mu\nu} =\frac 14 (\sigma +2\nabla^\rho \xi_\rho )
g_{\mu\nu} +(L\xi )_{\mu\nu} +
h^\perp_{\mu\nu} \,. \label{DYnew}
\end{equation}
Since the change of the variables $h\to \sigma$ does not introduce
any Jacobian factor we conclude that the path integral measure is
\begin{equation}
\mathcal{D}h_{\mu\nu} =\det_V (L^\dag L)^{\frac 12}
\mathcal{D}\sigma \mathcal{D}\xi \mathcal{D}h^\perp \,.
\label{measgr}
\end{equation}

To simplify the discussion we suppose that the background metric
$g_{\mu\nu}$ satisfies the Einstein equations
\begin{equation}
R_{\mu\nu}(g)=\Lambda g_{\mu\nu} \,. \label{Einsteq}
\end{equation}
The quadratic part of the action reads:
\begin{eqnarray}
&&\mathcal{L}_2=\frac 1{16\pi G} \int d^4x \sqrt {g} \left(\frac 14
h^{\perp \mu \nu}(-\Delta g_{\mu \rho } g_{\nu \sigma}
+2R_{\mu \rho \nu \sigma})h^{\perp \rho \sigma}\right. \nonumber \\
&&\qquad\qquad\qquad \left.
-\frac 3{32} \sigma \left(-\Delta -\frac R3 \right) \sigma \right) \,.
\label{S2grav}
\end{eqnarray}
Due to the gauge invariance (\ref{S2grav}) does not contain $\xi$.
Functional integration over $\xi$ produces an infinite constant
equivalent to the volume of the diffeomorphism group which
will be neglected.

The kinetic term for $\sigma$ has a wrong (negative) sign. This
represents the well known conformal factor problem of quantum
general relativity. Different explanations of this phenomenon
\cite{Gibbons:1978ac,Schleich:1987fm,Mazur:1990by,Novozhilov:1991tf}
suggest (roughly) the same remedy: the conformal mode must be rotated
to the imaginary axis, $\sigma\to i\sigma$.

The path integral can be written in terms of functional determinants
corresponding to vector fields (indicated by the subscript $V$),
scalars (indicated by $S$), and transverse traceless tensors
(indicated by $T\perp$): 
\begin{equation}
Z=\det_V (L^\dag L)^{\frac 12}\det_{S} \left(-\Delta -\frac R3 
\right)^{-\frac 12} \det_{T\perp}(-\Delta g_{\mu \rho } g_{\nu \sigma}
+2R_{\mu \rho \nu \sigma})^{-\frac 12}
\,.\label{Zgrav}
\end{equation}
This expression is, in principle,
suitable for calculations on some homogeneous spaces since the harmonic
expansion on such spaces usually respects separation of tensors
to transverse and longitudinal parts. However, we must warn the reader
again that we have neglected the presence 
of Killing vectors and of conformal Killing vectors. 
The ways to treat these vectors created a long discussion in the
literature 
\cite{Gibbons:1978ji,Christensen:1980iy,Fradkin:1984mq,Allen:1986ta,
Griffin:1989sq,Taylor:1990ua,Vassilevich:1993rk,Mottola:1995sj,
Volkov:2000ih}.

The representation (\ref{Zgrav}) is not convenient on generic
manifold since the tensor operator is restricted to the
transverse modes. To remove this restriction one can either
multiply (\ref{Zgrav}) by a compensating vector determinant, or
add a suitable gauge fixing term to the classical action
(\ref{gract}) and repeat the quantisation procedure right from
the beginning. The result can be found elsewhere in the
literature (see, e.g., 
\cite{Gibbons:1978ji,Christensen:1980iy,Fradkin:1984mq,Taylor:1990ua}).

Boundary conditions for one-loop Euclidean quantum gravity
must be diffeomorphism invariant and must lead to ``good''
(i.e. symmetric and elliptic) operators describing metric
and ghost fluctuations. The problem of finding a suitable set
of boundary conditions in gravity appeared to be much harder
that for the lower spin fields. There exist several
proposals on the market
\cite{Barvinsky:1987dg,Luckock:1990xr,Esposito:1995aj,Vassilevich:1995zk,
Marachevsky:1996dr,
Avramidi:1996ae,Avramidi:1997hy,Avramidi:1998sh,Moss:1997ip,Esposito:1998hp}.
However, neither of these proposals is fully satisfactory.
It is clear now that proper boundary conditions must depend on
tangential derivatives. Moreover, they probably must be non-local.

The material of this section can be generalised to the case
when an independent torsion field is present.
Explicit expressions for relevant operators acting on fields 
with different spin in the Riemann--Cartan space can be found in
\cite{Obukhov:1983mm,Cognola:1988qw,Yajima:1988eh}.


\section{Heat kernel expansion on manifolds without boundary}
\label{sec4}
\subsection{General formulae}\label{s4gen}
For physicists, the most familiar way to calculate the heat
kernel coefficients is the DeWitt iterative procedure which
will be briefly described in sec.\ \ref{s4iter}. Here
we take a different route following the method of Gilkey.
Advantages of this method are most clearly seen on manifolds
with boundaries. However, even if boundaries are absent
we use the same method for the following reasons:
(i) this will ensure a smooth transition to more
complicated material of the next section; (ii) we believe
that calculations of the coefficients $a_k(f,D)$ are a little
bit easier even without a boundary.
We are not going
to give complete proofs of all statements. Instead,
we concentrate on main ideas of the method. 
For the details an interested reader
can consult the original paper \cite{Gilkey1975}
and the monographs \cite{Gilkey:1995,Kirsten:2001wz}.

We start with very general properties of the heat
kernel coefficients. Let us consider 
a smooth compact Riemannian manifold $M$ without boundary.
To be able to define functions on $M$ which carry some discrete
(spin or gauge) indices  we need a vector bundle $V$ over $M$.
Let $D$ be an operator of Laplace type on $V$, 
and let $f$ be a smooth function on $M$.
There is an asymptotic expansion (\ref{asymptotex}) and

\noindent 1) Coefficients with odd index $k$ vanish,
$a_{2j+1}(f,D)=0$;

\noindent 2) Coefficients $a_{2j}(f,D)$ are locally
computable in terms of geometric invariants.

Already the existence of the power-law asymptotic expansion (\ref{asymptotex})
is a non-trivial statement. We postpone the discussion of this 
property to sec.\ \ref{s5other}.

The second statement above is very
important. It means that the heat kernel coefficients
can be expressed as integrals of local invariants:
\begin{equation}
a_k(f,D)= \ptr \iM \{ f(x) a_k (x;D) \} 
=\sum_I \ptr \iM \{ f u^I \mathcal{ A}_k^I(D) \} \,, \label{locinv}
\end{equation}
where $\mathcal{ A}_k^I$ are all possible independent invariants 
of the dimension $k$  constructed
from $E$, $\Omega$, $R_{\mu\nu\rho\sigma}$ and their derivatives.
We use usual assignments of the dimensions
when $E$ has dimension two, any derivative has dimension one, etc.
 $u^I$ are some constants. For example, if $k=2$ only two 
independent invariants exist. These are $E$ and $R$.
Note, that we can
always integrate by parts to remove all derivatives from $f$.
The first statement, $a_{2j+1}=0$, is clear now. 
One cannot construct an
odd-dimension invariant on a manifold without boundary.

Let us study further relations between the heat kernel coefficients
which will turn out to generate relations between the constants $u^I$.
Consider now the case when the manifold $M$ is a direct product
of two manifolds, $M_1$ and $M_2$, with coordinates $x_1$ and
$x_2$ respectively,
and the operator $D$ is a 
sum of two operators acting independently on $M_1$ and $M_2$,
$D=D_1\otimes 1 +1\otimes D_2$. This means that the bundle indices
are also independent. As an example, one can consider the vector Laplacian
on $M_1\times M_2$.  One can write symbolically
$\exp (-tD)=\exp (-tD_1)\otimes \exp (-tD_2)$. Next we multiply
both sides of this equation by $f(x_1,x_2)=f_1(x_1)f_2(x_2)$, take
the functional trace and perform the asymptotic expansion in $t$
to get
\begin{equation}
a_k(x;D)=\sum_{p+q=k} a_p(x_1;D_1)a_q(x_2;D_2)\,.
\label{product}
\end{equation}

The consequences of eq.\ (\ref{product})
are very far reaching. In particular, eq.\ (\ref{product}) allows to fix
the dependence of $u^I$ on the dimension of the manifold $M$.
Consider an even more specialised case when
one of the manifolds is a one-dimensional circle:  $M_1=S^1$, $0<x_1\le 2\pi$.
Let us make the simplest choice for $D_1$: 
$D_1=-\partial_{x_1}^2$. All geometric invariants associated with
$D$ are defined solely by the $D_2$-part. Moreover, all invariants are
independent of $x_1$.
Therefore, by the equation (\ref{locinv}),
\begin{eqnarray}
a_k(f(x_2),D)&=&\int_{S^1\times M_2}d^nx\sqrt{g}\sum_I
\ptr \{f(x_2) u^I_{(n)} 
\mathcal{ A}^I_k (D) \} \nonumber \\
&=&2\pi\int_{M_2}d^{n-1}x\sqrt{g}\sum_I
\ptr \{f(x_2) u^I_{(n)} \mathcal{ A}^I_k (D_2) \}\,.
\label{77hk}
\end{eqnarray}
Here dependence of the constants $u^I_{(n)}$ on the dimension $n$ of
the manifold is shown explicitly.
On the other hand, we can use (\ref{product}). Spectrum
of the operator $D_1$ is known. The eigenvalues are $l^2$,
$l \in {\mathbb {Z}}$. The heat kernel asymptotics for $D_1$
can be easily obtained
by using the Poisson summation formula
\begin{eqnarray}
K(t,D_1)&=&\sum_{l\in \mathbb{Z}} \exp (-tl^2)=\sqrt{\frac {\pi}t}
\sum_{l\in \mathbb{Z}} \exp (-\pi^2 l^2/t) \nonumber\\
&\simeq& \sqrt{\frac {\pi}t} +\mathcal{O}\left( e^{-1/t} \right)\,.
\label{eq:asPoi}
\end{eqnarray}
Since exponentially small terms have no effect on the heat kernel
coefficients, the only non-zero coefficient is
$a_0(1,D_1)=\sqrt{\pi}$. Therefore, we obtain by eq.\ (\ref{product})
\begin{equation}
a_k(f(x^2),D)={\sqrt \pi} \int_{M_2}d^{n-1}x\sqrt{g}\sum_I
\ptr
 \{f(x^2) u^I_{(n-1)} \mathcal{ A}^I_n (D_2) \} \,.  
\label{78hk}
\end{equation}
There are no restrictions on the 
operator $D_2$ or on the manifold $M_2$.
By comparing the equations (\ref{77hk}) and (\ref{78hk})
we obtain
\begin{equation}
u^I_{(n)}=\sqrt{4\pi} u^I_{(n+1)}. \label{uIn}
\end{equation}
This proves that the
constants $u^I$ depend on the dimension
$n$ only through the overall normalisation factor
$(4\pi )^{-n/2}$ .

To calculate the heat kernel coefficients we need also
the following variational equations (see \cite{Gilkey:1995},
Lemma 4.1.15):
\begin{eqnarray}
&&\frac{d}{d\epsilon} \vert_{\epsilon =0} a_k
(1, e^{-2\epsilon f} D)=(n-k) a_k (f,D)\,,\label{vara} \\
&&\frac{d}{d\epsilon} \vert_{\epsilon =0} a_k
(1,D-\epsilon F) =a_{k-2}(F,D)\,,\label{varb} \\
&&\frac{d}{d\epsilon} \vert_{\epsilon =0} a_{n-2}
(e^{-2\epsilon f}F,e^{-2\epsilon f} D)=0\,,\label{varc}
\end{eqnarray}
where $f$ and $F$ are some smooth functions.

To prove the first property (\ref{vara}) we note that
\begin{equation}
\frac{d}{d\epsilon} \vert_{\epsilon =0} {\rm Tr}
(\exp (-e^{-2\epsilon f}t D))=
{\rm Tr} (2ftD \exp (-tD)) =-2t\frac {d}{dt}
{\rm Tr} (f\exp (-tD))
\nonumber
\end{equation}
and expand both sides of this equation in the power series in $t$.
Eq.
(\ref{varb}) can be checked in a similar way. To prove (\ref{varc}) consider
the operator
\begin{equation}
D(\epsilon ,\delta )=e^{-2\epsilon f} (D -\delta F)\,.
\end{equation}
We use first (\ref{vara}) with $k=n$ to show:
\begin{equation}
0=\frac{d}{d\epsilon} \vert_{\epsilon =0} a_{n} (1,
D(\epsilon ,\delta )). \end{equation}
Then we vary the equation above with respect to $\delta$:
\begin{equation}
0= \frac{d}{d\delta}\vert_{\delta =0}
\frac{d}{d\epsilon} \vert_{\epsilon =0} a_{n} (1,
D(\epsilon ,\delta ))=\frac{d}{d\epsilon} \vert_{\epsilon =0}
\frac{d}{d\delta}\vert_{\delta =0} a_{n} (1,
D(\epsilon ,\delta ))\,. \end{equation}
Finally, eq.\ (\ref{varb}) yields (\ref{varc}).

Eq.\ (\ref{varb}) restricts dependence of the heat kernel coefficients
on the ``potential'' $E$ while (\ref{vara}) and (\ref{varc}) describe
properties of the heat kernel coefficients under local scale transformations.

To calculate the heat kernel coefficients we adopt the following
strategy. First we write down a general expression for $a_k$
containing all invariants $\mathcal{ A}_n^I$ of dimension $k$ with
arbitrary coefficients $u^I$. The constants $u^I$ are then
calculated by using the properties derived above. The first three
coefficients read \cite{Gilkey1975,Gilkey:1995}:
\begin{eqnarray}
&&a_{0}(f,D)=(4\pi)^{-n/2}\iM \ptr\{\alpha_0 f\} \label{a0gen} \\
&&a_2(f,D)
         =(4\pi)^{-n/2}\frac 16 \iM \ptr\{f(
\alpha_1E+\alpha_2R)\},\label{a2gen}\\
&&   a_4(f,D)=(4\pi)^{-n/2}\frac 1{360} \iM \ptr\{f(\alpha_3E_{;kk}
    +\alpha_4R E+\alpha_5E^2 \nonumber \\
&&\qquad\qquad
    +\alpha_6R_{;kk}+\alpha_7R^2+\alpha_8 R_{ij}R_{ij}
    +\alpha_9 R_{ijkl}R_{ijkl}+\alpha_{10}\Cur_{ ij}\Cur_{ij})\}.
\label{a4gen}
\end{eqnarray}
Instead of the $u^I$ we use rescaled constants $\alpha_I$. By (\ref{uIn})
the coefficients $\alpha_I$ are true constants, i.e. they do not depend on
$n$. One can check that indeed no more invariants exist. For example,
$R_{ij;ij}$ is proportional to $R_{;jj}$ due to the Bianchi identity.

The coefficient $\alpha_0$ follows immediately from the
heat kernel expansion for the ``free'' scalar Laplacian
on $S^1$ (see eq. (\ref{eq:asPoi})). We obtain
$\alpha_0=1$. Let us now use (\ref{varb}).
First take $k=2$. Then
\begin{equation}
\frac 16 \iM \ptr \left\{ \alpha_1 F \right\} =
\iM \ptr \{F\} \,. \label{4140}
\end{equation}
This gives $\alpha_1=6$. Take $k=4$ to see
\begin{equation}
\frac 1{360}\iM \ptr \{ \alpha_4 FR +2\alpha_5 FE \}
=\frac 16 \iM \ptr \{ \alpha_1 FE +\alpha_2 FR \}\,.
\label{4141}
\end{equation}
The equation (\ref{4141}) shows that $\alpha_5=180$,
$\alpha_4=60\alpha_2$. 

To proceed further we need local scale transformations defined in
(\ref{vara}) and (\ref{varc}). These scale transformations look
similar to the local Weyl transformations but are not exactly
the same. The scale transformations in (\ref{vara})
are designed in such a way that the operator $D$ always transforms
covariantly. This is not the case of the Weyl transformations of
an arbitrary operator of Laplace type. For example, the scalar Laplacian
(\ref{opDscalar}) is conformally covariant for a special value (\ref{confxi})
of the conformal coupling only.
Therefore, some of the basic quantities
are transformed in a somewhat unusual way. The metrics transforms
as $g_{\mu\nu}\to e^{2\epsilon f}g_{\mu\nu}$ thus defining standard
conformal properties of the Riemann tensor, the Ricci tensor and the
scalar curvature. The functions $a^\sigma$ and $b$ in eq. (\ref{D1})
transform homogeneously. Transformation properties of $\omega_\mu$
and $E$ are then defined through (\ref{oab}) and (\ref{Eab}).
One can obtain the following relations 
\cite{Gilkey1975,Gilkey:1995}
\begin{eqnarray}
&&\frac d{d\epsilon} \vert_{\epsilon =0} \sqrt{g}=nf\sqrt{g} \,,
\nonumber \\
&&\frac d{d\epsilon} \vert_{\epsilon =0} R_{ijkl}=-2fR_{ijkl}
+\delta_{jl}f_{;ik} +\delta_{;ik}f_{;jl}-\delta_{il}f_{;jk}
-\delta_{jk}f_{;il} \,,\nonumber \\
&&\frac d{d\epsilon} \vert_{\epsilon =0} E=-2fE +\frac 12 (n-2)f_{;ii}
\,,\nonumber \\
&&\frac d{d\epsilon} \vert_{\epsilon =0} R =-2fR -2(n-1)f_{;ii}
\,,\nonumber \\
&&\frac d{d\epsilon} \vert_{\epsilon =0} E_{;kk} =-4fE_{;kk}
-2f_{;kk}E +\frac 12 (n-2) f_{;iijj} +(n-6)f_{;k}E_{;k}
\,,\nonumber \\ 
&&\frac d{d\epsilon} \vert_{\epsilon =0} R E = -4f R E
+\frac 12 (n-2) f_{;ii}R -2(n-1)f_{;ii}E
\,,\nonumber \\
&&\frac d{d\epsilon} \vert_{\epsilon =0} E^2 =-4fE^2 
+(n-2) f_{;ii}E
\,,\nonumber 
\end{eqnarray}\begin{eqnarray}
&&\frac d{d\epsilon} \vert_{\epsilon =0} R_{;kk} =-4fR_{;kk}
-2f_{;kk}R -2(n-1)f_{;iijj} +(n-6)f_{;i}R_{;i}
\,,\nonumber \\
&&\frac d{d\epsilon} \vert_{\epsilon =0} R^2 =-4fR^2
-4(n-1) f_{;ii} R
\,,\nonumber \\
&&\frac d{d\epsilon} \vert_{\epsilon =0} R_{ij}R_{ij} =-4fR_{ij}R_{ij}
-2f_{;ii}R -2(n-2)f_{;ij}R_{ij}
\,,\nonumber \\
&&\frac d{d\epsilon} \vert_{\epsilon =0} R_{ijkl}R_{ijkl} 
=-4fR_{ijkl}R_{ijkl} -8f_{;ij}R_{ij}
\,,\nonumber \\
&&\frac d{d\epsilon} \vert_{\epsilon =0} \Omega^2=-4f\Omega^2
\label{cvar1}
\end{eqnarray}
Let us remind that the indices $i,j,k,l$ are flat, so we can put them all
down and sum up over the repeated indices by contracting them with the
Kronecker $\delta$. 

Let us apply (\ref{varc}) to $n=4$.
\begin{equation}
\nonumber \frac{d}{d\epsilon} \vert_{\epsilon =0} a_2
(e^{-2\epsilon f}F,e^{-2\epsilon f} D)=0
\,. \end{equation}
By collecting the terms with $\ptr\iM \{ Ff_{;jj} \} $ we obtain
$\alpha_1=6\alpha_2$ and consequently $\alpha_2=1$ and
$\alpha_4=60\alpha_2=60$. This completes calculation of $a_2$.

Let $M=M_1\times M_2$ with a product metric and let $D=(-\Delta_1)+
(-\Delta_2)$ where $\Delta_{1,2}$ are scalar Laplacians on $M_1$ and
$M_2$ respectively\footnote{More precisely, we assume that in 
(\ref{opDscalar}) the potential $U=0$, the conformal coupling is minimal
$\xi=0$, and there is no gauge coupling $G_\mu=0$ on both $M_1$
and $M_2$.}.
Eq.\ (\ref{product}) yields
\begin{eqnarray}
&&a_4(1,-\Delta_1-\Delta_2)=
a_4(1,-\Delta_1)a_0(1,-\Delta_2)+
a_2(1,-\Delta_1)a_2(1,-\Delta_2)+ \nonumber \\
&&\qquad\qquad\qquad +a_0(1,-\Delta_1)a_4(1,-\Delta_2)
\label{prod4}
\end{eqnarray}
It is clear from (\ref{omEscal}) that $E=0$ and $\Omega =0$. 
By collecting the terms
with $R_1R_2$ (where $R_1$ and $R_2$ are scalar curvatures on
$M_1$ and $M_2$ respectively) we obtain
$$
\frac 2{360} \alpha_7 = \left( \frac{\alpha_2}6 \right)^2 \,.
$$
Consequently, $\alpha_7=5$. 

Let us apply (\ref{varc}) to $n=6$.
We obtain with the help of the variational equations (\ref{cvar1})
\begin{eqnarray}
&&0=\ptr \iM \{ F( (-2\alpha_3-10\alpha_4+4\alpha_5)f_{;kk}E
\nonumber\\
&&\qquad +(2\alpha_3 -10\alpha_6)f_{;iijj}
\nonumber\\
&&\qquad +(2\alpha_4-2\alpha_6-20\alpha_7-2\alpha_8 )f_{;ii}R
\nonumber \\
&&\qquad +(-8\alpha_8-8\alpha_9 )f_{;ij}R_{ij} )\} \,.
\label{var1012}
\end{eqnarray}
The coefficients in front of independent invariants in (\ref{var1012})
must be zero. We determine $\alpha_3=60$, $\alpha_6=12$,
$\alpha_8=-2$ and $\alpha_9=2$.

The most elegant way to calculate the remaining constant $\alpha_{10}$
is based on the Gauss-Bonnet theorem 
(see \cite{Gilkey:1995}). We use here a more lengthy way 
\cite{Nepomechie:1985wt} which
however works perfectly on flat manifolds without boundary and
of trivial topology of $\mathbb{R}^n$. 
Note that $\mathbb{R}^n$ is non-compact.
To make
the heat kernel well defined we should suppose certain fall-off
conditions on the background fields and on the smearing function $f$.
The basis in the space of the square integrable
functions is given by the plane waves $\exp (ikx)$.
Therefore, for $M=\mathbb{R}^n$ with flat metric the heat kernel
reads:
\begin{eqnarray}
&&K(f;t)={\rm Tr}_{L^2} (f\exp (-tD))\nonumber \\
&&\qquad =\int d^nx \int 
\frac {d^nk}{(2\pi)^n} e^{-ikx}\ptr \{f(x) \exp (-tD) e^{ikx}\}
\nonumber \\
&&\qquad =\int d^nx \int \frac {d^nk}{(2\pi)^n} \ptr\{ f(x) 
\exp (t((\nabla^\mu +ik^\mu )^2+E))\} \,.
\label{flathk1}
\end{eqnarray}
The following integrals will be useful:
\begin{eqnarray}
&&\int \frac {d^nk}{(2\pi)^n} e^{-tk^2} =\frac 1{(4t\pi)^{n/2}}
\, , \nonumber \\
&&\int \frac {d^nk}{(2\pi)^{n}}  e^{-tk^2} k^\mu k^\nu
=\frac 1{(4t\pi)^{n/2}}\frac 1{2t} g^{\mu \nu} , \nonumber \\
&&\int \frac {d^nk}{(2\pi)^n} e^{-tk^2}  k^\mu k^\nu k^\rho k^\sigma
=\frac 1{(4t\pi)^{n/2}} \frac 1{4t^2} (g^{\mu\nu}g^{\rho \sigma}+
g^{\mu\rho}g^{\nu\sigma}+g^{\mu\sigma}g^{\nu\rho} ) .\label{eq:ex2}
\end{eqnarray}
Now we isolate $\exp (-tk^2)$ on the right hand side of (\ref{flathk1})
and expand the rest of the exponent in a power series of $t$.
\begin{eqnarray}
K(f;t) &=&\int d^nx \int \frac {d^4k}{(2\pi)^n} e^{-tk^2}
\ptr \left\{ f(x)\left( 1+t(\nabla^2 +E)
-\frac {t^2}2 4(k\nabla ) \right. \right. \nonumber \\
&+& \frac {t^2}2 (\nabla^2 \nabla^2 +\nabla^2 E +E\nabla^2 +E^2)
\nonumber \\
&-& \frac {4t^3}6 ((k\nabla )^2E+E(k\nabla )^2 +(k\nabla )E(k\nabla ))
\nonumber \\
&-& \frac {4t^3}6
((k\nabla )^2\nabla^2+\nabla^2(k\nabla )^2 +(k\nabla )\nabla^2(k\nabla ))
\nonumber \\
&+&\left. \left. \frac {16t^4}{24} (k\nabla )^4 +\dots \right) \right\}
\end{eqnarray}
We use the integrals (\ref{eq:ex2}) to obtain
\begin{eqnarray}
K(f;t)&=& \frac 1{(4\pi t )^{n/2}}   \int d^nx 
  \ptr \left\{ f(x) \left( 1 + t E \right.\right.
\nonumber\\
&+&\frac {t^2}2 (\nabla^2 \nabla^2 +\nabla^2 E+E\nabla^2+E^2)
\nonumber \\
&-& \frac {t^2}3 (\nabla^2 E +E\nabla^2 +\nabla^\mu E\nabla_\mu )
\nonumber \\
&-& \frac {t^2}3 (2 \nabla^2 \nabla^2 +\nabla^\mu \nabla^2 \nabla_\mu )
\nonumber \\
&+& \left. \left.
\frac {t^2}6 (\nabla^\mu \nabla^\nu \nabla_\mu \nabla_\nu
+\nabla^2 \nabla^2+\nabla^\mu \nabla^2 \nabla_\mu )+O(t^3)
\right) \right\} \,.
\end{eqnarray}
All derivatives combine into commutators. Finally we get:

\begin{eqnarray}
&&K(f;t)=\frac 1{(4\pi t)^{n/2}} \int d^nx \ptr \left\{ f(x)
\left( 1 +t E \right. \right. \nonumber \\
&&\qquad\qquad\qquad \left. \left. + t^2 \left(
\frac 12 E^2 +\frac 16 E_{;\mu\mu}+
\frac 1{12} \Omega_{\mu \nu}\Omega^{\mu \nu} \right) +O(t^3)\right)
\right\}\,.
\label{eq:ex3}
\end{eqnarray}
From the equation (\ref{eq:ex3}) we conclude that $\alpha_{10}=30$.
We have obtained also an independent confirmation for the values 
of $\alpha_0$, $\alpha_1$, $\alpha_3$ and $\alpha_5$.

The method we used above in eqs. (\ref{flathk1}) - (\ref{eq:ex3})
can be applied to more general manifolds and operators as well.
The key ingredient is a convenient basis which should be used
instead of the plane waves.  
For the case of a box with (anti-) periodic boundary condition
such a basis is rather obvious (see  
\cite{Andrianov:1984fg,Andrianov:1984qj}). On curved manifold
$M$ one has to use the so-called geodesic waves 
\cite{Fujikawa:1979ay,Fujikawa:1980eg,Yajima:1988pj,Yajima:1996jk,
Novozhilov:1991nm}
although calculations with ordinary plane 
waves are also possible \cite{Ceresole:1989hn}.

For calculation of the coefficient $a_6$ we refer to
\cite{Gilkey1975}. The results for the leading heat kernel
coefficients are summarised in the following equations
\begin{eqnarray}
&&a_{0}(f,D)=(4\pi)^{-n/2}\iM \ptr\{f\}. \label{a0nobou}\\
&&a_2(f,D)
         =(4\pi)^{-n/2}6^{-1}\iM \ptr\{f(6E+R)\}.\label{a2nobou}\\
&&   a_4(f,D)=(4\pi)^{-n/2}360^{-1}\iM \ptr\{f(60E_{;kk}
    +60R E+180E^2 \nonumber \\
&&\qquad\qquad
    +12R_{;kk}+5R^2-2 R_{ij}R_{ij}
    +2 R_{ijkl}R_{ijkl}+30\Cur_{ ij}\Cur_{ij})\}.\label{a4nobou} \\
&&a_{6}(f,D)=(4\pi)^{-n/2}\iM \ptr\bigl\{ \frac f{7!}(
    18R_{;ii jj}+17R_{;k}R_{;k}
    -2R_{ij;k}R_{ ij;k} \nonumber \\
&&\qquad\qquad -4R_{jk;n}R_{jn;k}
    +9R_{ij kl;n}R_{ij kl;n}+28RR_{;nn}
    -8R_{jk}R_{jk;nn} \nonumber \\
&&\qquad\qquad
    +24R_{ jk}R_{jn;kn}
    +12R_{ij kl}R_{ij kl ;nn}
    +35/9R^{3}
    -14/3RR_{ij}R_{ij} \nonumber \\
&&\qquad\qquad
    +14/3R R_{ijkl}R_{ijkl}
     -208/9R_{jk}R_{jn}R_{kn}
     -64/3R_{ ij}R_{kl}R_{ik jl} \nonumber \\
&&\qquad\qquad
     -16/3R_{jk}R_{jn l i}R_{kn l i}
     -44/9R_{ij kn}R_{ij l p}R_{kn l p} \nonumber\\
&&\qquad\qquad
     -80/9R_{ij kn}R_{il  kp}R_{jl  np})
     +360^{-1}f(
      8\Cur_{ij;k} \Cur_{ij;k}
     +2\Cur_{ij;j}\Cur_{ ik;k}\nonumber \\
&&\qquad\qquad
     +12\Cur_{ij;kk}\Cur_{ij}
     -12\Cur_{ij}\Cur_{jk}\Cur_{ki}
     -6R_{ij kn}\Cur_{ij}\Cur_{kn}
     -4R_{jk}\Cur_{jn}\Cur_{kn}
\nonumber \\
&&\qquad\qquad
       +5R\Cur_{kn}\Cur_{kn}
     +6E_{;ii jj}+60EE_{;ii}
      +30E_{;i}E_{;i}
     +60E^{3}
\nonumber \\
&&\qquad\qquad
     +30E\Cur_{ij}\Cur_{ij}
     +10R E_{;kk}+4R_{jk}E_{;jk}
     +12R_{;k}E_{;k}+30EER
     \nonumber \\
&&\qquad\qquad
      +12ER_{;kk}+5ER^2
     -2ER_{ij}R_{ij}+2ER_{ijkl}R_{ijkl})\}.\label{a6nobou}
\end{eqnarray}

Everyone who ever attempted calculations of the heat kernel
coefficients on curved background for arbitrary spin should
appreciate that the method presented here is a quite
efficient one. One should also take into account that some
of the universal constant were indeed calculated twice.

The coefficients $a_0$ -- $a_4$ are contained in 
\cite{DeWitt:1965,McKean:1967}.
$a_6$ was first computed by Gilkey \cite{Gilkey1975}.
The next coefficient $a_8$ has been calculated by Amsterdamski
{\it et al} \cite{Amsterdamski:1989bt} for the scalar Laplacian
and by Avramidi \cite{Avramidi:1990ug,Avramidi:1991je} for
the general operator of Laplace type.
The coefficient $a_{10}$ has been calculated 
by van de Ven\cite{vandeVen:1997pf}.
Higher heat kernel coefficients in flat space were studied in
\cite{Fliegner:1995zc,Fliegner:1998rk}.
\subsection{Examples}\label{s4examples}
Here we consider several simple physical systems in four dimensions
and calculate the heat kernel coefficient $a_4$ which defines the
one-loop divergences in the zeta function regularization. 
\subsubsection{Yang--Mills theory in flat space}\label{s4YM}
Our first example is pure Yang--Mills theory in flat space. We are 
interested in the ``total'' heat kernel coefficient $a_4^{\rm tot}$
defined by (\ref{atot}). Let us start with the first term describing
contribution from the vector fields. We choose the gauge (\ref{bg-gauge}).
The only non-vanishing invariants are $E$ and $\Omega$. The coefficient
$a_4$ is quadratic in these quantities. By using eqs.\ (\ref{EYM})
and (\ref{OmegaYM}) we obtain:
\begin{eqnarray}
&&\ptr \left( E^2\right)=E_{\nu\beta}^{\alpha\rho}E_{\rho\alpha}^{\beta\nu}=
4 F_{\rho\nu}^\delta F_{\rho\nu}^\gamma K_{\delta\gamma} \,,
\label{trE2vect} \\
&&\ptr \left( \Omega_{ij} \Omega_{ij} \right) = 
-4 F_{\rho\nu}^\delta F_{\rho\nu}^\gamma K_{\delta\gamma} \,,
\label{trO2vect}
\end{eqnarray} 
where 
\begin{equation}
K_{\delta\gamma}=c_{\alpha\beta}^{\delta}c_{\alpha\beta}^{\gamma} 
\label{Killform}
\end{equation}
is the Killing form of the gauge algebra.

Let us remind that the indices $i,j,k$ in (\ref{a4nobou}) refer to a local
orthonormal frame. In flat space they may be identified with the
vector indices $\mu,\nu,\rho$. In the case of vector fields the trace
in (\ref{trE2vect}) and (\ref{trO2vect}) is taken over {\it pairs}
consisting of a gauge index $\alpha$, $\beta$ or $\gamma$ and of a
vector index $\mu$, $\nu$ or $\rho$.

For the ghost operator we have to put $R=U=0$ in (\ref{omEscal})
and identify the connection with the background gauge field
according to (\ref{ghostcon}). Consequently, $E^{[{\rm gh}]}=0$ and
\begin{equation}
\ptr \left( \Omega_{ij}^{[{\rm gh}]} \Omega_{ij}^{[{\rm gh}]} \right)=
-F_{\rho\nu}^\delta F_{\rho\nu}^\gamma K_{\delta\gamma} \,.\label{trO2ghost}
\end{equation}
Next we substitute (\ref{trE2vect}), (\ref{trO2vect}) and 
(\ref{trO2ghost}) in (\ref{a4nobou}) to obtain
\begin{equation}
a_4^{[{\rm tot}]}=a_4^{[{\rm vec}]}-2a_4^{[{\rm gh}]}=
\frac{11}{96\pi^2} \int_M d^4x\, \sqrt{g}\, 
F_{\rho\nu}^\delta F_{\rho\nu}^\gamma K_{\delta\gamma}\,.
\label{YMa4}
\end{equation}

The Yang--Mills gauge group $\mathcal{G}$ has usually a direct product
structure, $\mathcal{G}=\mathcal{G}_1\times \mathcal{G}_2 \times\dots$,
so that on each of the irreducible components $\mathcal{G}_i$ the Killing
form $K$ is proportional to the unit matrix. Therefore, the one-loop
divergence (\ref{YMa4}) reproduces the structure of the classical action
with different charges for each $\mathcal{G}_i$. We also recover the
coefficient $11/3$ which is familiar from computations of
the Yang--Mills beta functions.
\subsubsection{Free fields in curved space}\label{s4free}
Consider free quantum fields on a Riemannian manifold without boundaries.
``Free'' means that we neglect all interactions except for the one
with the background geometry. The heat kernel coefficients can be
expressed then in terms of local invariants of the metric.
In particular,
\begin{equation}
a_4(x)=\frac 1{2880\pi^2} \left[
aC_{\mu\nu\rho\sigma}C^{\mu\nu\rho\sigma} +b\left( R_{\mu\nu}R^{\mu\nu}
-\frac 13 R^2 \right) +c {R_{;\mu}}^\mu +d R^2 \right]\,,
\label{a4in4dim}
\end{equation} 
where $C_{\mu\nu\rho\sigma}$ is the Weyl tensor,
\begin{equation}
C_{\mu\nu\rho\sigma}C^{\mu\nu\rho\sigma}=R_{\mu\nu\rho\sigma}
R^{\mu\nu\rho\sigma} - 2R_{\mu\nu}R^{\mu\nu} +\frac 13 R^2 \,.
\label{Weylsquare}
\end{equation}
$a$, $b$, $c$ and $d$ are some constants depending on the spin.
The first two structures in (\ref{a4in4dim}) which appear together
with $a$ and $b$ are conformally invariant in four dimensions.
This explains our choice of the basis in the space of invariants.

The constants $a$, $b$, $c$ and $d$ can be evaluated by substituting
particular expressions for $E$ and $\Omega$ obtained in sec.\ 
\ref{sec3} in (\ref{a4nobou}). Alternatively, one can use the
analysis of Christensen and Duff \cite{Christensen:1979md}
who calculated $a_4$ for arbitrary spin fields (see also
\cite{Birrell:1982ix}). The results are collected in Table \ref{a4table}.
Some comments are in order. Spin $1/2$ means 4-component Dirac spinors.
For spin 1 and spin 2 fields we took into
account contributions from corresponding ghost fields. Note, that
vector Yang-Mills fields and vector ghosts for gravity interact
{\em differently} with the background geometry. The cosmological
constant is taken to be zero. 
\begin{table}
\caption{$a_4$ for various spins}
\begin{tabular}{|l|c|c|c|c|}\hline \label{a4table}
Spin & $a$ & $b$ & $c$ & $d$ \\ \hline
$0$ &    $1$ & $1$ & $30\xi -6$ & $90 (\xi -1/6)^2$ \\
$1/2$&    $-7/2$ & $-11$ & $6$ & $0$ \\
$1$ &    $-13$  & $62$  & $18$ & $0$ \\
$2$ &    $212$  & $0$   & $0$  & $717/4$ \\ \hline 
\end{tabular}
\end{table}  

As a physical application of the heat kernel expansion in curved
space we may mention, for example, the asymptotic conformal invariance
phenomenon which was studied by using this technique in
\cite{Buchbinder:1984zx,Buchbinder:1985ba,Buchbinder:1984gj,Buchbinder:1988pm}.

A similar analysis can be performed also in the presence of boundaries
\cite{Moss:1994jj}.
\subsection{DeWitt iterative procedure}\label{s4iter}
The iterative method by DeWitt uses separation of the heat kernel
with non-coinciding arguments into a leading part (which is non-analytic
in $t$) and the power-law corrections. For a flat manifold this separation
has been described in sec.\ \ref{sec1} (cf.\ eqs.\ (\ref{simplestHK})
and (\ref{sHKexp})). Let us consider a massless scalar field on
a curved compact Riemannian manifold $M$. The DeWitt ansatz reads
in this case:
\begin{equation}
K(t;x,y;D)=(4\pi t)^{-n/2} \Delta_{VVM}^{1/2}(x,y) \exp 
\left( -\frac{\sigma(x,y)}{2t} \right) \Xi (t;x,y;D)\,,
\label{DWansatz}
\end{equation}
where $\sigma (x,y)$ is one half the square of the length of
the geodesic connecting $x$ and $y$. In Cartesian coordinates
on a flat manifold $\sigma_{\rm flat}(x,y)=\frac 12 (x-y)^2$.
$\Delta_{VVM}$ is the so called Van Vleck--Morette determinant
\begin{equation}
\Delta_{VVM}(x,y) =\frac{\det \left( -\frac{\partial}{\partial x^\mu}
\frac{\partial}{\partial y^\nu} \sigma (x,y) \right)}{\sqrt{g(x)g(y)}}
\,.\label{VVMdet}
\end{equation}

As a consequence of the heat equation (\ref{heateq}) the kernel
$\Xi$ should satisfy
\begin{equation}
\left( \partial_t +t^{-1} (\nabla^\mu \sigma )\nabla_\mu +
\Delta_{VVM}^{-1/2} D \Delta_{VVM}^{1/2} \right) \Xi =0
\label{eqforXi}
\end{equation}
with the initial condition
\begin{equation}
\Xi (0;x,y;D)=1 \,.\label{initXi}
\end{equation}

The essence of the DeWitt method is to look for a solution of the
equation (\ref{eqforXi}) in the following form
\begin{equation}
\Xi (t;x,y;D) = \sum_{j=0}^\infty t^{j}b_{2j}(x,y;D)\,.
\label{ansatzforXi}
\end{equation}
The initial condition (\ref{initXi}) yields
\begin{equation}
b_0=1\,.\label{b0DW}
\end{equation}
The recursion relation 
\begin{equation}
\left( j +(\nabla^\mu \sigma )\nabla_\mu \right) b_{2j}+
\Delta_{VVM}^{-1/2} D \Delta_{VVM}^{1/2} b_{2(j-1)}=0\,,
\label{recursionDW}
\end{equation}
which follows from (\ref{eqforXi}) and (\ref{ansatzforXi}),
allows, in principle, to find the coincidence limits $x=y$
of higher heat kernel coefficients $b_{2j}$. An important
ingredient of such calculations is the coincidence limits
of symmetrized derivatives of the geodesic interval $\sigma (x,y)$
\cite{Synge:1931,Barvinsky:1985an}. This method becomes
very cumbersome beyond $a_4$. A refined nonrecursive procedure
to solve the DeWitt equation (\ref{recursionDW}) was used
by Avramidi \cite{Avramidi:1991je} to calculate $a_8$
(see also \cite{Avramidi:2001ns} for a short overview).
The method of DeWitt can be naturally extended to treat
coincidence limits of the derivatives of $b_{2j}(x,y;D)$.

The recursion relations can be generalised to the case
of manifolds with boundaries \cite{McAvity:1991we,McAvity:1992rf}.
However, for a practical use the functorial methods of sec.\ 
{\ref{sec5}} seem to be more convenient.
\subsection{Non-minimal operators}\label{s4nonmin}
Quantisation of gauge theories oftenly leads to second order
differential operators which are not of the Laplace type.
For example, by taking $\kappa\ne 1$ in (\ref{Svgf}) 
one obtains the following operator acting on the gauge field
fluctuations:
\begin{equation}
D^{[{\rm nm}]}_{\mu\nu} =-\Delta g_{\mu\nu} +
\left( 1-\frac 1\kappa \right) \nabla_\nu\nabla_\mu \,,
\label{nonminop}
\end{equation}
where, for simplicity, we suppose that the manifold $M$ is flat
($R_{\mu\nu\rho\sigma}=0$) and the gauge group is abelian
($c_{\alpha\beta}^\gamma=0$). The leading symbol of the operator
(\ref{nonminop}) (the part with the highest derivatives) has a 
non-trivial matrix structure. Such operators are called {\it 
non-minimal}.

In some simple cases (see, e.g., 
\cite{GBF:1991,Alexandrov:1996gu,Vassilevich:1995bg}) the spectral
problem for non-minimal operators can be reduced to the Laplacians.
If the leading part of a non-minimal operator has a form similar to
(\ref{nonminop}) but the lower order part is more or less 
arbitrary\footnote{This case covers most of the physical applications.},
necessary generalisations of the DeWitt technique were suggested by
Barvinsky and Vilkovisky \cite{Barvinsky:1985an}. The technique was
further developed in \cite{Gusynin:1991ek,Cho:1995ic}. Complete
calculation of $a_4$ required extensive use of the computer algebra
\cite{Gusynin:1997qs}. Most general non-minimal operators were
considered in \cite{Avramidi:2001tx} where, because of great technical
complexity, only the first two heat kernel coefficients were analysed.
Some calculations in various physical systems with non-minimal operators
can be found in 
\cite{Fradkin:1982ts,Parker:1984pe,Guendelman:1994ke,Pronin:1997bu}.

\section{Heat kernel expansion on manifolds with boundaries}\label{sec5}
\subsection{Two particular cases}\label{s5cases}
We start our analysis of manifolds with boundaries with two simple
examples. First, let us consider a one-dimensional manifold
$M=[0,\pi ]$. Let $D=-\partial_x^2$. We consider both Dirichlet
(\ref{Dirbc}) and Neumann (\ref{Neubc}) boundary conditions (taking
$S=0$ in (\ref{Neubc}) to simplify the calculations). The eigenfunctions
of $D$ are:
\begin{eqnarray}
{\mbox{Dirichlet:}}& &\ \sin (lx),\ l=1,2,\dots ;\nonumber\\
{\mbox{Neumann:}}& &\ \cos (lx), \ l=0,1,2,\dots 
\label{1dimeigen}
\end{eqnarray}
The eigenvalues are $l^2$ in both cases, where $l$ is a positive integer
for Dirichlet boundary conditions, and $l$ is a nonnegative integer for
Neumann boundary conditions. The heat kernel asymptotics can be calculated
with the help of the Poisson summation formula (\ref{eq:asPoi}):
\begin{eqnarray}
&&K(t,-\partial_x^2,\mathcal{B}^-)=\sum_{l>0} \exp (-tl^2)=
\frac 12 \left( \sqrt{\frac{\pi}t} -1 \right) 
+\mathcal{O} \left( e^{-1/t} \right)
\,,\label{1dimDir}\\
&&K(t,-\partial_x^2,\mathcal{B}^+)=\sum_{l\ge 0} \exp (-tl^2)=
\frac 12 \left( \sqrt{\frac{\pi}t} +1 \right) 
+\mathcal{O} \left( e^{-1/t} \right)
\,,\label{1dimNeu}
\end{eqnarray}
where for the later use we explicitly mention the boundary operators
$\mathcal{B}^-$ and $\mathcal{B}^+$ which define Dirichlet (\ref{Dirbc})
and Neumann (\ref{Neubc}) boundary conditions respectively.

Let us modify a little bit the example above by allowing for a non-zero
$S$ at one of the components of the boundary:
\begin{equation}
\partial_x\phi \vert_{x=0}=0,\qquad
(\partial_x-S)\phi \vert_{x=\pi}=0 \label{1dimRob}
\end{equation}
Note, that at $x=\pi$ the derivative with respect to an {\it inward}
pointing unit normal is $-\partial_x$. The eigenfunctions  
\begin{equation}
\phi_k=\cos (kx) \label{1derob}
\end{equation}
satisfy the boundary conditions (\ref{1dimRob}) at $x=0$.
The spectrum is defined by the condition at $x=\pi$ which reads\footnote{
For $S>0$ also a negative mode $\cosh (kx)$ appears. Here we take
$S\le 0$.}:
\begin{equation}
-k \sin (k\pi)=S\cos (k\pi) \,.\label{1dspeceq}
\end{equation}
In this example we restrict ourselves to the linear order in $S$.
We suppose that $S$ is small and that the spectrum (\ref{1dimeigen})
is only slightly
perturbed: 
\begin{equation}
k_l=l+z_l \,.\label{1dpertspec} 
\end{equation}
The equation (\ref{1dspeceq}) gives in this approximation
\begin{eqnarray}
z_l=-\frac{S}{l\pi}&\qquad\mbox{for}\qquad& l>0\,,\nonumber \\
z_0^2=-\frac{S}{\pi}&\qquad\mbox{for}\qquad& l=0\,.\label{1dpertsp}
\end{eqnarray}
We expand also the heat kernel in a power series in $S$ and use
(\ref{1dpertsp}) and (\ref{1dimDir}) to obtain:
\begin{eqnarray}
K(t,-\partial_x^2,\mathcal{B}^+_S)&=&\sum_{l\ge 0} \exp (-tk_l^2)
\nonumber\\
&=&\sum_{l> 0} \exp (-tl^2) \left( 1+\frac{2St}{\pi} \right)
+1+\frac{St}{\pi} +\mathcal{O}(S^2)\nonumber\\
&=&\frac 12 \left( \sqrt{\frac{\pi}t} +1 \right) +
S\sqrt{\frac t\pi}  +\mathcal{O}(S^2) \,,\label{1dlinS}
\end{eqnarray}
where we also dropped exponentially small terms in $t$.

Our next example is the heat kernel expansion for scalar fields in
a ball with Dirichlet and Neumann boundary conditions
\cite{Stewartson:1971,Waechter:1972,Kennedy:1978jp,Moss:1989yf,Bordag:1996gm}.
The metric of the unit ball in $\mathbb{R}^n$ reads:
\begin{equation}
ds^2=dr^2+r^2 d\Omega^2, \quad 0\le r \le 1 , \label{ballmetric}
\end{equation}
where $d\Omega^2$ is the metric on unit sphere $S^{n-1}$.
The scalar Laplace operator has the form
\begin{equation}
\Delta \phi =
\left( \partial^2_r+\frac {n-1}r \partial_r+^{(n-1)}\Delta 
\right) \phi \,,
\label{ballLapl}
\end{equation}
where $^{(n-1)}\Delta$ is the Laplace operator on $S^{n-1}$.
The eigenfunctions
of the operator (\ref{ballLapl}) are well known
\begin{equation}
\phi_{l,\lambda }\propto r^{(2-n)/2 } J_{(n-2)/2 +l}
(r\lambda^{1/2} ) Y_{(l)}(x^a). \label{balleigenf}
\end{equation}
$J_p$ are the Bessel functions. The eigenvalues $-\lambda$
are defined by boundary conditions. $Y_{(l)}(x^i)$ are
$n$-dimensional scalar spherical harmonics. Their degeneracies
are
\begin{equation}
N_l=\frac {(2l+n-2)(l+n-3)!}{l!(n-2)!} . \label{spherdeg}
\end{equation}

To proceed further it is convenient to consider the zeta function
which may be presented through a contour integral 
\cite{Bordag:1996gm}\footnote{A general discussion of the representation
of the zeta function by contour integrals can be found in 
\cite{Kirsten:2003py}.}:
\begin{equation}
\zeta (s,D)=\sum_{l=0}^\infty N_l \int_\gamma
\frac{dk}{2\pi i} k^{-2s} \frac{\partial}{\partial k}
\Phi_{(n-2)/2 +l} (k) \,,\label{ballzeta}
\end{equation}
where $\Phi_\nu$ is a function which has zeros at the spectrum
$k=\sqrt\lambda$.
For Dirichlet boundary conditions this function reads:
\begin{equation}
\Phi_\nu =k^{-\nu} J_\nu (k) \,.\label{ballPhi}
\end{equation}
For Robin boundary condition $\Phi_\nu$ is given by a somewhat
more complicated combination of the Bessel functions. The contour 
$\gamma$ runs counterclockwise and encloses all the solutions
of  $\Phi_\nu =0$ on the positive real axis. Note the presence 
of $k^{-\nu}$ in (\ref{ballPhi}) which is included to avoid unwanted
contributions coming from the origin $k=0$.

The next step is to rotate the contour to the imaginary axis and
to calculate residues of $\Gamma (s)\zeta (s,D)$ (as prescribed
by (\ref{aRes})). It is interesting to note that the heat kernel
coefficients are defined by several leading terms in the uniform
asymptotic expansion of $\Phi_\nu (i\nu z)$ for large $\nu$ and
fixed $z$. For further details we refer to  \cite{Bordag:1996gm}.

Euclidean ball was frequently used in calculations of the heat kernel
coefficients and functional determinants. Apart from the papers
already quoted above also the computations for a scalar field 
\cite{Dowker:1995sp,Bordag:1996zc,Dowker:1996bh,
Bordag:1997ma,Nesterenko:1999bb}, spinors 
\cite{D'Eath:1991td,Kamenshchik:1993vj,Kamenshchik:1994tw,Dowker:1996ny,Dowker:1996sw,Kirsten:1996sy,Elizalde:1998hx,Esposito:2002vz},
abelian gauge fields 
\cite{Louko:1988ay,Esposito:1994bv,Esposito:1994xe,Vassilevich:1995we,Vassilevich:1995zn,Dowker:1996ny,Kirsten:1996sy,Esposito:1996hu,Esposito:1996gw,Elizalde:1997zw,Esposito:2000dy,Bernasconi:2003}\footnote{Some of these works consider contributions of the
so called physical modes only. As explained in \cite{Vassilevich:1995cz}
complete answer for the effective action and for the scaling behaviour
must include contributions from ghosts and non-physical modes. This
applies also to the spherical cap case considered below.}
should be mentioned.
A similar technique works also for more complicated
geometries as, e.g., the spherical cap 
\cite{Barvinsky:1992tr,Kamenshchik:1992gz,Barvinsky:1992dz,Kamenshchik:1993vj,
Dowker:1995sp}.
Note that here we consider local boundary conditions without tangential
derivatives only. For other types of boundary operators see sec.\
\ref{s5other}.
\subsection{Dirichlet and Neumann boundary conditions}\label{s5DN}
Let us now find analytic expressions for the heat kernel coefficients
in terms of geometric invariants. Here we follow the method of
Branson and Gilkey \cite{Branson:1990a,Gilkey:1995}. For both 
Dirichlet and modified Neumann (Robin) boundary conditions the
heat kernel coefficients are locally computable. This means
that $a_k$ may be represented as a sum of volume and boundary
integrals of some local invariants:
\begin{equation}
a_k(f,D,\mathcal{B})=\iM f(x) a_k(x,D)
+\sum_{j=0}^{k-1} \idM f^{(j)} a_{k,j}(x,D,\mathcal{B})\,,
\label{akb-loc}
\end{equation}
where $f^{(j)}$ denotes $j$-th normal derivative of the smearing
function $f$. Note, that we cannot now integrate by parts to remove
normal derivatives from $f$. This reflects the distributional nature
of the heat kernel asymptotics.

The volume terms $a_k(x,D)$ are the same as in the previous section
(see (\ref{a0nobou}) - (\ref{a6nobou})). Here we evaluate the
boundary terms $a_{k,j}(x,D,\mathcal{B})$. Obviously, the canonical
mass dimension of $a_{k,j}$ is $k-j-1$. In addition to the usual bulk
invariants $E$, $R$, etc, $a_{k,j}$ can also contain specific boundary
quantities as the extrinsic curvature $L_{ab}$ or $S$ (for Robin
boundary conditions). Note, that $L_{ab}$ and $S$ are defined on the
boundary only and, therefore, can be differentiated only 
tangentially\footnote{See sec.\ \ref{s2geo} for more information on
differential geometry of manifolds with boundary.}. Canonical mass dimension
of $L_{ab}$ and $S$ is $+1$.

In this section we explicitly calculate the first three heat kernel
coefficients $a_0$, $a_1$ and $a_2$. Basing on the considerations of the
preceding paragraph we may write:
\begin{eqnarray}
&&a_0(f,D,\mathcal{ B}^\pm)=(4\pi)^{-n/2}\iM {\rm tr}_V(f),\label{a0DN} \\
&&a_1(f,D,\mathcal{ B}^\pm)=(4\pi)^{-(n-1)/2}\idM \, b_1^\pm
     {\rm tr}_V ( f), \label{a1DN} \\
&&a_2(f,D,\mathcal{ B}^\pm)=\frac 16 (4\pi)^{-n/2}\left\{ \iM
     {\rm tr}_V(6fE+fR) \right. \nonumber \\
&&\qquad\qquad \left. +\idM {\rm tr}_V(b_2^\pm fL_{ aa}
     +b_3^\pm f_{ ;n}+b_4^\pm fS)
     \right\} ,\label{a2DN}
\end{eqnarray}   
where $\mathcal{B}^\pm$ denotes either Robin ($\mathcal{B}^+$) or
Dirichlet ($\mathcal{B}^-$) boundary operator. $b^\pm$ are some
constants. To keep uniform notations we formally included $S$ also in the
expression for Dirichlet boundary conditions. It should be assumed
that $S=0$ for that case.

Since the product 
formula (\ref{product}) is still valid with obvious modifications
for the boundary contributions, we can again consider the
case $M=M_1\times S^1$ and repeat step by step the calculations
(\ref{77hk}) - (\ref{uIn}) thus arriving to the same conclusion that
the constants $b^\pm$ do not depend on dimension of the manifold
(all explicit $n$-dependence of the heat kernel is contained in the power
of $4\pi$). This property is crucial for our calculations.

The one dimensional examples\footnote{For $M=[0,\pi ]$ the boundary
integral reduces to a sum of two contributions from $x=0$ and $x=\pi$.}
(\ref{1dimDir}) and (\ref{1dimNeu}) immediately give
\begin{equation}
b_1^+=-b_1^-=\frac 14 \,.\label{b1pm}
\end{equation}
Our next example (\ref{1dlinS}) controls linear terms in $S$ and
gives
\begin{equation}
b_4^+=12 \,.\label{b4p}
\end{equation}
For Dirichlet boundary conditions we have set $S=0$. Therefore,
$b_4^-$ plays no role.

To define the constants $b_2^\pm$ and $b_3^\pm$ we use the conformal
variation equation (\ref{vara}). Variations of $E$ and $R$ are given
by (\ref{cvar1}). Conformal properties of the second fundamental
form $L_{ab}$ are dictated by that of the metric:
\begin{equation}
\frac d{d\epsilon} \vert_{\epsilon =0} L_{aa}=-fL_{aa} 
-(n-1)f_{;n} \,.\label{cvarLaa}
\end{equation}
We wish to keep the boundary conditions invariant under the conformal
transformations. This implies that the boundary operators $\mathcal{B}^\pm$
must transform homogeneously. This property holds automatically for
Dirichlet boundary conditions (\ref{Dirbc}). In the case of modified
Neumann boundary conditions conformal transformation of $S$ should
cancel inhomogeneous term in the connection $\omega_n$. This yields
\begin{equation}
\frac d{d\epsilon} \vert_{\epsilon =0}S=-fS+\frac 12 (n-2)f_{;n}\,.
\label{cvarS}
\end{equation}
Next we substitute the variational formulae (\ref{cvar1}), (\ref{cvarLaa})
and (\ref{cvarS}) in (\ref{vara}) with $k=2$. We collect the
terms with $f_{;n}$ on the boundary to obtain
\begin{equation}
-(n-4)-(n-1)b_2^-=(n-2) b_3^- \label{cvara2b}
\end{equation}
for Dirichlet conditions and
\begin{equation}
-(n-4)-(n-1)b_2^+ +6(n-2)=(n-2) b_3^+ \label{cvara2N}
\end{equation}
for generalised Neumann (Robin) ones. Since the constants $b^\pm$ do
not depend on $n$ the two equations (\ref{cvara2b}) and (\ref{cvara2N})
are enough to define these constants:
\begin{equation}
b_2^+=b_2^-=2\,,\qquad b_3^+=-b_3^-=3 \,.\label{bpm23}
\end{equation}
This completes the calculation of $a_2$. 

The coefficients $a_3$ and $a_4$ can be obtained as particular
cases of more general formulae of the next subsection.
\subsection{Mixed boundary conditions}\label{s5mixed}
Let us now turn to mixed boundary conditions (\ref{mixbc}) which,
as we have seen in sections \ref{s3spin} and \ref{s3vect}, are
natural boundary conditions for spinor and vector fields.  
These boundary conditions depend on two complementary local
projectors $\Pi_-$ and $\Pi_+=1-\Pi_-$ which define subsets
of components of the field satisfying Dirichlet and Robin
boundary conditions respectively\footnote{Examples of such projectors
are given by eq.\ (\ref{proBG}) and below eq.\ (\ref{relbc}).}.
More precisely,
\begin{equation}
\Pi_-\phi \oB =0\,,\qquad (\nabla_n+S)\Pi_+\phi \oB =0\,.\label{mixagain}
\end{equation}
Consequently, there is one more independent entity (as compared to Robin
case, for example) on which the heat kernel coefficients for mixed 
boundary conditions can depend. This makes the calculations somewhat
more complicated. It is convenient to define
\begin{equation}
\chi = \Pi_+-\Pi_- \,.\label{defchi}
\end{equation}

Calculation of the coefficients $a_k$ up to $k=4$ can be found
in \cite{Branson:1990a} (see also \cite{Vassilevich:1995we} for
some corrections). The result reads:
\begin{eqnarray}
&&a_0(f,D,\mathcal{ B})=(4\pi)^{-n/2}\iM {\rm tr}_V(f).\label{a0bou} \\
&&a_1(f,D,\mathcal{ B})={\frac 14}(4\pi)^{-(n-1)/2}\idM
     {\rm tr}_V (\chi f). \label{a1bou} \\
&&a_2(f,D,\mathcal{ B})=\frac 16 (4\pi)^{-n/2}\left\{ \iM
     {\rm tr}_V(6fE+fR) \right. \nonumber \\
&&\qquad\qquad \left. +\idM {\rm tr}_V(2fL_{ aa}+3 \chi f_{ ;n}+12fS)
     \right\} .\label{a2bou} \\
&&a_3(f,D,\mathcal{ B})=\frac 1{384}(4 \pi )^{
       -(n-1)/2}  \idM \ptr\big\{ f(96 \chi E+16\chi  R \nonumber\\
&&\qquad\qquad+8f \chi R_{anan} 
       +(13 \Pi_{+}-7 \Pi_{ -})L_{ aa}L_{ bb}+(2 \Pi_{ +}+10
      \Pi_{-})L_{ab}L_{ ab}\nonumber\\
&&\qquad\qquad+96SL_{ aa}
      +192S^2 
       -12 \chi_{ :a} \chi_{:a})+f_{ ;n}
      ((6 \Pi_{ +}+30 \Pi_{ -})L_{
        aa}\nonumber\\
&&\qquad\qquad+96S)+24 \chi f_{ ;nn} \}.\label{a3bou}\\
&&a_4(f,D,\mathcal{ B})=\frac 1{360} (4 \pi )^{
       -n/2} \big\{ \iM \ptr
       \{ f(60E_{ ; ii}+60 R E+180E^2 \nonumber \\
&&\qquad\qquad +30 \Omega_{ij} 
      \Omega_{ij} +12 R_{;ii}+
      5 R^2-2R_{ij}R_{ij}+2R_{ijkl}R_{ijkl}) \} \nonumber \\
&&\qquad\qquad+ \idM \ptr
      \big\{ f \{ (240 \Pi_{ +}-120 \Pi_{ -})E_{ ;n}\nonumber\\
&&\qquad\qquad+(42 \Pi_{
     +}-18 \Pi_{ -}) R_{ ;n} 
     +24L_{ aa:bb}+0L_{ ab:ab}+120EL_{
       aa}\nonumber\\ 
&&\qquad\qquad +20 R L_{ aa}+4R_{ an    an}L_{bb} 
      -12R_{ an    bn}L_{ ab}+4R_{
       ab    cb}L_{ ac}\nonumber\\
&&\qquad\qquad +\frac 1{21} \{ (280 \Pi_{ +}+40 \Pi_{
-})L_{ aa}L_{ bb}L_{ cc} 
    +(168 \Pi_{ +}\nonumber\\
&&\qquad\qquad -264 \Pi_{
-})L_{ ab}L_{ ab}L_{ cc}+(224 \Pi_{
+}+320 \Pi_{ -})L_{ ab}L_{ bc}L_{ac} \} \nonumber \\
&&\qquad\qquad +720SE+120S R +0SR_{ an    an}+144SL_{
aa}L_{ bb}+48SL_{ ab}L_{ ab} \nonumber \\
 &&\qquad\qquad+480S^2L_{ aa}+480S^{
3}+120S_{ :aa}+60 \chi  \chi_{ :a} \Omega_{ an}-12 \chi_{ :a} \chi_{
:a}L_{ bb} \nonumber \\
&&\qquad\qquad -24 \chi_{ :a} \chi_{
:b}L_{ ab}-120 \chi_{ :a} \chi_{
:a}S \} +f_{ ;n}(180 \chi E+30 \chi  R +0R_{
an an} \nonumber \\
&&\qquad\qquad +\frac 17 \{ (84 \Pi_{
+}-180 \Pi_{ -})L_{ aa}L_{ bb}+(84 \Pi_{
+}+60 \Pi_{ -})L_{ ab}L_{ ab} \} \nonumber \\
&&\qquad\qquad +72SL_{ aa}+240S^2-18 \chi_{
:a} \chi_{ :a})+f_{ ;nn}(24L_{
aa}+120S) \nonumber \\
&&\qquad\qquad
\left. +30 \chi f_{ ;iin}  \big\} \right\} . \label{a4bou}
\end{eqnarray}
Dirichlet and modified Neumann (Robin) boundary conditions 
are recovered when $\Pi_+=0$
or $\Pi_-=0$ respectively.

The coefficients $a_k$ with $k=0,1,2,3,4$ were calculated
in  \cite{McKean:1967,Kennedy:1980ar,Smith:1981,Melmed:1988hm,Moss:1989mz,
Branson:1990a,McAvity:1991we,McAvity:1992rf,Vassilevich:1995we}.
A different algorithm was suggested in \cite{Cognola:1990kq}.
For ``pure'' (not mixed) boundary conditions the 
coefficient $a_5$ was calculated
by Branson, Gilkey and Vassilevich \cite{Branson:1997cm} for the
special case of a domain in flat space or of a curved domain with
totally geodesic boundaries. Kirsten \cite{Kirsten:1997qd} generalised
these results for arbitrary manifolds and boundaries. Branson, Gilkey,
Kirsten and Vassilevich \cite{Branson:1999jz} calculated the coefficient
$a_5$ for mixed boundary conditions.
\subsection{Other boundary conditions}\label{s5other}
From the technical point of view boundary conditions for a Laplace
type operator are needed to exclude infinite number of negative and
zero modes and to ensure self-adjontness. In principle, any linear
relation between the boundary data $\phi\oB$ and $\phi_{;n}\oB$
is admissible as long as it serves this purpose. Here we consider
two physically motivated examples of boundary conditions which
contain tangential derivatives on the boundary. These examples should
give the reader an idea of what can be expected for a more general 
boundary value problem.
\subsubsection{Boundary conditions with tangential derivatives and
Born--Infeld action from open strings}\label{s5tang}
The boundary condition
\begin{equation}
\left. \left( \nabla_n +\frac 12 ( \nabla_a \Gamma_a + \Gamma_a \nabla_a )
+S \right) \phi\right\oB=0 \label{obliquebc}
\end{equation}
is the simplest condition containing both normal and tangential
derivatives. $\Gamma_a$ and $S$ are some matrix valued functions
defined on the boundary. Such or similar structures appear in
open strings (cf. sec.\ \ref{s3string}) and in chiral bags
(cf. sec.\ \ref{s3spin}). They also describe photons with the
Chern-Simons interaction term concentrated on the boundary
\cite{Elizalde:1998ha,Bordag:1999ux}
and may be relevant for solid state physics applications.
The conditions (\ref{obliquebc}) appeared in the mathematical
literature \cite{Grubb:1974,Gilkey:1983}.

Several heat kernel coefficients for the boundary conditions
(\ref{obliquebc}) (which are called {\it oblique}) have been
calculated by McAvity and Osborn \cite{McAvity:1991xf} and
by Dowker and Kirsten \cite{Dowker:1997mn,Dowker:1998hm}.
Avramidi and Esposito 
\cite{Avramidi:1998sb,Avramidi:1998sh,Avramidi:1997hy,Avramidi:1998xj}
lifted some commutativity assumptions and proved a simple
criterion of strong ellipticity (see below).

Consider a differential operator $Q$, not necessarily of the second order.
Let is separate the part containing highest derivatives and replace
$\partial_\mu \to ik_\mu$, like in doing the Fourier transformation.
In this way we obtain an object $A_Q=a^{\mu\nu\dots\rho}(x)k_\mu k_\nu
\dots k_\rho$ which is called the leading symbol of $Q$. If the operator
$Q$ acts on fields with indices (i.e. in a vector bundle), the index structure
of $Q$ is inherited by $A_Q$. In other words, for fixed $k$ the leading symbol
$A$ is a matrix valued function (an endomorphism of the vector bundle).
If $A_Q(x)$ is non-degenerate for all $k\ne 0$, the operator $Q$ is called
elliptic. For the Dirac operator $Q=\Dir$ the leading symbol is 
$A_{Q} =\gamma^\mu k_\mu$. Therefore, $\Dir$ is obviously elliptic.
The Laplace type operators are also elliptic since $A_D=k^2$. 
Ellipticity means that ``at large momenta'' the operator $Q$ is dominated
by its' highest derivative part. This is the property which guarantees
that on a compact manifold  without boundaries Laplace operators have
at most finite number of negative and zero eigenvalues and which
ensures existence of the heat kernel. On manifolds with boundaries
just ellipticity is not enough to ensure nice properties of the spectrum.
There is an additional requirement, called {\it strong ellipticity},
which should be satisfied by the boundary operator (see 
\cite{Gilkey:1995} for details). Dirichlet and Robin boundary conditions
are always strongly elliptic. Oblique boundary conditions are strongly
elliptic if and only if $|\Gamma^2|<1$. If this inequality is violated,
infinite number of negative eigenmodes appears (a simple example can be
found in Appendix B of \cite{Kummer:2000ae}). Note that the boundary 
conditions (\ref{obliquebc}) correspond to a symmetric $D$ if 
the matrices $\Gamma_a$ are anti-hermitian. Therefore, 
$\Gamma^2=\Gamma_a\Gamma^a$ is typically negative.

Another complication stems from the fact that $\Gamma_a$ is dimensionless.
Consequently, arbitrary powers of $\Gamma_a$ may enter $a_k$, so that instead
of the undetermined constants (cf. (\ref{a0DN}) - (\ref{a2DN}))
one has to deal with undetermined functions of $\Gamma$. The problem
becomes more tractable if we suppose that $\Gamma_a$'s commute among
themselves, $[\Gamma_a,\Gamma_b]=0$. For this case, the coefficients
$a_1$ and $a_2$ have been calculated by McAvity and Osborn 
\cite{McAvity:1991xf} ($a_0$ is still given by (\ref{a0DN})):
\begin{eqnarray}
&&a_1(f,D)=(4\pi )^{-(n-1)/2} \idM \ptr (f\gamma (\Gamma ))\,,
\label{a1ob}\\
&&a_2(f,D)=\frac 16 (4\pi)^{-n/2}\left\{ \iM
     {\rm tr}_V(6fE+fR) \right. \nonumber \\
&&\qquad\qquad +\idM {\rm tr}_V(b_0(\Gamma) fL_{ aa}
     +b_1(\Gamma) f_{ ;n}\nonumber\\
&&\qquad\qquad \left. +b_2(\Gamma) fS+f\sigma (\Gamma) L_{ab}
     \Gamma_a\Gamma_b )
     \right\} ,\label{a2ob}
\end{eqnarray}   
where
\begin{eqnarray}
&&\gamma =\frac 14 \left[ \frac 2{\sqrt{1+\Gamma^2}} -1\right]\,,
\nonumber\\
&&b_0=6\left[ \frac 1{1+\Gamma^2}-\frac 1{\sqrt{-\Gamma^2}}
{\rm artanh} \left( \sqrt{-\Gamma^2} \right) \right]+2\,,\nonumber\\
&&b_1=\frac 6{\sqrt{-\Gamma^2}} {\rm artanh} \left( \sqrt{-\Gamma^2} \right)
-3 \,,\label{obliqueconst}\\
&&b_2=\frac {12}{1+\Gamma^2} \,,\nonumber\\
&&\sigma = \frac 1{\Gamma^2} (2-b_0) \,.\nonumber
\end{eqnarray}
We see, that if $\Gamma^2$ approaches $-1$ the heat kernel blows up
indicating violation of the strong ellipticity condition.

As an example, let us consider the open string sigma model of sec.\
\ref{s3string}. To simplify the subsequent calculations we put $B_{AB}=0$.
In the zeta function regularization divergent part of the effective
action (\ref{Wzpole}) reads $(1/s)a_2(D)=(1/s)a_2(1,D)$. On general grounds
we expect that $a_2$ repeats the structure of the classical action
(\ref{actsigma}):
\begin{equation}
a_2=\int_{\partial M}d\tau  \beta_A^{[A]} \partial_\tau \bar X^A +
\mbox{bulk terms},\label{divstring} 
\end{equation}
where $\beta_A^{[A]}$ is a beta 
function\footnote{In this case the subscript $A$
is a target space vector index, while the superscript $[A]$ indicates the
coupling. Here we do not consider other beta functions $\beta^{[G]}$,
$\beta^{[B]}$ etc.}. We put $L_{ab}=0$ (otherwise we have had to introduce
a dilaton coupling on the boundary to achieve renormalizability).
We also suppose that the target space metric $G_{AB}$ is trivial.
With these simplifying assumptions the beta function can be easily
calculated from (\ref{divstring}), (\ref{a2ob}), (\ref{obliqueconst}),
(\ref{bopsigma}):
\begin{equation}
\beta^{[A]}_C=-\frac 1{2\pi} (\partial_A F_{BC})
(1+F^2)^{-1}_{BA} \,.\label{betaAC}
\end{equation}
By a lengthy but straightforward calculation one can demonstrate
that the condition $\beta^{[A]}_C=0$ is equivalent to the
equations of motion following from the Born-Infeld action on the
target space:
\begin{equation}
\mathcal{L}_{\rm BI}=\int d\bar X \sqrt{\det (1+iF)}=
\int d\bar X \exp \left( \frac 14 {\rm tr} \ln
(1+F^2) \right)\,,\label{BIaction}
\end{equation}
where $i$ appeared due to our rule of the Euclidean rotation
for the gauge fields. 

The Born-Infeld action has been derived from the beta functions
of the open string sigma model in \cite{Abouelsaood:1987gd,Callan:1987bc}
confirming an earlier work \cite{Fradkin:1985qd} which used different 
methods. The heat kernel analysis was performed by Osborn 
\cite{Osborn:1991gm} and then repeated in \cite{Kummer:2000ae}
for more general couplings. 
\subsubsection{Spectral or Atiyah--Patodi--Singer (APS) boundary conditions}
\label{s5spec}
Spectral boundary condition were introduced by Atiyah, Patodi and
Singer in their study of the Index Theorem 
\cite{Atiyah:1975jf,Atiyah:1976jg,Atiyah:1980jh}. These boundary conditions
are {\it global}, i.e. they cannot be defined by using
local data only.

Consider a Dirac type operator 
\begin{equation}
\Dir =i \gamma^\mu \tilde \nabla_\mu +\mathcal{E}\,,\label{gDir}
\end{equation}
where $\tilde\nabla$ is a covariant derivative with a compatible
connection:
\begin{equation}
\tilde \nabla_\mu \gamma_\nu =0,\label{compaticon}
\end{equation}
i.e., the gamma matrices are covariantly constant. $\mathcal{E}$
is a zeroth order operator (a matrix valued function). We also 
suppose that the connection in $\tilde\nabla$ is unitary. This means
that the connection one-form is represented by an anti-hermitian
matrix in a suitable basis. We restrict ourselves to the case
when $\mathcal{E}^\dag =\mathcal{E}$, so that the operator
$\Dir$ is formally self-adjoint in the bulk. Note, that compatible
unitary connection is not unique. 
Consider a first
order differential operator on the boundary:
\begin{equation}
P=\gamma_n \gamma^a \tilde\nabla_a +\frac i2 \left(
\mathcal{E}\gamma_n -\gamma_n \mathcal{E} \right) +\Theta (x)\,,
\label{Poper}
\end{equation}
where $\Theta (x)$ is a hermitian matrix valued function on $\partial M$.
The operator $P$ is a self-adjoint operator of Dirac type on the boundary. 
All functions on $\partial M$ can be decomposed in positive, negative,
and zero modes of $P$. Let us define $\Pi_-$ as a projector on the
space spanned by {\it non-negative} eigenspaces of $P$. Then the
equation
\begin{equation}
\Pi_-\phi \oB=0 \label{APSPmin}
\end{equation}
defines the APS boundary conditions. 

Of course, relative complexity and non-locality limits physical
applications of spectral boundary conditions. However, they appeared
in a number of axial anomaly calculations 
\cite{Romer:1977qq,Hortacsu:1980kv,Hortacsu:1983iq,
Ninomiya:1985ge,Niemi:1986ht,Ma:1986qp,
Forgacs:1987ti,Falomir:1998as}, quantum cosmology \cite{D'Eath:1991sz},
and in works on the Aharonov-Bohm effect
\cite{Mishchenko:1992vm,Beneventano:1998ai}.
In brane models spectral boundary conditions describe T-selfdual
configurations which may be interpreted as mixtures of D-branes
and open strings \cite{Vassilevich:2001at}.

General form of the heat kernel expansion for spectral boundary
conditions was established by
Grubb and Seeley \cite{GS:1993,GS:1995,GS:1996}:
\begin{equation}
K(t,f,D)\simeq \sum_{k=0}^{n-1} a_kt^{(k-n)/2}
+\sum_{j=n}^\infty \left( a'_j \ln t +a''_j \right)
t^{(j-n)/2} \,.\label{APSasymp}
\end{equation}

In the contrast to all previous cases the expansion (\ref{APSasymp}) contains
logarithms of the proper time $t$. Although, such terms appear ``typically''
\cite{GG98} for pseudo-differential operators, they may lead to rather
unpleasant physical consequences. As follows from (\ref{heatzeta}),
non-zero $a'_n$ means that the zeta function has a pole at $s=0$, and,
therefore, the expression (\ref{Wzren}) for the renormalised
effective action does not make sense. Fortunately, for the APS
boundary conditions $a'_n=0$ if $f=1$ near the boundary. 
Consequently, the integrated
zeta function is regular at $s=0$ and eq.\ (\ref{Wzren}) still can be
used. However, all calculations involving localised heat kernel
coefficients remain problematic. Many logarithmic terms vanish if
the manifold $M$ has a product structure near $\partial M$.
In non-product cases strong criteria of ``partial vanishing of
logarithms'' have been found recently \cite{Grubb:2003ys}.

Another problem with the heat kernel expansion (\ref{APSasymp})
is that the coefficients may have a more complicated dependence
on $n$ than just a power\footnote{Let us remind that
to prove the simple dependence on $n$ we used the product
formulae (\ref{77hk}), (\ref{eq:asPoi}). We assumed that the
spectral problem can be ``trivialised'' in one direction.
In the present case all tangential coordinates enter the operator
(\ref{Poper}) on equal footings. Therefore, the proof does not go through.}
 of $(4\pi )$. After this long list of troubles it is not a surprise
that {\it only} the coefficients $a_k$ with $k<n$ are locally
computable.
The coefficient $a_0$ is given by (\ref{a0nobou}).
$a_1$ and  $a_2$ have been calculated in \cite{Dowker:2000sy}:
\begin{eqnarray}
&&a_1=(4\pi )^{(1-n)/2} \frac 14 (\beta (n)-1) \idM \ptr (f)\,,
\nonumber\\
&&a_2=(4\pi )^{-n/2}\left[ \iM \ptr \left( f\left(\frac 16 R+E\right) \right)
\right.\nonumber\\
&&\qquad\qquad +\idM \ptr \left( f \mathcal{E} +\frac 13 \left( 1-\frac 34 
\pi \beta (n) \right)L_{aa}f \right. \label{hkAPS} \\
&&\qquad\qquad \left.\left. -\frac{n-1}{2(n-2)} \left( 1-\frac 12 \pi
\beta (n)\right) f_{;n} \right) \right] \,,\nonumber
\end{eqnarray}
where
\begin{equation}
\beta (n)=\Gamma \left (\frac n2 \right) \Gamma \left( \frac 12 \right)^{-1}
\Gamma \left( \frac{n+1}2 \right)^{-1} \,.\label{betaAPS}
\end{equation}
The coefficient  $a_3$ can be found in \cite{Gilkey:2000bc}.

Some string theory applications suggest 
\cite{Vassilevich:2001at,Vassilevich:2001wf} that spectral boundary conditions
can be defined directly for a second order differential operator.
Existence of the asymptotic expansion (\ref{APSasymp}) and vanishing of
leading logarithms for such problems have been stated in
\cite{Grubb:2003yr}. 

\section{Manifolds with singularities}\label{sec6}
All results on the heat kernel expansion formulated in the
previous sections are valid on
smooth manifolds only. If there are boundaries, they also
have to be smooth. As well, any singularities in the potential
term or in the field strength are strictly speaking forbidden.
However, many physical models deal with singular backgrounds.
Even if such backgrounds may be represented through certain limiting
procedures from smooth configurations, the heat kernel coefficients
are not given by limits of their ``smooth'' values. The most visible
manifestation of failure of the smooth field approximation is that $a_k$
with sufficiently large $k$ are divergent. Usually, the presence of
singularities changes even the structure of the heat kernel
kernel expansion as compared to the smooth case. 

\subsection{Non-integrable potentials}\label{s6sinpot}
According to (\ref{divWLam}) divergences in the effective action
are defined by {\it integrated} heat kernel coefficients.
Although the formulae (\ref{a0nobou}) - (\ref{a6nobou}) for 
the {\it localised} heat kernel coefficients are valid on
non-compact manifolds (provided the smearing function $f$
falls off sufficiently fast), transition to the integrated
heat kernel is not that straightforward. Already the coefficient
$a_0(D)=a_0(1,D)$, which is proportional to the volume,
is divergent. This divergence is usually removed by replacing
$\det (D)$ in (\ref{pathdet}) by
\begin{equation}
\det (D)/\det (D_0) \,,\label{referdet}
\end{equation}
where the operator $D_0=-\partial^2+m^2$ describes a free particle
propagation in an ``empty'' space. It is argued that since $D_0$ does
not depend on ``essential'' variables division by $\det (D_0)$ does not
change physical predictions of the theory. In all subsequent formulae
the heat kernel $K(t;x,y;D)$ is then replaced by the subtracted
heat kernel
\begin{equation}
K_{\rm sub} (t;x,y)=K(t;x,y;D)-K(t;x,y;D_0) \,.
\label{Ksub}
\end{equation}
In flat space the coefficient $a_0$ corresponding to $K_{\rm sub}$
is identically zero\footnote{On a curved manifold the subtraction
procedure is more subtle. On has to define a reference metric which
differs from the physical one on a compact submanifold.}. If the field
strength $\Omega_{\mu\nu}$ and the (subtracted) potential $E+m^2$ 
have a compact support or decay sufficiently fast at the infinity,
the small $t$ asymptotic expansion of $K_{\rm sub}(t;x,x)$ is
integrable on the whole $M$. If not, the very structure of the
global heat kernel may be changed.

As an example of non-integrable potentials consider the harmonic
oscillator in one dimension. The Schr\"{o}dinger operator reads
\begin{equation}
D=-\partial_x^2 +\nu^2 x^2 \,.\label{harmos}
\end{equation}
If we consider the problem on the whole real axis, $M=\mathbb{R}$,
the potential term is not integrable. Already the expression
(\ref{a2nobou}) for $a_2(1,D)$ diverges. Therefore, analytic expressions
of sec.\ \ref{sec4} cannot be used in this case. However, the
(integrated) heat kernel can be easily calculated. Eigenvalues of the
operator (\ref{harmos}) are contained in almost any textbook on quantum
mechanics:
\begin{equation}
\lambda_j=\nu (2j+1),\qquad j=0,1,2,\dots \label{harmspec}
\end{equation}

The integrated heat kernel reads
\begin{equation}
K(t;D)=\sum_{j=0}^\infty e^{-t\nu (2j+1)} =
\frac 12 [\sinh (\nu t)]^{-1} \,.\label{harmheat}
\end{equation}
As $t\to 0$ it behaves like $1/t$ while for smooth rapidly decaying
potentials in one dimension the leading singularity in the heat kernel
is $1/\sqrt{t}$. This statement may be generalised to higher dimensions.
If $D=-\partial^2 +P_{\mu\nu}x^\mu x^\nu$ with a non-degenerate 
matrix $P_{\mu\nu}$ on $M=\mathbb{R}^n$, the leading
term in $K(t;D)$ is $(2t)^{-n} (\det P)^{-1/2}$ \cite{Avramidi:1995ik}.

\subsection{Conical singularities}\label{s6consin}
Conical space is defined as $M=[0,1]\times N$ where $N$ is an
$n-1$-dimensional manifold called the base. The metric of the
cone has the form
\begin{equation}
(ds)^2=dr^2 +r^2 d\Omega^2 \,,\label{conmetr}
\end{equation}
where $r\in [0,1]$ and $d\Omega^2$ is the line element on the
base $N$. This metric is, in general, singular at $r=0$. However, if we 
take the unit sphere $S^{n-1}=N$ with standard round metric,
the singularity disappears and we obtain the $n$-dimensional unit ball
(\ref{ballmetric}). If a manifold has singular points where
the metric can be approximated by (\ref{conmetr}) we say that this
manifold has conical singularities. 

Conical singularities appear in many physical applications.
First of all, with $N=S^1$ the metric (\ref{conmetr}) is the Euclidean
version of the Rindler metric. Conical singularities appear in
classical solutions of the Einstein equations 
\cite{Sokolov:1977,Frolov:1987dz,Barriola:1989hx}
and in the supermembrane theory 
\cite{Vassilevich:1991vy}. Gravitational field of a point mass in 
three dimensional gravity is a conical space 
\cite{Deser:1984tn,Deser:1988qn}. There are evidences \cite{Schleich:1993bs}
that ``conifolds'' dominate the path integral for quantum gravity in
topological sectors.

Sommerfeld \cite{Sommerfeld:1894}
was probably the first to consider the heat kernel in the presence
of conical singularities. The mathematical theory of the heat kernel
asymptotics with conical singularities was developed almost 100 years
later \cite{Cheeger:1983,Bruening:1984,Bruening:1987}. There two peculiar
features of these asymptotics. First, the heat kernel expansion
contains in general both integer and half-integer powers of $t$ even
without boundaries. Second, a non-standard $\ln t$ term may be contained in
the asymptotic series\footnote{The present author is not aware of
any simple example where the $\ln t$ terms actually appear.}.

On a manifold with conical singularities no closed analytical expression
for the heat kernel coefficients is available. However, usually it is
possible to disentangle contributions of the singularities from the
smooth part. For example, if $N=S^1$, $d\Omega^2=d\varphi^2$ with
$\varphi \in [0,\alpha ]$ only $a_2$ receives a contribution from the
tip of the cone:
\begin{equation}
a_2({\rm tip})=\frac{4\pi^2 -\alpha^2}{24\pi \alpha} \,.\label{a2tip}
\end{equation} 

In many particular cases of conical singularities
a very detailed analysis of the heat kernel
expansion has been performed 
\cite{Dowker:1977zj,Dowker:1989pp,Chang:1993fu,Dowker:1994jv,
Fursaev:1994qk,Fursaev:1994in,Cognola:1994qg,Cognola:1994ps,
Fursaev:1995ef,Bordag:1996fw,DeNardo:1997kp,Shtykov:1995fe,
Dowker:1998sc,BezerradeMello:2000mb}. One-loop computations on
general orbifolds were considered recently in 
\cite{GrootNibbelink:2003gd}.
\subsection{Domain walls and brane world}\label{s6dom}
Delta function is an example of an extremely sharp background potential.
Let us consider a manifold $M$ and a submanifold $\Sigma$ of the
dimension $n-1$. Let
\begin{equation}
D[\dvxi ]=D+\dvxi \delta_\Sigma \,.\label{singop}
\end{equation}
$D$ is an operator of Laplace type (\ref{laplaceb}). 
Let $h$ be
the determinant of the induced metric on $\Sigma$. Then $\delta_\Sigma$ is a
delta function defined such that
\begin{equation}
\int_M dx\sqrt{g} \delta_\Sigma f(x) =\int_\Sigma dx \sqrt{h}
f(x) \,.\label{deltaSig}
\end{equation}

The spectral problem
for $D[\dvxi ]$ on $M$ as it stands is ill-defined 
owing to the discontinuities (or
singularities) on $\Sigma$. It should be replaced by a pair of spectral
problems on the two sides
$M^\pm$ of
$\Sigma$ together with suitable matching conditions on $\Sigma$. In order to
find such matching conditions, we 
consider an eigenfunction $\phi_\lambda$ of the operator 
(\ref{singop}):
\begin{equation}
D[v]\phi_\lambda =\lambda \phi_\lambda \,.
\label{eigen}
\end{equation}
Let us choose the coordinate system such that $e_n$ is a unit normal
to $\Sigma$ and $x^n=0$ on $\Sigma$.
It is clear that $\phi_\lambda$ must be continuous on $\Sigma$:
\begin{equation}
\phi\vert_{x^n=+0}=\phi\vert_{x^n=-0} \,.\label{mc1}
\end{equation}
Otherwise, the second normal derivative of $\phi_\lambda$ would create
a $\delta'$ singularity on $\Sigma$ which is absent on the right
hand side of (\ref{eigen}). Let us integrate (\ref{eigen}) over
a small cylinder $\mathcal{C}=C^{n-1}\times [-\epsilon ,+\epsilon]$
\begin{equation}
\int_{\mathcal{C}} d^nx\sqrt{g} \left( -\nabla_n^2 \phi_\lambda
-\left[ \nabla_a^2 \phi_\lambda +(E+\lambda)\phi_\lambda \right]\right)
+\int_C d^{n-1}x\sqrt h \dvxi\phi_\lambda =0\,. \label{intcyl}
\end{equation} 
We now take the limit as $\epsilon\to 0$. Since the expression in the
square brackets in (\ref{intcyl}) is bounded, the contribution
that this term makes vanishes in the limit. We obtain
\begin{equation}
0=\int_C d^{n-1}x\sqrt h \left( -\nabla_n \phi_\lambda\vert_{x^n=+0}
+\nabla_n \phi_\lambda\vert_{x^n=-0} +\dvxi\phi_\lambda \right)\,.
\label{int2cyl}
\end{equation}
Since $C$ and $\lambda$ are arbitrary, we conclude that a proper
matching condition for the normal derivatives is
\begin{equation}
-\nabla_n \phi \vert_{x^n=+0}
+\nabla_n \phi \vert_{x^n=-0} +\dvxi\phi =0 \,.\label{dermatch}
\end{equation}

Physically this problem corresponds to two domains separated by a
penetrable membrane $\Sigma$ (a domain wall). In many cases penetrable
membranes are better models of physical boundaries then just boundary
conditions which are imposed on each side of $\Sigma$ independently
and thus exclude any interaction between the domains
\cite{Jaekel:1992bn,Actor:1995vc}. $\delta$-potentials
are being used in quantum mechanics \cite{Albeverio:1988}
where one studies the Schr\"{o}dinger equation (which is
nothing else than the imaginary time heat equation).
The Casimir energy calculations have been performed e.g. in
\cite{Scandurra:1998xa}.
In the formal limit
$\dvxi\to\infty$ one obtains Dirichlet boundary conditions 
on $\Sigma$, although the heat kernel coefficients are divergent in this
limit (see below).

Further generalisations are suggested by the brane world scenario
\cite{Randall:1999ee,Randall:1999vf} which assumes that our world
is a four dimensional membrane in a five dimensional space\footnote{
A similar scenario was proposed earlier in \cite{Rubakov:1983bb},
see \cite{Rubakov:2001kp} for a review.}. According to the Israel
junction condition \cite{Israel:1966rt} the metric in such models
cannot be smooth on $\Sigma$. Typical form of the metric near 
$\Sigma$ is 
\begin{equation}
(ds)^2=(dx^n)^2 + e^{-\alpha |x^n|} (ds_{n-1})^2, \label{branemetr}
\end{equation}
where $\alpha$ is a constant and where $(ds_{n-1})$ is a line element
on the $(n-1)$-dimensional hypersurface $\Sigma$. Due to the presence of the
absolute value of the $n$-th coordinate in (\ref{branemetr}), the normal
derivative of the metric jumps on $\Sigma$.
One can think of two smooth manifolds
$M^+$ and $M^-$ glued together along their common boundary $\Sigma$.
Neither Riemann tensor, nor matrix potential $E$ must be continuous
on $\Sigma$. Also, the extrinsic curvatures $L_{ab}^+$ and $L_{ab}^-$
of $\Sigma$ considered as a submanifold in $M^+$ and in $M^-$ 
respectively are, in general,
different. All geometric quantities referring to $M^-$ (respectively,
$M^-$) and their limiting values on $\Sigma$ will be supplied by
a superscript ``$+$'' (respectively, ``$-$'').  

For the case at hand there is still an asymptotic expansion (\ref{asymptotex})
for the heat kernel. The heat kernel coefficients can be decomposed as
\begin{equation}
a_k(f,D[\dvxi ])=a_k^+(f,D) + a_k^-(f,D)^- +a_k^\Sigma (f,D,\dvxi )\,,
\label{apmSigma}
\end{equation}
where $a_k^\pm (f,D)$ are known volume contributions corresponding
to $M^\pm$ (cf. (\ref{a0nobou}) - (\ref{a6nobou})). The coefficients
$a_k^\Sigma$ are given by integrals over $\Sigma$ of some local
invariants. Note, that 
\begin{equation}
\tilde\omega_a=\nabla_a^+-\nabla_a^- \label{tiloma}
\end{equation}
being a difference of two connection is a (pseudo-) vector with respect to
all space-time and gauge symmetries. Therefore, $\tilde\omega_a$ can be
used for constructing the surface invariants.

To make the formulae more symmetric we introduce {\it two} inward pointing
unit normals $\nu^+$ and $\nu^-$ to $\Sigma$ in $M^+$ and $M^-$ respectively.
We do not suppose that the smearing function $f$ is smooth on $\Sigma$
(therefore, there is no relation between $f_{;\nu^+}$ and $f_{;\nu^-}$),
but we assume continuity of $f$: $f^+=f^-=f$ on $\Sigma$.  

The surface invariants can be constructed from $L_{ab}^\pm$,
$R_{ijkl}^\pm$, $E^\pm$, $\dvxi$, $\tilde\omega_a$ and their
derivatives. This gives much more invariants than we usually have
for a boundary value problem. There are, however, some properties
of $a_k^\Sigma$ which simplify the calculations considerably.
First of all, $a_k^\Sigma$ must be invariant with respect to
interchanging the roles of $M^+$ and $M^-$. Also, $a_k^\Sigma$
must vanish when the singularity disappears. The first property
excludes, for example, the term $f(E^+-E^-)$ which changes sign
under $M^+\leftrightarrow M^-$. The second requirement excludes
the invariant $f(E^++E^-)$ because it survives even if there is no
singularity on $\Sigma$. These simple arguments show that $a_3^\Sigma$
does not contain $E^\pm$ even though such terms are allowed on dimensional
grounds. It is also very helpful that in some particular cases the
problem in question can be reduces to a sum of Dirichlet and Robin
boundary value problems \cite{Bordag:1999ed,Gilkey:2001mj}.

The coefficients $a_k^\Sigma$, $k=0,1,2,3$, read
\begin{eqnarray}
&&a_0^\Sigma(f,D,\dvxi)=0.\nonumber\\
&&a_1^\Sigma(f,D,\dvxi)=0.\nonumber\\
&&a_2^\Sigma(f,D,\dvxi)=
     (4\pi)^{-n/2}\frac16\int_\Sigma d^{n-1}x\sqrt{h}
       \ptr \{2f(L_{aa}^++L_{aa}^-)-6f\dvxi\}.\label{asig03}\\
&&a_3^\Sigma(f,D,\dvxi)=(4\pi)^{(1-n)/2}\frac1{384}
\int_\Sigma d^{n-1}x\sqrt{h}
\ptr\left\{\frac32f(L_{aa}^+L_{bb}^+
+L_{aa}^-L_{bb}^-\right.  \nonumber\\
&&\qquad +2L_{aa}^+L_{bb}^-)
    +3f(L_{ab}^+L_{ab}^++L_{ab}^-L_{ab}^-+2L_{ab}^+L_{ab}^-)
\nonumber\\
&&\qquad    +9(L_{aa}^++L_{aa}^-)(f_{;\nu^+}^++f_{;\nu^-}^-)
+48f\dvxi^2+24f\tilde\omega_a\tilde\omega_a \nonumber\\
&&\qquad \left.   -24f(L_{aa}^++L_{aa}^-)\dvxi
    -24(f_{;\nu^+}^++f_{;\nu^-}^-)\dvxi\right\}.\nonumber
\end{eqnarray}
The coefficients $a_4^\Sigma$ and $a_5^\Sigma$ are too long to be
presented here in full generality. Therefore, we restrict ourselves to 
the case of smooth geometry ($R_{ijkl}^+=R^-_{ijkl}$, $L_{ab}^+=-L_{ab}^-$),
smooth connection ($\tilde\omega_a=0$), and smooth smearing function
($f_{;\nu^+}=-f_{;\nu^-}=f_{;n}$). In other words, the only singularity comes
from the surface potential $\dvxi$.
\begin{eqnarray}
&&a_4^\Sigma (f,D,\dvxi)=(4\pi )^{-n/2}
\int_\Sigma d^{n-1}x\sqrt{h}\ptr\left\{ -\frac 16 f\dvxi^3 -\frac 1{6}
fR \dvxi -fE\dvxi \right. \nonumber \\
&&\qquad \left. -\frac 16 f\dvxi_{:aa} +\frac 16 f_{;n}\dvxi L_{aa} 
-\frac 16 f_{;nn}\dvxi
\right\} \nonumber \\
&&a_5^\Sigma (f,D)= (4\pi )^{-(n-1)/2} \int_\Sigma d^{n-1}x\sqrt{h}\ptr
\left\{\frac 1{64} f\dvxi^4 +\frac 1{48} fR \dvxi^2
 \right. \label{asig45}\\
&&\qquad +\frac 1{192}R_{nn}\dvxi^2 +\frac 18 f\dvxi^2 E 
-\frac 1{256} f\dvxi^2 L_{aa}L_{bb} 
+\frac 1{128}f\dvxi^2 L_{ab}L_{ab}  \nonumber \\
&&\qquad \left. +\frac 1{24}f\dvxi_{:aa}\dvxi
+\frac 5{192} f\dvxi_{:a}\dvxi_{:a} -\frac 5{384}f_{;n} \dvxi^2L_{aa}
+\frac 1{64}f_{;nn}\dvxi^2 \right\}
\nonumber
\end{eqnarray}
The heat kernel coefficients for $\Sigma =S^{n-1}\subset\mathbb{R}^n$
and $\dvxi =const.$ were calculated in \cite{Bordag:1998vs}.
Generic $\Sigma$ with arbitrary $\dvxi$ was considered in
\cite{Bordag:1999ed}. Moss \cite{Moss:2000gv} added a non-smooth
connection (see also \cite{Drozdov:2002um}).
Calculations on a particular brane-world background can be found
in \cite{Nojiri:2000bz}.
The heat kernel coefficients in the general setting
described here were calculated in \cite{Gilkey:2001mj}. This latter
paper also considered renormalization of the brane-world scenario
and predicted a non-standard Higgs potential on the brane.
Related calculations in 
wormhole models were done in 
\cite{Khusnutdinov:2002qb,Khusnutdinov:2003ii}. The $\zeta$
function for brane-world geometries with matching conditions
(\ref{mc1}), (\ref{dermatch}) was considered recently in
\cite{Elizalde:2002dd,Moss:2003zk} 
(where one can find some further references).

It is very well known \cite{Albeverio:1988}
that the conditions (\ref{mc1}),
(\ref{dermatch}) (which we call {\it transmittal}) are not the most
general matching conditions which can be defined on a surface. 
In general, boundary values of a function and of its' normal
derivatives are related by a $2\times 2$ {\it transfer} matrix:
\begin{equation}
0=
     \left(\begin{array}{cc}
          \nabla_{\nu^+}^++S^{++},\qquad&S^{+-}\\
          S^{-+},\qquad&\nabla_{\nu^-}^-+S^{--}\end{array}\right)
      \left(\begin{array}{l}\phi^+\\\phi^-\end{array}\right)
\bigg|_\Sigma\,.\label{BBT}
\end{equation} 
Note, that the transfer conditions (\ref{BBT}) do not assume
identification of $\phi^+$ and $\phi^-$ on $\Sigma$. In other
words, there is no ad hoc relation between the restrictions
of the vector bundles $V^+|_\Sigma$ and $V^-|_\Sigma$. We can even
consider the situation when we have ${\rm dim}V^+\ne{\rm dim}V^-$,
i.e. the fields on $M^-$ and $M^-$ can have different structures
with respect to space-time and internal symmetries. $S^{\pm\pm}$
are some matrix valued functions on $\Sigma$ (one can even consider
the case when they are differential operators). It is clear from
the notations on which spaces they act. For example,
$S^{+-}: V^-|_\Sigma\to V^+|_\Sigma$.

The matching conditions (\ref{BBT}) arise in heat transfer problems
\cite{Carslaw:1986}, some problems of quantum mechanics 
\cite{Grosche:1994uv}, and in conformal field theory
\cite{Bachas:2001vj}. In a formal limiting case 
$S^{++}-S^{-+}=S^{--}-S^{+-} \to \infty$ while 
$\dvxi =2(S^{++}+S^{+-})$ is kept finite one arrives at transmittal
boundary conditions (\ref{mc1}), (\ref{dermatch}). The heat kernel
coefficients are divergent in this limit.

In a particular case of spherical $\Sigma$ the heat kernel
expansion with transfer boundary conditions was evaluated in 
\cite{Bordag:2001ta}. General expressions for $a_k$ with
$k=0,1,2,3,4$ were obtained in \cite{Gilkey:2002nv}.
Somewhat surprisingly, the calculations for (\ref{BBT})
are easier than for the singular particular case (\ref{mc1}),
(\ref{dermatch}).

We should note that not all singular limiting cases of
(\ref{BBT}) are described by the transmittal conditions
(\ref{mc1}), (\ref{dermatch}). The heat kernel expansion
for a generalisation of transmittal condition is known
in the spherically symmetric case only \cite{Bordag:2001ta}.  
Very little is known about the heat kernel if the transfer
matrix contains differential operators on $\Sigma$ (conformal
walls of ref.\ \cite{Bachas:2001vj} belong to this class
of problems).

The case when the singular potential is located on a surface
of co-dimension larger than one (i.e. when 
${\rm dim}\Sigma <n-1$) is rather complicated. Even a careful
translation of this problem to the operator language
was done only in 1960's \cite{Berezin:1961}. Direct calculations
show \cite{Albeverio:1994,Solodukhin:1997xn} that
the heat kernel asymptotics may contain very unusual $(\ln t)^{-1}$
terms. More references can be found in \cite{Camblong:2001um}.
\subsection{Non-smooth boundaries}\label{s6nsbou}
Rectangular region in a plane is probably the simplest
manifold with boundaries as far as eigenvalues of the Laplacian  
are concerned. However, the formulae (\ref{a0bou}) - 
(\ref{a4bou}) are not valid for this case because of the presence
of corners. The heat kernel expansion on manifolds
with piecewise smooth boundaries was considered by Kac in his
famous paper \cite{Kac:1966}. He demonstrated that for a 
region in $\mathbb{R}^2$ each corner with the inside
facing angle $\alpha$ contributes 
\begin{equation}
a_2({\rm corner})= \frac {\pi^2 -\alpha^2}{24\pi\alpha}
\label{a2Kac}
\end{equation}
to the coefficient $a_2$ while $a_0$ and $a_1$ are still defined
by their ``smooth'' expressions.

The formula (\ref{a2Kac}) looks similar to the contribution of a
conical singularity (\ref{a2tip}). The reason for this similarity
is that the cone can be obtained from the wedge by gluing the sides
together and imposing the periodicity conditions.

The study of boundary discontinuities was continued by Apps and
Dowker who calculated the coefficients $a_3$ \cite{Dowker:1995sp}
and $a_4$ \cite{Apps:1998zr} for piecewise smooth boundaries.
We also refer to \cite{Aurell:1994} where functional determinants
on simplicial complexes were analysed, and to
\cite{Dowker:2000fs} where divergences in the Casimir energy
found in \cite{Nesterenko:2000eb} were attributed to non-smoothness
of the boundaries. A recent study \cite{Nesterenko:2002ng}
should also be mentioned.

When the angle $\alpha$ goes to $0$ we obtain a cusp. In this limit
(\ref{a2Kac}) is divergent. Presence of the cusp is an essential
singularity which modifies {\it powers} of the proper
time $t$ which appear in the asymptotic expansion of the heat
kernel \cite{Stewartson:1971}.
\subsection{Dielectric bodies}\label{s6diel}
Calculations of the Casimir energy of a dielectric body 
have attracted much attention and created many controversial
results. A (rather large) literature on this subject is reviewed
in \cite{Bordag:2001qi,Milton:2001yy}. Quantum field theory formulation
of this problem is known for a long time already (cf. 
\cite{Schwinger:1979pa,Milton:1980yx,Candelas:1982qw}). However, the
heat kernel analysis of divergences in the Casimir energy in
dielectric is a relatively new subject. 

Wave propagation with variable speed of light $c(x)$
is described by the operator $D=-c(x)^2 \nabla^2 +\dots$,
where we neglected the lower order terms. In a dielectric medium
$c$ is expressed in terms of the dielectric permittivity $\varepsilon$
and of the magnetic permeability $\mu$:
$c(x)^2=1/(\varepsilon (x)\mu (x))$. 
For a smooth distribution of $\varepsilon$ and $\mu$
the operator $D$
is a particular case of curved space Laplacian with
an effective metric defined by $c(x)$. The heat kernel
coefficients can be calculated in the standard way \cite{Bordag:1998fh}.

Consider now a dielectric body bounded by $\Sigma$. Typically, 
$c(x)$ (and the effective metric!) jumps on $\Sigma$.
This singularity is much stronger than the one considered
in sec.\ \ref{s6dom}. Thus the geometric interpretation of this
problem is very difficult. Very little is known about the heat kernel
expansion in a dielectric body of an arbitrary shape.
There are calculations for a dielectric ball 
\cite{Bordag:1998vs} and for a dielectric cylinder \cite{Bordag:2001zj}.
These calculations exhibit a puzzling property of the heat kernel
expansion in dielectrics in dilute approximation ($\varepsilon \sim 1$):
for a dielectric body the Casimir energy
in the ultra violet limit behaves 
better than for the ``smooth'' case.
The heat kernel expansion for a frequency dependent $\varepsilon$
was considered in refs.\ \cite{Falomir:2001uv,Bordag:2001fq}.

\section{Anomalies} \label{sec7}
The most immediate application of the technique
developed in the previous sections is calculation of
quantum anomalies which are defined as (non-zero) variations of
quantum effective action with respect to symmetry transformations
of the classical theory. In this section we consider two most
important examples of quantum anomalies. These are conformal (scale)
and chiral anomalies. We also discuss briefly the Index Theorem.

There exists also a broader view on the
anomalies which includes any qualitatively new phenomena of a quantum
theory which are absent in its' classical counterpart. An example is the
so-called dimensional reduction anomaly \cite{Frolov:1999an}.
For two operators $D_1$ and $D_2$ the quantity $\det (D_1) \det (D_2)/
\det (D_1D_2)$ is called the multiplicative anomaly 
\cite{Elizalde:1998nd}
since $\det (D_1D_2)\ne \det (D_1)\det(D_2)$  is an ``anomalous''
property of infinite dimensional operators. It is interesting to note
that the heat kernel is also useful for treatment of these non-standard
anomalies. 

To simplify the discussion in this section we work on manifolds without
boundaries.  
\subsection{Conformal anomaly}\label{s7conf}
Conformal invariance is one of the symmetries which are usually broken
by quantisation. This phenomenon (called conformal or trace anomaly)
is known since mid 1970's 
\cite{Capper:1974ic,Deser:1976yx,Dowker:1976tf,Brown:1977pq,Tsao:1977tj}
(see \cite{Birrell:1982ix,Duff:1994wm} for more extensive literature).

The vacuum polarisation induced by quantum effects is described
by the energy-momentum tensor
\begin{equation}
T_{\mu\nu}=\frac 2{\sqrt{g}} \frac{\delta W}{\delta g^{\mu\nu}}
\,,\label{EMT}
\end{equation}
where $W$ is the quantum effective action calculated on the background
with the metric tensor $g_{\mu\nu}$. Consider the
conformal transformation 
\begin{equation}
g_{\mu\nu}\to e^{2\rho(x)}g_{\mu\nu} \label{contra}
\end{equation}
for an infinitesimal value of the parameter:
$g_{\mu\nu}\to (1+2\delta\rho )g_{\mu\nu}$,
$g^{\mu\nu}\to (1-2\delta\rho )g^{\mu\nu}$. According to (\ref{EMT})
variation of the effective action reads:
\begin{equation}
\delta W = \frac 12 \iM T_{\mu\nu}\delta g^{\mu\nu}
=-\iM T_\mu^\mu \delta \rho \,.\label{convarW}
\end{equation}
It is clear from the equation above that the trace of the energy-momentum
tensor measures conformal non-invariance of the theory. If the classical
action $\mathcal{L}$ is conformally invariant, classical energy-momentum
tensor is traceless. However, even in this case conformal invariance is
typically broken by quantum effects. For this reason, quantum $T_\mu^\mu$
is called {\it trace} or {\it conformal} anomaly.

If the classical action $\mathcal{L}$ is conformally {\it invariant},
the fluctuation operator $D$ is conformally {\it covariant}. This means
that $D$ transforms homogeneously $D\to e^{-2\rho}D$ under 
(\ref{contra})\footnote{Strictly speaking this property holds up to a
similarity transformation $D\to e^{\alpha\rho}De^{-\alpha\rho}$ which
does not change the functional determinant.}.

We restrict ourselves to the one-loop level and employ the zeta function
regularization in which the effective action is expressed through the
zeta function of the operator $D$ (see (\ref{Wzren})). Hence, we have to
study the conformal properties of $\zeta (s,D)$.
One can prove that the variation of the zeta function with
respect to variation of the operator $D$ reads\footnote{\label{ft}
To derive this formula rigorously we have to use the method
of \cite{Atiyah:1980jh,Romanov:1979sc}. We first assume that $s$ is 
sufficiently large to keep us away from the singularities, then
use the Mellin transformation (\ref{zetaheat}), perform the variation,
then perform the transformation back, and then continue the result
to $s=0$. This is a perfectly standard procedure which allows us to 
work with variations of positive integer powers of $D$; see 
\cite{Esposito:1997wt} for further detail.}
\begin{equation}
\delta \zeta (s,D)=-s {\rm Tr}((\delta D)D^{-s-1})\,.
\label{varizeta}
\end{equation}   

Next we use that under infinitesimal conformal transformations
the operator $D$ transforms as
\begin{equation}
\delta D=-2(\delta \rho) D \,.\label{infcontraD}
\end{equation}
This equation yields
\begin{equation}
\delta \zeta (s,D) =2s \zeta (s,\delta \rho,D) \,.\label{convarizeta}
\end{equation}
For the operators which we consider in this section the zeta function
is regular at $s=0$. Consequently, the variation of the effective action
(\ref{Wzren}) reads:
\begin{equation}
\delta W=-\zeta (0,\delta\rho,D)=-a_n(\delta\rho,D)\,,
\label{cvWzren}
\end{equation}
where we have used the relation (\ref{anz0}) between the zeta function
at zero argument and the heat kernel coefficient $a_n$. 

We compare (\ref{convarW}) with (\ref{cvWzren}) to see that
\begin{equation}
T_\mu^\mu (x)= a_n(x,D) \,.\label{Tan}
\end{equation}
Note, that $\zeta (0,D)$ (as well as the
pole term in (\ref{Wzpole})) is conformally invariant. 
Consequently, the conformal anomaly is
not divergent, and (\ref{Tan}) does not contain the normalisation
scale $\mu$.

The anomaly (\ref{Tan}) is defined by the same coefficient $a_n$ as the
divergent part of the effective action in the zeta function regularization.
Important difference is that the divergence is given by the integrated
coefficient $a_n(D)$, while the conformal anomaly is defined by the
localised coefficient $a_n(x,D)$. The use of the integrated coefficient in 
(\ref{Tan}) is insufficient to recover total derivatives in the anomaly.  

As an example, consider quantum scalar field $\phi$ 
in two dimensions\footnote{Much work on conformal anomalies
on two-dimensional manifolds was done in the context of string
theory. Here we like to mention the papers 
\cite{Alvarez:1983zi,Friedan:1984xq,Friedan:1985ii,
Callan:1985ia,Alvarez:1987vr}.}
coupled
to the background dilaton $\Phi$:
\begin{equation}
\mathcal{L}=\int d^2x\sqrt{g} e^{-2\Phi} (\partial_\mu\phi)
(\partial_\nu\phi)g^{\mu\nu} \,.\label{nonminact}
\end{equation}
On dimensional and symmetry grounds the inner product may also
contain an arbitrary function $\Psi$ of the dilaton:
\begin{equation}
\langle \phi_1,\phi_2 \rangle =\int d^2x\sqrt{g} e^{-2\Psi}
\phi_1(x)\phi_2(x) \,.\label{dilaprod}
\end{equation} 
Such couplings and inner products appear, for example, after the
spherical reduction of higher dimensional theories to two
dimension (see the review paper \cite{Grumiller:2002nm} for
more details and further references).

The rescaled field $\tilde\phi =e^{-\Psi}\phi$ possesses the standard
dilaton-independent inner product (\ref{inn-prod-scal}). In terms 
of $\tilde\phi$ the action (\ref{nonminact}) reads  
\begin{eqnarray}
 &  & \mathcal{L}
=\int d^{2}x\sqrt{g}\tilde\phi D \tilde\phi \, ,\label{ehnac} \\
 &  & D=-e^{2(\Psi -\Phi )}g^{\mu \nu }(\nabla _{\mu }\nabla _{\nu }+2(\Psi _{,\mu }-\Phi _{,\mu })\partial _{\nu }\nonumber \\
&  & \qquad \qquad \qquad \qquad 
+\Psi _{,\mu \nu }+\Psi _{,\mu }\Psi _{,\nu }-2\Psi _{,\mu }\Phi _{,\nu }),
\label{ehnA} 
\end{eqnarray}
where comma denotes covariant differentiation with $\nabla$:
$\Psi_{,\nu}=\nabla_\nu\Psi$.
We can bring $D$ to the standard form (\ref{laplaceb})
\begin{equation}
D=-(\hat g^{\mu\nu} \hat\nabla_\mu\hat\nabla_\nu +E) \label{hatlalpl}
\end{equation}
by introducing the effective metric 
$\hat g^{\mu\nu}=e^{2(\Psi -\Phi)}g^{\mu\nu}$ and the covariant derivative
\begin{equation}
\label{ehno}
\hat{\nabla }_{\mu }=\partial _{\mu }+\hat{\Gamma }_{\mu }+\hat{\omega }_{\mu }\, ,\qquad \hat{\omega }_{\mu }=\Psi _{,\mu }-\Phi _{,\mu }\, ,
\end{equation}
 where \( \hat{\Gamma } \) is the Christoffel connection for the metric \( \hat{g} \).
Here the potential $E$ reads 
\begin{equation}
\label{ehnEh}
E=\hat{g}^{\mu \nu }(-\Phi _{,\mu }\Phi _{,\nu }+\Phi _{,\mu \nu })\, .
\end{equation}
Now we combine (\ref{Tan}) with (\ref{a2nobou}) and the definitions
given above to obtain:
\begin{equation}
\label{ehnT}
T_{\mu }^{\mu }=\frac{1}{24\pi }(R-6(\nabla \Phi )^{2}+4\nabla ^{2}\Phi +2\nabla ^{2}\Psi )\,. 
\end{equation}
In the case $\Psi =\Phi$ the expression (\ref{ehnT}) was first obtained in
\cite{Mukhanov:1994ax}, and in the general case in 
\cite{Kummer:1997jr}\footnote{Different values for the numerical coefficients
in the conformal anomaly (\ref{ehnT}) were reported in 
\cite{Bousso:1997cg,Nojiri:1997hx,Mikovic:1998xq,Nojiri:1998sr}.
The reason for these discrepancies was clearly stated by Dowker
\cite{Dowker:1998bm} who confirmed the result (\ref{ehnT}).
More extensive literature can be found in \cite{Grumiller:2002nm}.}.

In four dimensions the conformal anomaly for different spins can be read off
from (\ref{a4in4dim}) with the numerical coefficients given 
in Table \ref{a4table}.
\subsection{Chiral anomaly}\label{s7chiral}
Chiral anomaly was discovered in 1969 by Adler, Bell
and Jackiw \cite{Adler:1969gk,Bell:1969ts} and since
that time plays a crucial role in understanding of the
low energy hadron physics. A detailed introduction
to the field and extensive literature can be found
in \cite{Bertlmann:1996,Zinn-Justin:2002vj}.

The spinor action (\ref{spinS}) with the Dirac operator
given in (\ref{dirop}) is invariant under the gauge
transformations
\begin{eqnarray}
&&\delta_\lambda \psi = -\lambda\psi\,,\qquad
\delta_\lambda \bar \psi = \bar\psi \lambda \,,\nonumber \\
&&\delta_\lambda A_\mu^5 =[A_\mu^5,\lambda ]\,,\nonumber \\
&&\delta_\lambda A_\mu = \partial_\mu \lambda +[A_\mu ,\lambda ]
\label{gautr}
\end{eqnarray}
and local (Euclidean) chiral transformations
\begin{eqnarray}
&&\tilde\delta_\varphi \psi =-i\varphi \gamma^5\psi \,,\qquad
\tilde\delta_\varphi \bar \psi =-i\bar\psi \varphi \gamma^5
\,,\nonumber \\
&&\tilde\delta_\varphi A_\mu^5 =\partial_\mu \varphi +
[A_\mu ,\varphi ] \,,\nonumber \\
&&\tilde\delta_\varphi A_\mu = -[A_\mu^5,\varphi ] 
\label{chitr}
\end{eqnarray}
with anti-hermitian local matrix parameters $\lambda$ and
$\varphi$. The Dirac operator transforms as:
\begin{equation}
\delta_\lambda \Dir=[\Dir ,\lambda ]\,,\qquad
\tilde\delta_\varphi \Dir =i\{ \varphi \gamma^5,\Dir \} \,.
\label{transP}
\end{equation}
The Lie algebra structure of the transformations
(\ref{gautr}) and (\ref{chitr}) is encoded in the following
relations:
\begin{eqnarray}
&&\delta_{\lambda_1}\delta_{\lambda_2} -
\delta_{\lambda_2}\delta_{\lambda_1} 
=\delta_{[\lambda_1 , \lambda_2]} \label{gau-gau}\\
&&\delta_{\lambda}\tilde\delta_{\varphi} -
\tilde\delta_{\varphi} \delta_{\lambda}=
\tilde\delta_{[\lambda ,\varphi ]} \label{gau-chi}\\
&&\tilde\delta_{\varphi_1} \tilde\delta_{\varphi_2}-
\tilde\delta_{\varphi_2}\tilde\delta_{\varphi_1}
=-\delta_{[\varphi_1,\varphi_2]}\,, \label{chi-chi}
\end{eqnarray}
where all transformation parameters are taken at the same
space-time point. If the matrices $\lambda (x)$ and
$\varphi (x)$ belong to a finite-dimensional compact
Lie algebra of some Lie group $\mathcal{G}$, the transformations
(\ref{gautr}) and (\ref{chitr}) generate locally
the group $\mathcal{G}\times \mathcal{G}$ with gauge transformations
belonging the diagonal sub-group.

The gauge invariance can (and should) be retained in quantum
theory while the chiral invariance is typically broken by the
quantisation. Let us study these effects in the zeta function
regularization. Define determinant of the Dirac operator
as a square root of the determinant of the associated
Laplace operator:
\begin{equation}
\log \det \Dir =\frac 12 \log \det \Dir^2
=\frac 12 \log \det D
\label{defdetP}
\end{equation}
The effective action $W$ corresponding to the path integral
(\ref{pathspin}) reads
\begin{equation}
W=-\log Z =-\frac 12 \log \det D
=\frac 12 \zeta (0,D)' +\frac 12 \ln (\mu^2) \zeta (0,D)
\,,
\label{Wspin}
\end{equation}
where $\mu$ is a normalisation scale.

By virtue of (\ref{varizeta})
\begin{equation}
\delta_\lambda \zeta (s,D) =-
\left( s {\rm Tr} ( [D,\lambda ] D^{-s-1}) \right)
=\left( {\rm Tr} ( [D^{-s},\lambda ]) \right)
=0\,.
\end{equation}
This proves gauge invariance of the effective action.

For the chiral transformation we have:
\begin{equation}
\mathcal{A}(\varphi ):=\tilde\delta_{\varphi} W=
-2i{\rm Tr} (\gamma^5\varphi D^{-s})|_{s=0}\,.
\label{chian1}
\end{equation}
$\mathcal{A}(\varphi )$ measures non-invariance of the
effective action with respect to the chiral transformations. It 
is called the {\it chiral anomaly}. One can express $\mathcal{A}$
through the heat kernel coefficients:
\begin{equation}
\mathcal{A}(\varphi )=-2i a_n(\gamma^5\varphi , D) \,,
\label{Ahk}
\end{equation}
where $a_n(\gamma^5\varphi , D)$ is defined as in 
(\ref{asymptotex}) but with a matrix-valued smearing
function instead of the scalar one.

The chiral anomaly $\mathcal{A}(\varphi )$ should satisfy
certain consistency conditions following from the
Lie algebra identities (\ref{gau-gau})-(\ref{chi-chi})
and gauge invariance of the effective action:
\begin{eqnarray}
&&\delta_\lambda \mathcal{ A}(\varphi )=
\mathcal{A}([\lambda ,\varphi ] ) \,, \label{WZcc1}\\
&&\tilde\delta_{\varphi_1} \mathcal{A}(\varphi_2 )-
\tilde\delta_{\varphi_2} \mathcal{A}(\varphi_1 )=0 \,.\label{WZcc2}
\end{eqnarray}
The relations (\ref{WZcc1}) and (\ref{WZcc1}) are called the
Wess-Zumino consistency conditions \cite{Wess:1971yu}.
One can check by a direct calculation that the anomaly
defined in this section indeed satisfies these conditions.
For this reason $\mathcal{A}$ is called the {\it consistent}
anomaly (as opposed to the {\it covariant} anomaly which
we do not consider here).

Let us turn now to calculation of the chiral anomaly
$\mathcal{A}$. Since the smearing function $\gamma^5\varphi$
in the heat kernel coefficient (\ref{Ahk}) is matrix-valued
we need more information than we possess at the moment.
To recover the missing terms one can adopt the strategy of
the paper \cite{Branson:1997ze}. Consider 
${\rm Tr}(Q\exp (-tD))$ with arbitrary matrix-valued
function $Q$ and arbitrary Laplace type operator $D$.
There is an asymptotic expansion
\begin{equation}
{\rm Tr}_{L^2} (Q\exp (-tD)) 
\cong\sum_{k\ge0}t^{(k-n)/2}a_k(Q,D)
\label{Qasymptotex}
\end{equation}
where the coefficients $a_k$ are locally computable. This
means that they can be represented as integrals of local
invariants constructed from $Q$ and local invariants
of the operator $D$. These local invariants enter with
some coefficients (universal constants) which are to be
defined. Since $Q$ does not commute with $E$ and $\Omega$
there are more invariants than before. In the particular
case $Q=I_Vf$ (where $I_V$ is the unit matrix and $f$
is a function) we should recover the old result
(\ref{a0nobou}) - (\ref{a6nobou}). For $k=0,2,4$ this
last requirement is strong enough to recover 
$a_k(Q,D)$ completely. Therefore, calculation of the chiral
anomaly in the dimensions $n=2$ and $n=4$ is quite simple:
take the expression (\ref{a2nobou}) or (\ref{a4nobou})
and substitute $2i\gamma^5\varphi$ for $f$ and
(\ref{oEdir}), (\ref{Omdir}) for $\omega_\mu$, $E$
and $\Omega_{\mu\nu}$. Computation of some $\gamma$-matrix
traces is still required, but nevertheless calculations 
are considerably easier than presented in many papers.
In flat space and with
$A^5=0$ the result is particularly simple:
\begin{eqnarray}
&&n=2:\qquad \mathcal{A}(\varphi )= -\frac 1{2\pi}
\int d^2x {\rm tr}(\varphi \epsilon^{\mu\nu}F_{\mu\nu}) 
\label{Am2} \\
&&n=4:\qquad \mathcal{A}(\varphi )= \frac i{16\pi^2}
\int d^4x {\rm tr}(\varphi \epsilon^{\mu\nu\rho\sigma}
F_{\mu\nu}F_{\rho\sigma} )
\label{Am4}
\end{eqnarray}
where ${\rm tr}$ denotes trace over internal (flavour
or colour) indices. We can easily generalise this result
for arbitrary even dimension $n$. First we observe that
the only way to construct a pseudoscalar density of
appropriate dimension (for $A^5=0$) is to contract
$n/2$ tensors $F_{\mu\nu}$ with the Levi-Civita
tensor $\epsilon^{\mu_1\mu_2\dots\mu_n}$. Such tensor
structure can be produced only by a trace of $\gamma^5$ and
the maximal number
($n$) of the gamma matrices. 
The only invariant having the required  form is
\begin{equation}
\int d^n\,x {\rm tr}(QE^{n/2})\,.  
\label{1917}
\end{equation}
This term does not contain derivatives. Under the trace
$Q$ and $E$ commute. Therefore, we can calculate
the coefficient in front of (\ref{1917}) by considering the particular
case $E\sim {I}_V\times {const.}$. For this simple case
the dependence of the heat kernel on $E$ is given simply by $e^{tE}$.
By picking up an appropriate term in the expansion of the
exponential we find that the coefficient in front of (\ref{1917})
in $a_n$ is $(4\pi )^{-n/2} ((n/2)!)^{-1}$. Next we substitute
$E=\frac 12 \gamma^\mu \gamma^\nu F_{\mu\nu}$ 
and take trace over the spinorial indices to obtain:
\begin{equation}
\mathcal{A}(\varphi )=-\frac{2i (-i)^{n(n-1)/2}}{(4\pi)^{n/2}(n/2)!}
\int d^nx {\rm tr}\left(\varphi \epsilon^{\mu_1\mu_2\dots\mu_n}
F_{\mu_1\mu_2}\dots F_{\mu_{n-1}\mu_n} \right)
\label{Amarb}
\end{equation}
For global chiral transformations ($\varphi = const.$) this formula
was obtained in \cite{Wipf:1995dy}.

As soon as we know the anomaly (\ref{Amarb}) for $A_\mu^5=0$, many
more terms (containing $A_\mu^5$) can be restored by using the 
Wess-Zumino condition (\ref{WZcc2}).

Some comments on the Fujikawa approach 
\cite{Fujikawa:1979ay,Fujikawa:1980eg} 
to the chiral anomaly are in order.
Consider the path integral (\ref{pathspin}).
Since the action is invariant under the chiral rotations,
the only source of the chiral anomaly could be non-invariance
of the path integral measure. The Jacobian factor appearing due
to the change of the spinorial variables (\ref{chitr}) 
can be formally represented as
\begin{equation}
J=\det (1-2i\varphi\gamma^5 )\,. \label{Jac1}
\end{equation}
Then, to the first order of $\varphi$,
\begin{equation}
\mathcal{ A}(\varphi )=
\tilde\delta_\varphi W =-\log J^{-1} \simeq -2i{\rm Tr}(\varphi\gamma^5 ) 
\label{chian2}
\end{equation}
where $J^{-1}$ has appeared due to the negative homogeneity
of the fermionic measure. The operator $\varphi\gamma^5$ is not
trace class on the space of square integrable spinors (though
formally its' trace is zero at every point). The right hand side
of (\ref{chian2}) is therefore ill defined. Fujikawa suggested
to replace (\ref{chian2}) by a regularized expression
\begin{equation}
\mathcal{ A}(\varphi )= -2i \lim_{M\to \infty}
{\rm Tr}\left( \varphi\gamma^5 e^{-D/M^2} \right) \,.
\label{Fuji}
\end{equation}
The chiral anomaly (\ref{Fuji}) has now the heat-kernel
form with the identification $t=M^{-2}$. If we suppose that
all positive powers of the regularization parameter $M$ in
the small $t$ (large $M$) asymptotic expansion are somehow
absorbed in the renormalization, we arrive at 
\begin{equation}
\mathcal{ A}(\varphi ) =-2i a_n (\gamma^5\varphi, D)
\label{Fuji2} 
\end{equation}
that is just the expression (\ref{Ahk}) obtained above
in the zeta function regularization.
This method, based on calculations of the regularized
Jacobians, can be applied to conformal anomaly as well
\cite{Fujikawa:1980vr}\footnote{Applications of this method
to topological anomalies can be found in \cite{Fujikawa:2003gi}.
}.

The most essential ingredient of the anomaly calculation
presented in this section is the homogeneous transformation
low (\ref{transP}) for the Dirac operator. This homogeneity
allowed us to restore the power $-s$ in the transformation
law (\ref{varizeta}) for the zeta function and to obtain a
simple expression for the anomaly (\ref{chian1}). It is clear
therefore that as long as the operators transform homogeneously
we shall obtain relatively simple local expressions for
corresponding anomalies (understood as variations of the
effective action with respect to infinitesimal transformation
of the background fields). This suggests to consider extensions
of the chiral group. For example, one of such extensions 
\cite{Novozhilov:1995he}
identifies the group parameters with the diquark fields.

\subsection{Remarks on the Index Theorem}\label{s7index}
This report is mainly devoted to local aspects of the heat kernel
expansion. There is however one ``global'' 
application of the heat kernel which cannot be ignored. This is the
Index Theorem. In this section we briefly sketch formal
mathematical aspects of the index construction and its' relation
to the heat kernel. For more details we refer to 
\cite{Eguchi:1980jx,Gilkey:1995,Esposito:1997mw,Bertlmann:1996,
Zinn-Justin:2002vj}. Physical applications of the index theorem
to gravity, gauge theories and strings are so numerous that
we cannot even mention them all.

The index theorem was first formulated by Atiyah and Singer
\cite{Atiyah:1963a}, and the heat kernel approach appeared
later in \cite{Gilkey:1973,AtiyahBP:1973}. Roughly speaking,
their construction is as follows.
Consider two vector bundles $V_1$ and $V_2$ over a manifold $M$.
Let the operator $P$ map $V_1$ to $V_2$, and let $V_{1,2}$
have non-degenerate inner products $(\ ,\ )_{1,2}$ so that one
can define an adjoint $P^\dag$ by the equation
$(\phi_2,P\phi_1)_2=(P^\dag \phi_2,\phi_1)_1$. As an example,
one can keep in mind $V_1$ (respectively, $V_2$) describing
positive (respectively, negative) chirality spinors. In this
example $P$ is a part of the Dirac operator (see eq.\ (\ref{adiagdir})
below). Let us define two operators $D_1=P^\dag P$ and
$D_2=PP^\dag$ acting on (smooth sections of) $V_1$ and
$V_2$, and let us suppose that $D_1$ and $D_2$ are elliptic.
In this case we deal with an {\it elliptic complex}.

Since elliptic operators may have only a finite number of zero
modes (say, $N_1$ and $N_2$ for $D_1$ and $D_2$ respectively),
we may define the index by the following equations:
\begin{eqnarray}
{\mbox{index}}(P)&=& N_1 - N_2 \nonumber \\
&=& \dim \ker D_1 - \dim \ker D_2 \nonumber \\
&=& \dim \ker P - \dim \ker P^\dag \,.\label{defind}
\end{eqnarray}
We also have the intertwining relations
\begin{equation}
D_1P^\dag =P^\dag D_2 \,,\qquad PD_1=D_2P \,,\label{intertwin}
\end{equation}
which tell us that non-zero eigenvalues of $D_1$ and $D_2$
coincide. Consider now the heat kernels,
\begin{equation}
K(t,D_1)-K(t,D_2)=\sum_{(\lambda_1)} e^{-t\lambda_1}
-\sum_{(\lambda_2)} e^{-t\lambda_2} =
N_1 - N_2 ={\mbox{index}}(P) \label{hkindex}
\end{equation}
(where $\lambda_{1,2}$ are eigenvalues of $D_{1,2}$).

To be more specific, let us suppose that $D_1$ and $D_2$ are of
Laplace type and that boundary conditions and singularities (if any)
are such that the power-law expansion (\ref{asymptotex}) exists.
Then, expanding both sides of (\ref{hkindex}) in a power series
of $t$ one obtains:
\begin{eqnarray}
&&a_k(D_1)-a_k(D_2)=0 \qquad \mbox{for}\quad k\ne n,\nonumber\\
&&a_n(D_1)-a_n(D_2)=\mbox{index}(P) \,.\label{anindex}
\end{eqnarray}
The last equation provides a simple way to calculate the
index from known heat kernel expansion. It also allows to understand
a very important property of the index, namely, that it is a
{\it homotopy invariant} under quite general assumptions.
Indeed, suppose that $P$ depends on a parameter $\alpha$ in 
such a way that all geometric quantities (including
the metric, connection, the matrix potential, the boundary
conditions etc.)
corresponding to $D_1$ and $D_2$ are smooth functions of $\alpha$.
Clearly, no ``essential'' deformations, like changing the order
of the operator, or adding higher derivative terms to the boundary conditions, 
or including
new types of the singularities, are allowed. Under these smoothness
assumptions $a_n(D_1)$ and $a_n(D_2)$ are smooth functions of
$\alpha$ as well. Therefore, the index is also a smooth function
of $\alpha$. Since the index is an integer, it can only be a constant.
Hence, ${\mbox{index}}(P)$ is invariant under the deformations
described above.

The relations (\ref{anindex}) are also useful for calculations
of the heat kernel coefficients as they give restrictions on the
universal constants appearing in front of independent invariants
(see, e.g., \cite{Gilkey:1995}).

Let us consider an example of the {\it spin complex}. Let $M$ be
an even dimensional manifold admitting a spin structure (so that
one can define spinor fields on $M$). In a suitable basis
the chirality matrix may be presented in the diagonal form
(at this point it may be useful to consult sec.\ \ref{s3spin}):
\begin{equation}
\gamma_5 =
\left( \begin{array}{cc}
I & 0 \\
0 & -I \end{array} \right)\,.
\label{diag5} \end{equation}
Let $V_1$ correspond the $+1$ eigenvalue of $\gamma_5$ (positive
chirality), and $V_2$ -- to $-1$. The Dirac operator (\ref{dirop})
anti-commutes with $\gamma_5$ and, therefore, may be represented
in the form:
\begin{equation}
\Dir = \left( \begin{array}{cc} 0 & P^\dag \\ P & 0
\end{array} \right) \,. \label{adiagdir}
\end{equation}
We also have:
\begin{equation}
\Dir^2 = \left( \begin{array}{cc} P^\dag P & 0\\ 0 & PP^\dag 
\end{array} \right) \,. \label{dirsq}
\end{equation}
We conclude that $\mbox{index}(P)$, which now measures difference
between dimensions of the null subspaces of $\Dir$ with positive and
negative chirality, is equal to global chiral anomaly:
\begin{equation}
\mbox{index}(P)={\rm Tr} \left( e^{-tP^\dag P} \right)
-{\rm Tr} \left( e^{-tPP^\dag }\right) =
{\rm Tr} \left( \gamma_5 e^{-t\Dir^2}\right)=a_n (\gamma_5,\Dir^2)
 \label{Dindex}
\end{equation}
Therefore, the index can be calculated by the methods of the previous
section. In particular, the index of the Dirac operator in two and
four dimensions on flat manifolds without background axial vector
fields ($A_\mu^5=0$) can be read off from (\ref{Am2}) and (\ref{Am4})
with $\varphi :=i/2$.
We also see that chiral anomaly provides very important
information on topology of the background.

One can associate an index with any elliptic complex which must not
necessarily contain only two vector bundles. In general,
one has a sequence of bundles $V_p$ and a family of operators
$P_p$ which map (smooth sections of) $V_p$ to (smooth sections of)
$V_{p+1}$. One requires that $P_{p}P_{p+1}=0$ and that the corresponding
``Laplacians'' $D_p=P^\dag_p P_p + P_{p-1}P_{p-1}^\dag$ are elliptic.
Important examples include the de Rham complex (for which $V_p$ consists
of $p$-forms on $M$ and $P$ is the exterior derivative), the
Dolbeaux complex which deals with the forms on a complex manifold,
and the signature complex which treats selfdual and anti-selfdual
forms. More details can be found in \cite{Eguchi:1980jx}.
Another very important construction of this type is the
Witten index \cite{Witten:1982df} which constraints the supersymmetry
breaking. 

\section{Resummation of the heat kernel expansion}\label{sec8}
The heat kernel coefficients define the one-loop counterterms
in the background field formalism. In many cases the heat kernel
can also give a useful information on the finite part of the
effective action. Just one of the examples is the large mass
expansion (\ref{largemass}) which is valid when all background
fields and their derivatives are small compared to the mass of
the quantum field. In order to get the effective action in other
limiting cases one has to re-arrange the heat kernel 
expansion\footnote{A diagrammatic technique which can be used in resummations
of the heat kernel expansion is described in \cite{Moss:1999wq}.}.
\subsection{Modified large mass expansion}\label{s8finite}
In many physical applications there is a quantity $\mathcal{M}^2$
which is large compared to the rest of the background fields
and their derivatives. Therefore, it is a well motivated problem
to construct an expansion of the effective action in a
power series of $\mathcal{M}^{-1}$. To do so, one has to re-express
the heat kernel as
\begin{equation}
K(t;D)=e^{-t\mathcal{M}^2} \sum_k t^{(k-n)/2} \tilde a_k \,.
\label{modifHK}
\end{equation}
To obtain the effective action one has to integrate the heat kernel
over $t$ (cf eq.\ (\ref{WKtD})). To simplify the argumentation
we assume the cut-off regularization (\ref{WLamreg}), although in other
regularization schemes the result will be essentially the same.
The divergent and finite parts of the effective action are given
by (\ref{divWLam}) and (\ref{largemass}) respectively up to the obvious
replacements: $m\to \mathcal{M}$, $(4\pi )^{-n/2}b_k \to \tilde a_k$.
After a suitable renormalization, this procedure indeed gives a large
$\mathcal{M}$ expansion of the effective action.

Therefore, the problem is reduced to calculation of 
$\tilde a_k$. If $\mathcal{M}$ commutes with $D$ the coefficient
$\tilde a_k$ are the heat kernel coefficients for
the operator $D-\mathcal{M}^2$. This case returns us to the
standard large mass expansion. If $\mathcal{M}^2$ does not commute
with $D$ then
\begin{equation}
e^{-tD}\ne e^{-t(D-\mathcal{M}^2)}e^{-t\mathcal{M}^2} \,.\label{noteqexps}
\end{equation}
To achieve an equality the right hand side of (\ref{noteqexps}) must
be corrected by commutator terms. In this case calculation of
$\tilde a_k$ requires some amounts of extra work.

Heavy particles of non-equal masses described by the mass matrix
$\mathcal{M}^2$ is probably the most immediate example of a physical
system to which the modified large mass expansion should be applied.
Corresponding technical tools were developed recently 
\cite{Osipov:2000rg,Osipov:2001bj,Osipov:2001nx}.

The next example is a scalar field in curved space (cf. sec\ 
\ref{s3scalar}). Parker and Toms suggested 
\cite{Parker:1985dj,Parker:1985mv} to use the modified large
mass expansion with $\mathcal{M}^2=m^2+\xi R$ which partially
sums up contributions of the scalar curvature $R$ to the effective
action. This formalism was developed further in 
\cite{Jack:1985mw,Sushkov:2002rf}.

In some cases the term-by-term integration of the heat kernel
expansion gives good estimates of the vacuum energy even if no
large parameter is explicitly present in the model (see, e.g.
calculations of quantum corrections to the mass of two dimensional
solitons \cite{Dunne:1999du,AlonsoIzquierdo:2002eb}). 
\subsection{Covariant perturbation theory}\label{s8covpt}
Suppose that the matrix potential $E$, the Riemann curvature,
and the field strength $\Omega$ are small but rapidly varying.
Can we get any information on the heat kernel coefficients
containing a fixed power of the quantities listed above and
{\it arbitrary} number of derivatives? For the linear and
quadratic orders on a manifold without boundaries the answer
is positive and the results can be obtained either by the
functorial methods of sec.\ \ref{s4gen} 
\cite{Gilkey:1988,Branson:1990b}
or by solving the DeWitt equation 
\cite{Avramidi:1989er,Avramidi:1990ug,Avramidi:1991je}
(cf. sec.\ \ref{s4iter}). Retaining only the leading terms
we have for $k\ge 1$:
\begin{eqnarray}
&&a_{2k}(x,D) =(4\pi )^{-n/2} \left[ \alpha_1(k) 
\Delta^{k-1} R +\alpha_2 (k) \Delta^{k-1} E \right.\label{leadhk}\\
&&\qquad\qquad\qquad\qquad \left. 
+\mbox{higher order terms} \right]\,, \nonumber
\end{eqnarray}
where \cite{Gilkey:1979,Gilkey:1988}
\begin{equation}
\alpha_1(k)=\frac{k!k}{(2k+1)!} \,,\qquad
\alpha_2(k)=\frac{2k!}{(2k)!}\,.\label{al12k}
\end{equation}

One can sum up the expansion (\ref{leadhk}) and corresponding higher
order terms to obtain an information on the behaviour of the full
heat kernel in the limit described above. There is, however, a more
straightforward method to control non-localities which is called
the covariant perturbation theory 
\cite{Barvinsky:1987uw,Barvinsky:1990up,Barvinsky:1990uq}. To make the
idea of this method most transparent we consider a simple example
of flat $M$ and zero connection $\omega$. The exponent
$\exp (-tD)=\exp (t(\Delta + E))$ with $\Delta =\partial_\mu^2$
can be expanded in a power series
in $E$:
\begin{equation}
e^{-tD}=e^{t\Delta} + \int\limits_0^t ds e^{(t-s)\Delta} E e^{s\Delta}
+\int\limits_0^t ds_2 \int\limits_0^{s_2} ds_1
e^{(t-s_2)\Delta}E e^{(s_2-s_1)\Delta} E e^{s_1\Delta} +\dots
\label{pertex}
\end{equation}
The heat trace can be also expanded,
\begin{equation}
K(t,D)={\rm Tr}\left( e^{-tD}\right)=
\sum_{j=0}^\infty K_j(t),\label{pexhtr}
\end{equation}
where $K_j$ contains the $j$th power of $E$.

Covariant perturbation theory approach prescribes to take the
$0$th order heat kernel in the free space form (cf. eq.\ 
(\ref{simplestHK}) for $m=0$):
\begin{equation}
K_0(x,y;t)=(4\pi t)^{-n/2} \exp
\left( -\frac{(x-y)^2}{4t} \right) \,,\label{Kzero}
\end{equation}
which is an exact kernel on $M=\mathbb{R}^n$ only\footnote{The
original paper \cite{Barvinsky:1987uw} contained a curved space
generalisation of (\ref{Kzero}). This, however, does not improve
the global issues discussed below.}. This formula neglects all
global contributions and, therefore, is valid only for sufficiently
close $x$ and $y$ and for small $t$. We have:
\begin{equation}
K_0(t)=\iM {\rm tr}_V K_0(x,x;t)=(4\pi t)^{-n/2} \iM {\rm tr}_V (I_V)
\,,\label{trKzero}
\end{equation}
where $\ptr I_V$ simply counts discrete (spin and internal) 
indices of $D$. This formula reproduces the $a_0$ contribution
to the heat kernel.

In the next order, we have:
\begin{eqnarray}
K_1(t)=&&{\rm Tr} \left( \int_0^t ds e^{(t-s)\Delta}Ee^{s\Delta}
\right)=
{\rm Tr}\left( \int_0^t ds e^{t \Delta} E \right) \nonumber \\
=&& t\, \ptr \iM K_0 (x,x;t) E(x) \nonumber\\
=&&\frac t{(4\pi t)^{n/2}}
\iM \ptr E(x) \,,\label{Kone}
\end{eqnarray}
where we used the cyclic property of the trace and the expression
(\ref{Kzero}) for the $0$th order heat kernel. This expression is
consistent with the $k=1$ term of (\ref{leadhk}). The terms with
$k>1$ are total derivatives, and, therefore, they do not contribute
to the integrated heat kernel. 

The quadratic order of the heat kernel reads
\begin{eqnarray}
K_2(t)=&&{\rm Tr} \left( \int\limits_0^t ds_2 \int\limits_0^{s_2} ds_1
e^{(t-s_2)\Delta}E e^{(s_2-s_1)\Delta} E e^{s_1\Delta} \right) \nonumber \\
=&&\ptr \int_M dy \int_M dz 
\int\limits_0^t ds_2 \int\limits_0^{s_2} ds_1 
K_0 (z,y;t-s_2+s_1)E(y)\nonumber\\
&&\qquad\qquad\qquad\qquad\qquad \times K_0(y,z;s_2-s_1)E(z) \,.\label{K2a}
\end{eqnarray}
Next we introduce the rescaled variables $\xi =s/t$ and get rid of redundant
integrations to obtain:
\begin{eqnarray}
K_2(t)=&&\frac {t^2}2 \ptr \int_M dy \int_M dz \int\limits_0^1 d\xi
K_0(z,y;t(1-\xi )) E(z) \nonumber\\
&&\qquad\qquad\qquad\qquad\qquad 
\times K_0(y,z;t\xi)E(z) \,.\label{K2b}
\end{eqnarray}
Now we use the identity
\begin{equation}
K_0(z,y;t(1-\xi )) K_0(y,z;t\xi)=(4\pi t)^{-n/2}
K_0(z,y;t\xi(1-\xi)) \label{KKKide}
\end{equation}
and relate the heat kernel on the right hand side of (\ref{KKKide})
to a matrix element of $\exp (t\xi(1-\xi )\Delta )$ to obtain
the final result:
\begin{equation}
K_2(t)=\frac{t^2}{(4\pi t)^{n/2}}
\iM E f(-t\Delta ) E \,,\label{Ktwo}
\end{equation}
where the non-local form-factor $f$ reads:
\begin{equation}
f(q)=\frac 12 \int\limits_0^1 d\xi e^{-q\xi (1-\xi)}=
\frac 12 e^{-q/4} \sqrt{\pi /q} {\rm Erfi} \left[ \sqrt{q}/2 \right]\,.
\end{equation}
As we have already discussed above, applicability of this formula is
limited by our choice of $K_0(x,y;t)$. Namely, the potential $E(x)$
should have a compact support and $t$ should be reasonably small.
Note, that the small $t$ approximation in the expansion (\ref{pertex})
is self-consistent: if $t$ is small, the integration variables $s_i$
are even smaller.

The main difficulty in constructing an expansion in powers
the Riemann curvature and of the field strength is to organise
the procedure in a {\it covariant} way. The details of the
construction and higher order form-factors can be found in  
\cite{Barvinsky:1987uw,Barvinsky:1990up,Barvinsky:1990uq,Barvinsky:1994hw,
Barvinsky:1994ic}. From further developments of this method
we mention the work of Gusev and Zelnikov \cite{Gusev:1999cv}
who demonstrated that in two dimension one can achieve considerable
simplifications in the perturbation expansion by using the dilaton
parametrisation of the potential (cf. eq.\ (\ref{ehnEh})).
Recently Barvinsky and Mukhanov \cite{Barvinsky:2002uf} suggested
a new method for calculation of the non-local part of the effective
action based on the resummation of the perturbation series for the
heat kernel. This method was extended in \cite{Barvinsky:2003rx} to
include late time asymptotics of the heat kernel in curved space.

Let us stress that the expansion (\ref{pertex}) can be used also
for singular potentials. For example, it is very effective for 
the calculation of the heat kernel for $\delta$-potentials
concentrated on a co-dimension one subsurface 
\cite{Gaveau:1986,Bordag:1999ed,Moss:2000gv} (cf.\ sec.\ \ref{s6dom}).
If the $\delta$-potential has its support on a submanifold
of co-dimension greater than one, the expansion diverges 
\cite{Bordag:1999ed}. 
With a suitable choice of the zeroth order heat kernel $K_0(x,y;t)$
and of an operator to replace $E$ in (\ref{pertex}) one can treat
manifolds with boundaries \cite{Bordag:2001fj} where the perturbative
expansion takes the form of the multiple reflection expansion
\cite{Balian:1970,Balian:1977za,Hansson:1983xt}.
\subsection{``Low energy'' expansion}\label{s8low}
Let us now turn to the opposite case when the derivatives are less
important when the potential and the curvatures. In this case
one has to collect the terms which are of a fixed order in the
derivatives, but contain all powers of $E$, $R$ and $\Omega$.
Since the derivatives are sometimes identified with the energy,
this approximation is being called the low energy expansion.
This scheme goes back to Schwinger's calculations
\cite{Schwinger:1951} in constant electromagnetic 
fields\footnote{We like to mention also the paper
\cite{Heisenberg:1936} which treated the effective action
in external electromagnetic field from a different point view.}.

Let us consider a simple example \cite{Blau:1991iz}. Let
$M=\mathbb{R}^2$. Consider a scalar particle in constant
electromagnetic field with the field strength $F_{12}=-F_{21}=B$.
As a potential we choose $A_1=0$, $A_2=Bx^1$. Then the operator
acting on quantum fluctuations is
\begin{equation}
D=-\partial_1^2-(\partial_2 -iBx^1)^2 +m^2 \,.\label{cmagD}
\end{equation}
This operator commutes with $\partial_2$. Therefore, we can look
for the eigenfunctions of $D$ in the form 
$\phi_k (x)=\exp (ikx^2) \tilde\phi (x^1)$. 
\begin{equation}
D\phi_k (x) = \left( -\partial_1^2 +B^2 (x^1 - k/B)^2 +m^2 \right)
\phi_k (x) \,.\label{cmDk}
\end{equation}
In the $x^1$ direction we have the one-dimensional harmonic
oscillator potential (cf. (\ref{harmos})). Therefore, the 
eigenvalues are $\lambda_{k,p}= (2p+1)|B|+m^2$. These eigenvalues
do not depend on $k$. For this reason the heat kernel
\begin{equation}
K(t,D)=\sum_{k,p} \exp (-t\lambda_{k,p}) \label{cmagKt}
\end{equation}
is ill defined. To overcome this difficulty it was suggested 
\cite{Blau:1991iz} to put the system in a box (without specifying
any boundary conditions, however) and replace the sum over $k$
by the degeneracy factor $(|B| {\rm Vol})/(2\pi )$, where ${\rm Vol}$
is volume of the box. The degeneracy factor is chosen in such a way
that the resulting heat kernel
\begin{equation}
K(t,D)= \frac{B{\rm Vol}}{4\pi } e^{-m^2t} [{\rm sinh} (Bt)]^{-1}
\label{cmagKB}
\end{equation}
behaves as ${\rm Vol}/(4\pi t)$ in the limit $B\to 0$.

Calculations of the effective action in covariantly constant
background gauge fields by the spectral theory methods have 
been performed by many authors 
\cite{Batalin:1977uv,Dittrich:1983ej,Elizalde:1984yb,
Elizalde:1985zv,Blau:1991iz}. Such calculations were motivated, at least
partially, by various models of quark confinement (see 
\cite{Pagels:1978dd,Adler:1982rk} and references therein).

There exists an algebraic method \cite{Avramidi:1993yf,Avramidi:1995ik} 
which allows to evaluate the low energy heat kernel by using 
exclusively the commutator algebra. Let us briefly formulate the
results of \cite{Avramidi:1993yf,Avramidi:1995ik}. Consider a flat
manifold ($R_{\mu\nu\rho\sigma}=0$) with the background fields
satisfying the ``low-energy conditions'':
\begin{equation}
\nabla_\mu \Omega_{\mu\nu}=0,\qquad
\nabla_\mu\nabla_\nu\nabla_\rho E=0 
\label{lowecon}
\end{equation}
with usual definitions of $E$ and $\Omega$ (see (\ref{D1}) - 
(\ref{Omega})). Moreover, let us suppose that the background
is ``approximately abelian'', i.e. that $\Omega_{\mu\nu}$,
$E$ and all their covariant derivatives are mutually commuting matrices. 

With all these assumptions a closed expression for the heat kernel
may be obtained:
\begin{equation}
K(t;x,x;D)=(4\pi t)^{-n/2} \exp \left(
tE +\Phi (t) +\frac {t^3}4 E_{;\mu} \Psi^{\mu\nu} E_{;\nu} \right)\,,
\label{avrahk}
\end{equation}
where $\Phi$ and $\Psi$ are complicated functions of $t$, $E$ and $\Omega$
\cite{Avramidi:1995ik}. If $\nabla_\mu\nabla_\nu E=0$,
\begin{eqnarray}
&&\Phi (t)=-\frac 12 \ln \det \left( \frac{\sinh t \Omega}{t\Omega}
\right) \,,\nonumber\\
&&\Psi (t) = (t\Omega )^{-2} \left( t\Omega \coth (t\Omega )-1
\right) \,,\label{PhiPsi1}
\end{eqnarray}
where $\Omega$ has to be understood as a space-time matrix
$\Omega_{\mu\nu}$, so that multiplication in (\ref{PhiPsi1})
is the matrix multiplication, and $\det$ is the determinant
of an $n\times n$ matrix. These formulae generalise the equation
(\ref{cmagKB}) and justify the choice of the degeneracy factor made
to derive it.

If $\Omega =0$,
\begin{eqnarray}
&&\Phi (t) = -\frac 12 \ln \det \left( 
\frac{\sinh (2t\sqrt{P}}{2t\sqrt{P}} \right) \,,\nonumber\\
&&\Psi (t) = -\left( t\sqrt{P} \right)^{-3}
\left( \tanh (t\sqrt{P}) -t\sqrt{P}) \right) \,,\label{PhiPsi2}
\end{eqnarray}
where $P_{\mu\nu}=-(1/2)\nabla_\mu\nabla_\nu E$. In the particular
case of one-dimensional harmonic oscillator these formulae
reproduce (\ref{harmheat}). 

In curved space the best one can do in the framework of the
low-energy expansion is to consider
locally symmetric manifolds, i.e. vanishing $\Omega$,
covariantly constant Riemann curvature and constant $E$. 
In this case formulae similar to the ones presented above
are available \cite{Avramidi:1994zc,Avramidi:1996zp}.
Covariantly constant curvature means that locally 
the manifold $M$ is a symmetric space. Various approaches
to the heat kernel on such manifolds are described in detail
in monographs and survey articles 
\cite{Hurt:1983,Camporesi:1990wm,Bytsenko:2001xs}. In particular,
very detailed information may be obtained for group manifolds
(see, for example, \cite{Dowker:1971vu}) and for hyperbolic
spaces \cite{Bytsenko:1996bc}.
\subsection{Heat kernel on homogeneous spaces}\label{s8homo}
In this section we briefly explain how one can find
the spectrum of some ``natural'' differential
operators on homogeneous spaces
by purely algebraic methods. We start with
some basic facts from differential geometry and harmonic analysis
\cite{Kobayashi:1969,Helgason:1984}. 
Consider a homogeneous space $G/H$ of
two compact finite-dimensional Lie groups $G$ and $H$.
The Lie algebra $\mathcal{G}$ of $G$ can be decomposed as
\begin{equation}
\mathcal{G}=\mathcal{H}\oplus \mathcal{M} \,,\label{Gdecom}
\end{equation}
where $\mathcal{H}$ is the Lie algebra of $H$ and $\mathcal{M}$
is the complement of $\mathcal{H}$ in $\mathcal{G}$ with
respect to some bi-invariant metric. We have:
\begin{equation}
[\mathcal{H},\mathcal{M}] \subset \mathcal{M} \,,\label{HMM}
\end{equation}
where $[\ ,\ ]$ is the Lie bracket on $\mathcal{G}$. If, moreover,
$[\mathcal{M},\mathcal{M}]\subset \mathcal{H}$, then $G/H$ is a
symmetric space. We do not impose this restriction.

$\mathcal{M}$ can be identified with the tangent space to $M=G/H$
at the origin (i.e. at the point which represents the unit element of $G$).
Eq.\ (\ref{HMM}) tells us that $\mathcal{H}$ acts on $\mathcal{M}$
by some (orthogonal) representation. This action defines the embedding
\begin{equation}
\mathcal{H}\subset so(n) \,.\label{Hinso}
\end{equation}
All physical fields are classified according to certain representations
of the Lie algebra $so(n)$. Restrictions of these representations to
$\mathcal{H}$ define transformation properties of the field with
respect to $\mathcal{H}$. From now on we work with each irreducible
representation of $\mathcal{H}$ separately.

The field $\Phi^A$ belonging to
an irreducible representation $\mathcal{T}(H)$ can be expanded as 
(see, e.g., \cite{Salam:1982xd}):
\begin{equation}
 \Phi_A (x)=({\rm Vol})^{-\frac 12} \sum_{j, \xi ,q}
\sqrt {\frac {d_j}{d_{\mathcal{T}}}} \mathcal{T}^{(j)}_{A\xi ,q}
(g_x^{-1}) \phi^{(j)}_{q,\xi }\ , \label{eq:(A.2)}
\end{equation}
where ${\rm Vol}$ is the volume of $G/H$, 
$d_{\mathcal{T}}={\rm dim}\mathcal{T}(H)$.
We sum over the representations $\mathcal{T}^{(j)}$ of $G$
which give $\mathcal{T}(H)$ after reduction to $H$. $\xi$ labels
multiple components  $\mathcal{T}(H)$ in the branching
$\mathcal{T}^{(j)}\downarrow H$, $d_j=\dim \mathcal{T}^{(j)}$. 
$q$ runs from $1$ through $d_j$. The matrix
elements of $\mathcal{T}^{(j)}$ have the following orthogonality
property
\begin{equation} \iM
\mathcal{T}^{(j) \dag}_{A\zeta ,q} (g_x^{-1})
\mathcal{T}^{(j')}_{A\xi ,p} (g_x^{-1})
=({\rm Vol})d_j^{-1} d_{\mathcal{T}} \delta_{\zeta \xi} \delta_{p q}
\delta_{jj'} \label{eq:(A.3)}\end{equation}
Therefore, to construct the harmonic expansion on $G/H$ it is
necessary to have powerful methods for reduction of the representations 
from $G$ to $H$. There are several standard 
\cite{Barut:1977,Slansky:1981yr}
and less standard  \cite{Lyakhovsky:1989cr,Lyakhovsky:1991vv}
techniques which may be used depending on the particular
homogeneous space.

It is important that not only the harmonic expansion but also
the spectrum of the invariant operators can be analysed by the group
theoretical methods. The covariant derivative on $G/H$
reads \cite{Kobayashi:1969}:
\begin{equation}
\nabla_\mu = \nabla_\mu^{[c]}+\Gamma^{[R]}_\mu \,.\label{hscovdiv}
\end{equation}
Here $\nabla^{[c]}$ is the canonical covariant derivative on
$G/H$. At the origin $\nabla^{[c]}$ can be identified with 
the tangent space generators from $\mathcal{M}$ taken in
the representation $\mathcal{T}^{(j)}$. The part $\Gamma^{[R]}_\mu$
depends on the invariant metric on $G/H$ and on the
structure constants of $\mathcal{G}$ restricted to $\mathcal{M}$.
On symmetric spaces such structure constants are zero and,
therefore, the Laplace-Beltrami operator has a particularly
simple form:
\begin{equation}
D\simeq \nabla^{\mu[c]}\nabla_\mu^{[c]} \simeq C_2 (G)-C_2 (H)\,,
\label{LapCas}
\end{equation}
where $C_2$ are quadratic Casimir operators of $G$ and $H$ which
depend on the representations $\mathcal{T}^{(i)}$ and 
$\mathcal{T}(H)$ respectively.
On general homogeneous spaces the expressions are a bit more
complicated (see, e.g., 
\cite{Vassilevich:1988em,Lyakhovsky:1991vv} for explicit
examples). In any case, eigenvalues of $D$ are given by
a second order polynomial $Q(m_1,\dots,m_k)$ of several
natural numbers $m_l$. These eigenvalues are in general
degenerate with multiplicities defined essentially 
by dimensions of the representations of $G$ and $H$.
They are also polynomials\footnote{The spectrum of differential 
operators on coset spaces can be calculated exactly even if one adds
a homogeneous gauge field (see \cite{Dolan:2003bj} and references 
therein).}
 in $m_l$. The heat kernel is then
represented as an infinite sum
\begin{equation}
K(t,D)=\sum_{m_l} N(m_1,\dots,m_k) \exp (-t Q(m_1,\dots,m_k))\,.
\label{hsheatk}
\end{equation}
There exist several tricks which can be used to evaluate
the $t\to 0$ asymptotics of such sums. For example, one may use
the Poisson summation formula (\ref{eq:asPoi}). By taking derivatives
with respect to $t$ one obtains 
\begin{equation}
\sum_{l\in \mathbb{Z}} l^{2r} \exp \left( -tl^2\right)
\simeq \sqrt{\pi} 2^{-r} (2r-1)!! t^{-(2r+1)/2} +
\mathcal{O}\left( e^{-1/t} \right)\label{diffPoi}
\end{equation}
for $r\in\mathbb{N}$.

More general polynomials of $l$ may be treated by the 
Euler--Maclaurin formula which reduces sums to the integrals.
Let $F(\tau )$ be a function defined on $0 \le \tau < \infty$. If the
$2m$-th derivative
$F^{(2m)}(\tau)$ is absolutely integrable on $(0,\infty )$ 
\begin{eqnarray}
&&\sum_{i=0}^k F(i) -\int_0^\infty F(\tau )d\tau =\frac 12 [F(0)+F(k)]
\nonumber \\
&&\qquad\qquad\qquad +\sum_{s=1}^{m-1} \frac {B_{2s}}{(2s)!}
[F^{(2s-1)}(k)-F^{(2s-1)}(0)]+{\rm Rem}_m(n), \label{eq:asEM}
\end{eqnarray}
where the reminder ${\rm Rem}_m$ satisfies:
\begin{equation}
|{\rm Rem}_m(n)| \le (2-2^{1-2m}) \frac {|B_{2m}|}{(2m)!}
\int_0^n |F^{(2m)}(\tau )| d\tau .
\label{eq:asrem}
\end{equation}
$B_s$ are the Bernoulli numbers. If $F(i)$ is taken to be
$N_i \exp (-t \lambda_i )$, one can take limit $k\to \infty$
in (\ref{eq:asEM}) and restrict the summation to some finite
$m$ to obtain asymptotic series for the heat kernel.
For example, one can easily recover the expansion (\ref{diffPoi}).

In many cases calculation can be done by means of the Mellin transform.
This method will be clear from the following example.
Consider
\begin{equation}
K(t)=\sum_{k=1}^\infty k\exp (-tk^2) \label{(A.1)}
\end{equation}
The Mellin transform of $K(t)$ is
\begin{equation}
\tilde K(s)=\sum_k k \int_0^\infty dx\ x^{s-1}
\exp (-xk^2) =\sum_{k=1}^\infty k^{1-2s}\Gamma (s)
=\zeta_R(2s-1) \Gamma (s), \label{(A.2)}
\end{equation}
where $\zeta_R$ is the Riemann zeta-function. Inverse
Mellin transformation gives
\begin{equation}
K(t)=\frac 1{2\pi i} \int_C t^{-s} \zeta_R(2s-1)
\Gamma (s) ds. \label{(A.3)}
\end{equation}
The contour $C$ covers all poles of the integrand
at $s=1, 0, -1, -2,\dots$. By calculating the residues we
obtain the desired expansion:
\begin{equation}
K(t)=\frac 1{2t} -\frac 1{12} -\frac t{120}
-\frac {t^2}{504} -\frac {t^3}{1440} +O(t^4).
\end{equation}

Actually, the techniques introduced above should only be
used for complicated multi-parameter sums (see, e.g.,
\cite{Lyakhovsky:1991vv}).
For a simple one-parameter sum it is easier to
use combinations of known asymptotic series (cf. Appendix
of Ref. \cite{Birmingham:1987cy}).

In this method complexity of calculations of the heat kernel
coefficients $a_k$ is almost independent of $k$. Therefore,
the algebraic techniques were applied to higher dimensional theories
where one needs higher heat kernel coefficients to
perform renormalization or to calculate the anomalies.
Rather naturally, the most simple toroidal spaces were
considered first \cite{Appelquist:1983zs,Appelquist:1983vs,Rubin:1983zz,
Kogan:1983fp,Fradkin:1983kf,Buchbinder:1988np}, and then
computations on spheres were performed
\cite{Candelas:1984ae,Kikkawa:1985qc,Kantowski:1987ct,Birmingham:1987cy,
Birmingham:1988iv,Gleiser:1987mg,BuchOd,Buchbinder:1989cd,Elizalde:1996nb}.
Other homogeneous spaces were considered for example in 
\cite{Lyakhovsky:1991vv,Vassilevich:1991zi}. In the same approach
non-minimal operators on homogeneous spaces were treated
in \cite{Vassilevich:1992tb,Vassilevich:1993yt}. We refer to
\cite{Camporesi:1990wm} for a more extensive literature survey.

\section{Exact results for the effective action}\label{sec9}
In the previous section we have seen that the heat kernel can be
calculated exactly for all values of $t$ if the background satisfies certain
symmetry or smoothness conditions. In this section it will be
demonstrated that for some classes of the operators the effective
action can be calculated exactly without imposing any restrictions
on behaviour of the background fields. These are the cases when the
variation of $\zeta'(0)$ with respect to the background
can be reduced to the heat kernel coefficients which are locally
computable. Then the variation is integrated to the effective action.

The most immediate example of the variations admissible for this scheme
is quantum anomalies (cf.\ sec.\ \ref{sec7}). Historically, the first
action obtained by integration of an anomaly was the chiral Wess-Zumino
action. We do not consider this action here since such a consideration
would require a lot of additional technical devices. The reader can
consult excellent original papers 
\cite{Wess:1971yu,Witten:1983tw}\footnote{Calculation of the effective
action in two dimensional QED by integration of the anomaly may
be found in \cite{Hortacsu:1979fg}.}. 
The next subsection is devoted to another well know example,
which is the Polyakov action obtained through integration of the
conformal anomaly in two dimensions.

In sec.\ \ref{s9dual} below we consider a more complicated situation when
relevant variation of the effective action leads to a linear combination
of the heat kernel coefficients of several different operators. Typically
in this case one obtains exact relations between the effective actions
rather than the effective actions themselves.
\subsection{The Polyakov action}\label{s9polyakov}
Let us consider a two-dimensional Riemannian manifold $M$
without boundary and
a scalar field $\phi$ minimally coupled to the geometry.
This means that we choose $\Phi =0$ in the action
(\ref{nonminact}) and $\Psi =0$ in the inner product
(\ref{dilaprod}). In two dimensions any metric is conformally
flat, i.e. by a suitable choice of the coordinates one may
transform it to the conformal gauge
\begin{equation}
g_{\mu\nu}=e^{2\rho (x)} \eta_{\mu\nu},\qquad
\eta_{\mu\nu}={\rm diag} (1,1).\label{cgauge}
\end{equation} 
In this gauge
\begin{equation}
\sqrt{g}R=-2\eta^{\mu\nu} \partial_\mu\partial_\nu \rho
\,.\label{confR}
\end{equation}

The effective action $W$ depends on $\rho$ only (since there is no other
external field in this problem). We substitute (\ref{ehnT}) with
$\Phi =\Psi =0$ in (\ref{convarW}) to obtain
\begin{equation}
\delta W = \frac 1{12\pi} \int_M d^2x (\delta\rho) \eta^{\mu\nu}
\partial_\mu\partial_\nu \rho \,.\label{cgcvar}
\end{equation}
This equation can be integrated with the initial condition
$W(\rho =0)=0$. 
\begin{equation}
W=\frac 1{24\pi} \int_M d^2x\,\rho \eta^{\mu\nu}
\partial_\mu\partial_\nu \rho \,.\label{caction}
\end{equation}
By returning to the covariant notations we obtain the famous
Polyakov action \cite{Polyakov:1981rd} (see also 
\cite{Luscher:1980fr}):
\begin{equation}
W=\frac 1{96\pi} \int_M d^2x \sqrt{g} R\frac 1\Delta R \,.
\label{Polaction}
\end{equation}

Note, that on a compact manifold one has to take into account contributions
from the zero modes which lead to an additional term in the Polyakov action
\cite{Dowker:1994rt}.

The problem of finding an action whose conformal variation reproduces
the conformal anomaly can be posed
in higher dimension as well \cite{Deser:1976yx} and some physical consequences
of such actions can be studied \cite{Frolov:1979tu,Frolov:1981mz}.
Complete expressions for the conformal anomaly induced effective action
in four dimensions were obtained in \cite{Riegert:1984kt,Buchbinder:1986dc}.
Conformal action in the presence of boundaries was constructed in
\cite{Dowker:1990ue}. One should keep in mind that unless the
background is conformally flat the conformal action represents only
a part of the full effective action. Applicability and limitations
of the conformal action approach are discussed in 
\cite{Erdmenger:1997yc,Balbinot:1999vg,Gusev:1999cv}.
\subsection{Duality symmetry of the effective action}\label{s9dual}
In this section we deal with variations of the effective action
which can be expressed through zeta functions of several operators
(in contrast to genuine anomalous transformations of sec.\ \ref{sec7}
which involve a single zeta function each).

A rather simple example can be found in the paper \cite{Kummer:1998dc}
which considered two operators of Dirac type:
\begin{equation}
\Dir =i\gamma^\mu e^{\Psi}\partial_\mu e^{-\Phi}\,,\qquad
\Dir^\dag =i\gamma^\mu e^{-\Phi}\partial_\mu e^{\Psi} \label{twoDir}
\end{equation}
on flat two-dimensional manifold without boundary
depending on two scalar functions $\Phi$ and $\Psi$. 
Let us now consider small variations
of $\Phi$ and $\Psi$. According to (\ref{varizeta}) the zeta function
of corresponding Laplacian changes as
\begin{equation}
\delta \zeta (s,\Dir\Dir^\dag )=
-2s {\rm Tr}\left[ (\Dir\Dir^\dag )^{-s} \delta\Psi -
(\Dir^\dag\Dir )^{-s} \delta \Phi \right] \,.\label{dzPP}
\end{equation}
Variation of $\zeta'(0)$ reads:
\begin{eqnarray}
\delta \zeta'(0,\Dir\Dir^\dag )&=&
-2\left[ \zeta(0,\delta \Psi,\Dir\Dir^\dag )-
\zeta (0,\delta\Phi,\Dir^\dag\Dir)\right] \nonumber\\
&=&-2\left[ a_2(\delta\Psi,\Dir\Dir^\dag)
-a_2(\delta\Phi,\Dir^\dag \Dir )\right]\,.\label{dzpPP}
\end{eqnarray} 
Here we used that $\Dir\Dir^\dag$ and $\Dir^\dag\Dir$ are operators
of Laplace type, and, therefore, the corresponding zeta functions
are regular at $s=0$. In the last line we replaced $\zeta (0)$ 
by the heat kernel coefficients by means of (\ref{anz0}).

Next we have to rewrite $\Dir\Dir^\dag$ and $\Dir^\dag\Dir$ in the
canonical form (\ref{laplaceb}) with the help of (\ref{oab}),
(\ref{Eab}). For $\Dir\Dir^\dag$ we have:
\begin{eqnarray}
&&g^{\mu\nu}=e^{2(\Psi -\Phi )}\,,\qquad \omega_\mu =
\Psi_{,\mu} -\gamma^\nu\gamma_\mu \Phi_{,\nu} \,,\nonumber\\
&&E={\Phi_{,\mu}}^\mu \,,\qquad R =-2(\Phi -\Psi {)_{,\mu}}^\mu\,,
\label{DDdef}
\end{eqnarray}
where comma denotes ordinary partial derivatives, all contractions
are performed with the effective metric $g^{\mu\nu}$. Note, that in
two dimensions $\Gamma^\rho_{\mu\nu}g^{\mu\nu}=0$ in the conformal
coordinates. To obtain relevant geometric quantities for $\Dir^\dag\Dir$
one has to replace $\Phi$ by $-\Psi$ and vice versa. Equation
(\ref{a2nobou}) gives:
\begin{equation}
\delta \zeta'(0,\Dir\Dir^\dag )=-\frac 1{6\pi}\int d^2x
\left[ \delta\Psi (\Delta \Psi +2\Delta\Phi )
+\delta\Phi (\Delta\Phi +2\Delta\Psi )\right]\,,\label{zDDvar}
\end{equation}
where $\Delta$ is the flat space Laplacian. Let us recall, that 
$\zeta'(0,D)$ defines the functional determinant of $D$ (see
eq.\ (\ref{detzeta})). The variation (\ref{zDDvar}) can be integrated
to give\footnote{The heat kernel coefficients $a_2(x,\Dir\Dir^\dag )$
and $a_2(x,\Dir^\dag\Dir )$ are total derivatives. Therefore,
the second term on the right hand side of (\ref{detzeta}) does not
contribute.}
\begin{equation}
\ln\det (\Dir\Dir^\dag)=-\zeta'(0,\Dir\Dir^\dag)=
\frac 1{12\pi} \int d^2x (\Psi \Delta \Psi +4\Psi\Delta\Phi
+\Phi\Delta\Phi )\,.\label{detDDfin}
\end{equation}
We see from (\ref{detDDfin})
that $\ln \det (\Dir\Dir^\dag)=\ln \det (\Dir^\dag\Dir )$
which is obviously true 
up to possible topological contributions from zero modes which have
been neglected in this calculation. The result (\ref{detDDfin}) was
confirmed in \cite{Gusev:1999cv} by the methods of covariant perturbation
theory.

This example is clearly of (at least an academic) 
interest since it adds up to very
few cases discussed above in this section when a closed analytic
expression for the determinant may be obtained without any assumptions 
on smoothness of the background fields.    
The operator $\Dir\Dir^\dag$ shares some similarities with 
the kinetic operator
for non-minimally coupled scalars in two dimensions (\ref{ehnA}).
There are good grounds to believe 
\cite{Kummer:1999zy} the functional determinant (\ref{detDDfin})
may describe the spherically symmetric part of the Hawking flux in four
dimensions.

Let us consider the dilatonic operator (\ref{ehnA}).
We take $\Psi =\Phi$ for simplicity\footnote{Variation of $\det D$
with respect to $\Psi$ may be expressed through the scale anomaly.
Therefore, it is not a problem to find exact dependence of $\det D$
on $\Psi$ by using the methods of sec.\ \ref{sec7}.}. We shall be
interested in properties of $\det D$ under reflection of the dilaton,
$\Phi\to -\Phi$. The quantity $\ln \det D (\Phi) - \ln \det D(-\Phi)$
defines the dilaton shift under the $T$-duality transformations
in string theory. In the string theory context this problem was
solved in \cite{Schwarz:1993te} basing on earlier results of
\cite{Schwarz:1979ae}. Here we follow presentation of 
the paper \cite{Vassilevich:2000kt}.

Let us introduce two first order differential operators
\begin{equation}
L_\mu =e^{-\Phi}\partial_\mu e^{\Phi}\,,\qquad
L_\mu^\dag =-e^{\Phi}\partial_\mu e^{-\Phi}
\label{LLdag}
\end{equation}
and associated second order operators
\begin{equation}
D_+=L_\mu^\dag L_\mu \,,\qquad 
D_-=L_\mu L_\mu^\dag \,. \label{DpDm}
\end{equation}
We restrict ourselves to flat two-dimensional manifolds.
Therefore, position of the indices (up or down) plays no role.

We calculate again variation of the $\zeta$ function with respect
to variation of $\Phi$:
\begin{equation}
\delta \zeta (s,D_+)=-2s {\rm Tr} \left[
(\delta \Phi ) \left( L_\mu^\dag L_\mu D_+^{-s-1}
-L_\mu D_+^{-s-1} L_\mu^\dag \right)\right]\,.\label{dudz}
\end{equation}
The operators in the first term here recombine in $D_+^{-s}$. The situation
with the second term is more subtle. Strictly speaking, to treat this
term one has to perform the Mellin transform and use analytic continuation
in $s$  (see footnote \ref{ft} in sec.\ \ref{s7conf}). However, the
result of such manipulations is almost obvious:
\begin{equation}
{\rm Tr} \left[ (\delta \Phi) L_\mu D_+^{-s-1} L_\mu^\dag \right]=
{\rm Tr} \left[ (\delta \Phi) H^{-s} \right] \,,\label{duvarH}
\end{equation}
where
\begin{equation}
H_{\mu\nu}=L_\mu L_\nu^\dag \label{dudefH}
\end{equation}
is a matrix operator acting on the space $V_+$ of the vector fields
which can be represented as $v_\nu = L_\nu f$ with some scalar
function $f$. The operator (\ref{dudefH}) is not of Laplace type
as it contains a complicated differential projector on $V_+$.
Therefore, even though the variation (\ref{dudz}) looks very
similar to (\ref{dzPP}), for example, it cannot be used directly
to evaluate $\zeta'(0,D_+)$ since we cannot even guarantee that
(\ref{duvarH}) is regular at $s=0$ (in fact, it is {\it not}
regular, see \cite{Vassilevich:2000kt}).  

We also have:
\begin{equation}
\delta \zeta (s,D_-)=2s {\rm Tr}\left[ (\delta\Phi) \left(
D_-^{-s} - \bar H^{-s} \right) \right] \,,\label{dudmin}
\end{equation}
where the operator
\begin{equation}
\bar H_{\mu\nu}=\epsilon_{\mu\mu'}\epsilon_{\nu\nu'} L_{\mu'}^\dag
L_{\nu'} \label{defbarH}
\end{equation}
acts on the space $V_-$ of the vector fields of the form $\bar v_{\mu}=
\epsilon_{\mu\mu'}L_{\mu'}^\dag$. All epsilon tensors cancel after
taking the trace in (\ref{dudmin}) and we arrive at the same formula
as for $D_+$ but with the replacement $\Phi\to -\Phi$. The spaces
$V_+$ and $V_-$ are orthogonal, and $V_+ + V_-=V$ is the space of
all vector fields on $M$ (we neglect possible zero modes). Therefore,
\begin{equation}
\delta \left[ \zeta (s,D_+)-\zeta (s,D_-) \right]
=2s \left[ \zeta (s,\delta\Phi,H+\bar H )-
\zeta (s,\delta\Phi,D_+) -\zeta (s,\delta\Phi,D_-) \right]\,.
\label{duzz}
\end{equation}
The operator
\begin{equation}
(H+\bar H)_{\mu\nu}=\delta_{\mu\nu} D_+ +\Phi_{,\mu\nu}
\label{HbarH}
\end{equation}
is of Laplace type. We can now act in the standard manner to
obtain
\begin{equation}
\delta \left[ \zeta' (0,D_+)-\zeta' (s,D_-) \right]
=2\left[ a_2(\delta\Phi,H+\bar H )-
a_2(\delta\Phi,D_+) -a_2(\delta\Phi,D_-) \right]\,.
\label{duzzpr}
\end{equation}
The right hand side can be evaluated by using (\ref{a2nobou}).
We leave it is an exercise to show that the variation (\ref{duzzpr})
is zero (alternatively, one may look up in \cite{Vassilevich:2000kt}).  
This leads us to the conclusion that
\begin{equation}
\zeta'(0,D_+) -\zeta'(0,D_-)=-\ln\det D_+ +\ln \det D_-=0.
\label{detdet}
\end{equation}

In the one-dimensional case a similar relation may be obtained by 
methods of supersymmetric quantum mechanics. In this simplest case,
and in higher dimensions if $\Phi$ depends on one of the coordinates only,
$D_+$ and $D_-$ are isospectral up to zero modes and (\ref{detdet})
follows immediately. In two dimensions, $D_+$ and $D_-$ are
{\it not} isospectral in general. Although $D_+$ and $D_-$ have
coinciding determinants, other spectral functions can be different.

Extensions of (\ref{detdet}) to the case of matrix-valued $\Phi$,
manifolds with boundaries, and the dilaton-Maxwell theory in
four dimensions can be found in \cite{Vassilevich:2000kt}.
A generalisation to arbitrary dimension has been achieved recently 
\cite{Gilkey:2002qd}. Let $A_p$ be a $p$-form field with the field
strength $F=dA_p$. Consider the classical action
\begin{equation}
{\mathcal{L}}=\int_M e^{-2\Phi} F\wedge \star F \,,\label{pformact}
\end{equation}
where $\star$ is the Hodge duality operator. 
Such actions appear, for example, in extended supergravities
\cite{Cremmer:1978tt} and bosonic M-theory \cite{Horowitz:2000gn}.
Instead of the dilaton, also a tachyon coupling may appear
\cite{Klebanov:1998yy}. We are interested in the symmetry properties
of the effective action under the transformation 
$p\to n-p-2$, $\Phi\to -\Phi$. In higher dimensional supergravity
theories this is a part of the $S$-duality transformation.

It is convenient to define
the twisted exterior derivative
\begin{equation}
d_\Phi :=e^{-\Phi} d e^{\Phi} \label{twider}
\end{equation}
and the associated twisted coderivative and twisted Laplacian
\begin{equation}
\delta_\Phi :=e^\Phi \delta e^{-\Phi}\,,\qquad
\Delta_\Phi :=\left( d_\Phi + \delta_\Phi \right)^2\,.\label{twicoL}
\end{equation}
Since $d_\Phi^2=0$ we have an elliptic complex. The restriction
of $\Delta_\Phi$ on the space of $p$ forms will be denoted by
$\Delta_\Phi^p$. This twisted de Rham
complex was introduced by Witten in the context of Morse theory and
supersymmetric quantum mechanics \cite{Witten:1982im}. 

Any $p$-form can be decomposed as the sum of a twisted exact,
twisted co-exact, and twisted harmonic form:
\begin{equation}
A_p=d_\Phi A_{p-1} +\delta_\Phi A_{p+1} +\gamma_p\,,
\qquad \Delta_\Phi \gamma_p=0 \,.\label{twi-decomp}
\end{equation}
The projections on the spaces of twisted exact and twisted co-exact
forms will be denoted by the subscripts $\parallel$ and $\perp$
respectively.

We assume that the fields $\tilde A_p=e^{-\Phi}A_p$ have a standard
Gaussian measure and are to be considered as fundamental fields in the
path integral. The action given in eq. (\ref{pformact}) is invariant
under the gauge transformation which sends $\tilde A_p$ to $\tilde A_p +
d_\phi \tilde A_{p-1}$. This means that the $p$-forms which are
$d_\phi$ exact have to be excluded from the path integral, but that
a Jacobian factor corresponding to the ghost fields $\tilde A_{p-1}$
has to be included in the path integral measure. Next we note that
$d_\phi$-exact $(p-1)$-forms do not generate a non-trivial transformation
of $\tilde A_p$. Hence, such fields must be excluded from the ghost
sector. Then we have to include ``ghosts for ghosts''. This goes on
until the zero forms have been reached. By giving these arguments
an exact meaning, one arrives at the Faddeev--Popov approach to
quantisation of the $p$-form actions 
\cite{Duff:1980qv,Siegel:1980jj,Obukhov:1982dt,Barbon:1995zh}. 
We note that the procedure of 
\cite{Obukhov:1982dt,Barbon:1995zh} is valid also in the
presence of a dilaton interaction if one simply replaces the ordinary
derivatives by the twisted ones. As a result, we have the following 
expression for the
effective action: 
\begin{equation}
W_p(\Phi )=\frac 12 \sum_{k=0}^p (-1)^{p+k} \ln \det (\Delta^k_\Phi
\vert_{\perp}) +W_p^{\rm top} \,.\label{twiefa}
\end{equation}
$W_p^{\rm top}$ depends on certain topological characteristics
of the manifold (the Betti numbers). We shall neglect $W_p^{\rm top}$
and some other topological contributions 
in what follows.

By combinatorial arguments, similar to the presented above,
one can show that \cite{Gilkey:2002qd}:
\begin{equation}
\delta \left( W_p(\Phi )-W_{n-p-2}(-\Phi )\right)=
\sum_{k=0}^m (-1)^{p+k}  a_n(\delta\Phi,\Delta_{\Phi}^k)\,.
\label{twipvar}
\end{equation}
One should distinguish between $\delta \Phi$ (which is a variation
of the dilaton) and $\delta_\Phi$ (which is a twisted co-derivative).
Therefore, we have related the variation of the effective actions
with respect to $\Phi$ to a combination of the heat kernel coefficients
which is called the supertrace of the twisted
de Rham complex. A somewhat surprising feature of the supertrace 
is that it can be calculated for any $n$, with or without boundaries
\cite{Gilkey:2002qd}. For example, the volume term in (\ref{twipvar})
does not depend on $\Phi$ and, therefore, is the standard
Euler density which is given by
\begin{equation}
{\rm \bf E}_n:=(4\pi)^{-\bar n}\textstyle\frac1{2^{\bar n}\bar n!}
\epsilon^{i_1...i_m}\epsilon^{j_1...j_n}R_{i_1i_2j_2j_1}...
R_{i_{n-1}i_nj_{n}j_{n-1}}
\label{Pengui}\end{equation}
(with $\bar n=n/2$) for $n$ even. ${\rm \bf E}_n=0$ for $n$ odd.

It is known \cite{Fradkin:1985ai,Barbon:1995zh} that for $\Phi =0$
the dual theories are quantum equivalent:
\begin{equation}
W_p(0)-W_{n-p-2}(0)=0 \,.\label{Phi0}
\end{equation}
By using this equation as an ``initial condition'' we can integrate
the variation (\ref{twipvar}) to obtain:
\begin{equation}
W_p(\Phi )-W_{n-p-2}(-\Phi )=(-1)^p \iM \Phi {\rm \bf E}_n \,.\label{Phidual}
\end{equation}

Further terms $\sum_p (-1)^p a_k (x,\Delta_\Phi^p)$ are of some
interest in mathematics. They can be calculated 
\cite{Gilkey:2002hk,Gilkey:2002ts} for 
$k\le n+2$ and any $n$.


\section{Conclusions}\label{sec10}
Here we present a short guide in this report. In other words, we
are going to answer the following question. What should one do if
one likes to calculate one-loop counterterms, anomalies, an
expansion term in the effective action, or something else which is
defined by the heat kernel expansion? The first step is to find
the bulk part of the variation of the classical action
(\ref{actquad}) and corresponding operator $D$. Next one has to
bring this operator to the canonical form (\ref{laplaceb}) with
the help of (\ref{D1}), (\ref{oab}) and (\ref{Eab}). However, one
can first consult sec.\ \ref{sec3}. Probably, relevant expressions
can be found there. If the problem in question contains neither
boundaries nor singularities, one can look in sec.\ \ref{sec4} for
an expression for the heat kernel coefficient, or for a reference,
or for a method. In the case of the boundaries one has to proceed
with sec.\ \ref{sec5}, in the case of singularities -- with sec.\
\ref{sec6}. Relations between the heat kernel coefficients and
quantum anomalies can be found in sec.\ \ref{sec7}. The results
going beyond the standard heat kernel expansion are collected in
sec.\ \ref{sec8}. In particular, in this section we explain how
one can extract leading non-localities from the effective action,
and what the heat kernel looks like if the background is
approximately covariantly constant (in this context we also consider
invariant operators on homogeneous spaces). Exact results for the
effective action which can be obtained with the help of the heat
kernel expansion are reviewed in sec.\ \ref{sec9}.

The Casimir energy is one of the most ``classical'' applications
of the heat kernel and zeta function technique (cf. 
\cite{Blau:1988kv}). It follows 
from the locality of the heat kernel expansion that the divergences
in the Casimir energy are given by volume and surface integrals
of some local invariants. Therefore, if the boundaries are being
moved in such a way that the boundary values of the background fields 
remain unchanged, the boundary contributions
to the divergences also remain constant. This leads to the well
known conclusion \cite{Bordag:2001qi} that there are no boundary
divergences in the Casimir force (which is defined, roughly speaking
as a variation of the Casimir energy under infinitesimal translations
of the boundary). Consequently, one can assign a unique value to
the Casimir force (see, e.g., \cite{Milton:2002vm}). This observation,
however does not mean that the quantum field theory on a manifold
with boundary is finite. In general, some surface counterterms
are required (at one-loop order they may be read off from 
sec.\ \ref{sec5}). Moreover, if the background field are non-trivial,
the boundary divergences will not be constant. Similar arguments
created certain scepticism towards reliability of the Casimir energy
calculation \cite{Graham:2002xq}. This point has not been settled
so far for a ``generic'' theory. We may add that in supersymmetric 
theories cancellations between divergences in the bosonic and
fermionic sectors appear if the boundary terms are considered
together with the volume terms \cite{Bordag:2002dg}, therefore,
separation of boundary and volume contributions is not always natural
for that theories. Renormalization of self-interacting theories
on manifolds with boundaries was considered in 
\cite{Symanzik:1981wd,Odintsov:1990ny,
McAvity:1993fq,Wiesendanger:1994mw,Barvinsky:1996dp}
where one can find further references. Some aspects of the
relationship between the Casimir energy calculations and the
heat kernel coefficients have been clarified recently by
Fulling \cite{Fulling:2003zx}.

Of course, not everything can be found in this report.
Several topics are very close to the subject of this review,
but are not included.
\begin{enumerate}
\item
The heat kernel expansion can be successfully applied to quantum field
theory at finite temperature \cite{Dowker:1978md}. A new interesting
development in this field is related to the so called non-linear
spectral problem \cite{Fursaev:2002vi} (see \cite{Fursaev:2001yu}
for an overview).
\item
The heat kernel expansion has interesting applications to
integrable models and, in particular, to the
KdV hierarchies (see \cite{Das:1989fn} for an elementary introduction).
\item
Recently some attention has been attracted 
\cite{Dowker:2000bi,Dowker:2000vm,Seeley:2001,Avramidi:2001aj}
to the so called N/D or Zaremba problem which appears when one
defines Neumann and Dirichlet boundary conditions on two
(intersecting) components of the boundary. It is unclear whether
this problem may have applications to quantum theory.
\item
Instead of considering asymptotics of the heat trace ${\rm Tr} (fe^{-tD})$
one can also consider an asymptotic expansion for individual matrix
elements of the heat kernel $\langle f_1, e^{-tD} f_2 \rangle$ (which are
called the heat content asymptotics since they remind short time
asymptotics of the total heat content in a manifold with the specific heat
$f_1$ and the initial temperature distribution $f_2$). Such asymptotics do
not contain negative powers of $t$. More details can be found in
\cite{Kirsten:2001wz}.
\item
Many results on the heat kernel asymptotics can be extended
to higher-order differential operators (see, e.g., 
\cite{Gilkey:1980,Gusynin:1990zt}) and to differential operators
in superspace \cite{Buchbinder:1986im}.
\end{enumerate}

Although quite a lot is already known about the heat kernel expansion,
many interesting problems still remain open. There are many 
opportunities to extend and generalise the results presented in this
report. This refers especially to the material of sec.\ 
\ref{sec5}-\ref{sec9} where one could add new types of the operators,
boundary conditions, geometries, and singularities. 
There is a completely new field of research related to the
heat kernel expansion where very little has been done so far.
This is 
an extension to non-commutative geometry. This  problem
is an especially intriguing one since one can expect very unusual
properties of the spectral functions because of very unusual properties
of corresponding field theories in the ultra violet asymptotics.

\section*{Acknowledgements}
I am grateful to my collaborators S.~Alexandrov, M.~Bordag,
T.~Branson, E.~Elizalde, H.~Falomir, P.~Gilkey, D.~Grumiller,
K.~Kirsten,
W.~Kummer, H.~Liebl, V.~Lyakhovsky, V.~Marachevsky,
Yu.~Novozhilov, M.~Santangelo, N.~Shtykov, P.~van~Nieuwenhuizen
and A.~Zelnikov who contributed in many ways to the material
presented in this report. I have benefited from enlightening discussions
with I.~Avramidi, A.~Barvinsky, G.~Esposito, D.~Fursaev, G.~Grubb,
A.~Kamenshchik, V.~Nesterenko and 
with my friends and colleagues at St.~Petersburg,
Leipzig and Vienna.

I am grateful to all readers who suggested their comments on the
previous version of the manuscript. I also thank the referee
for useful critical remarks.

This work has been supported by Project BO 1112/12-1 of the
Deutsche Forschungsgemeinschaft. I am grateful to E.~Zeidler
for his kind hospitality at the Max Planck Institute for 
Mathematics in the Sciences where a part of this work has
been done.


\lhead{REFERENCES}
\bibliographystyle{fullcream} 
\bibliography{hk}

\end{document}